\documentclass[preprint,aps,prc,showpacs,nofootinbib]{revtex4-1}
\usepackage{amsmath}
\usepackage{amssymb}
\usepackage{graphics}
\usepackage{epsfig}

\usepackage{subfigure}

\newcommand\nn{\nonumber} 
\newcommand\ba{\begin{eqnarray}}
\newcommand\ea{\end{eqnarray}} 
\newcommand\beq{\begin{equation}}
\newcommand\eeq{\end{equation}}

\begin{document}

%%%%%%%%%%%%%%%%%%%%%%%%%%%%%%%%%%%%
\title{Polarization phenomena in the reaction $e^+ + e^- \to p + \bar p +\pi^0$ in frame of  the non-resonant mechanism }

%%%%%%%%%%%%%%%%%%%%%%%%%%%%%%%%%%%%

%% \tnotetext[label1]{}
\author{G. I. Gakh}
\email{gakh@kipt.kharkov.ua}
\affiliation{\it National Science Centre, Kharkov Institute of
Physics and Technology, Akademicheskaya 1, and V. N. Karazin
Kharkov National University, Dept. of
Physics and Technology, 31 Kurchatov, 61108 Kharkov, Ukraine}

\author{M.I. Konchatnij}
 \email{konchatnij@kipt.kharkov.ua}
 \affiliation{\it National Science Centre, Kharkov Institute of
Physics and Technology, Akademicheskaya 1, and V. N. Karazin
Kharkov National University, Dept. of
Physics and Technology, 31 Kurchatov, 61108 Kharkov, Ukraine}

\author{N.P. Merenkov}
\email{merenkov@kipt.kharkov.ua}
\affiliation{\it National Science Centre, Kharkov Institute of
Physics and Technology, Akademicheskaya 1, and V. N. Karazin
Kharkov
National University, Dept. of
Physics and Technology, 31 Kurchatov, 61108 Kharkov, Ukraine}

\author{E. Tomasi--Gustafsson}
\email{egle.tomasi@cea.fr}
\affiliation{\it IRFU, CEA, Universit\'e Paris-Saclay, 91191
Gif-sur-Yvette, France}

\begin{abstract}
The dependence of the nucleon polarization  in the reaction $e^+ + e^- \to N + \bar{N} +\pi^0 $
over different invariant variables in frame of the non-resonant mechanism, has been derived.
The nucleon polarization is expressed in terms of six invariant complex amplitudes, assuming the conservation of the hadron electromagnetic
currents and the P-invariance of the hadron electromagnetic interaction. 
An inclusive experimental setup when the proton (or the antiproton) and the pion are detected
in coincidence is considered. Numerical estimations were performed for the so called normal polarization in the energy range from threshold up to $s=16$ GeV$^2$,  using a specific  parametrization of the nucleon electromagnetic form factors and
taking into account the unpolarized differential cross section of  the non-resonant mechanism, as previously calculated.
\end{abstract}

\maketitle

\section{Introduction}

The interation between electrons and nucleons is considered the 'cleanest' probe to investigate the non perturbative aspects of QuantumChromoDynamics (QCD), the theory of the strong interaction. 
In a previous work \cite{Gakh:2022fad} we considered the reaction
\begin{equation}\label{eq:1}
e^+(k_1) + e^-(k_2) \to N(p_1) + \bar{N}(p_2) +\pi^0(k), \ \ N = p,\,n
\end{equation}
% $e^+ + e^- \to p + \bar p +\pi^0 $ and
%$e^+ + e^- \to n + \bar n +\pi^0$
that is currently accessible at BESIII \cite{Yuan:2019zfo}. This reaction is the most simple 'inelastic  annihilation' reaction, and is very sensitive to the electromagnetic structure of the hadron current. Similarly to the crossing symmetry related reactions induced by antiprotons, that will be investigated at PANDA (FAIR) \cite{PANDA:2021ozp} and by pions, that is object of study at HADES \cite{HADES:2020kwd}, these experiments bring strong constraints on nucleon models. At electron accelerators, as JLAB, MAMI, ELSA,  huge experimental programs are based on  the reaction $e^-+N\to e^-+N+\pi$  to determine the properties of baryon resonances and transition electromagnetic form factors.

The general formalism for the analysis of these processes was derived in Ref. \cite{Gakh:2022fad}, assuming  the conservation of the hadron electromagnetic currents and the P-invariance of the hadron electromagnetic interaction. The matrix element, which is the convolution of lepton and hadron currents,
in these conditions can be expressed by six independent complex invariant amplitudes.  This statement remains true also for different possible charge states of the pion, nucleon and antinucleon. 
The differential cross sections and different polarization observables including single-spin beam asymmetry, the nucleon (antinucleon) polarization  and
the correlation between electron and nucleon polarization states
can be expressed, in general case, in terms of the bilinear combinations of these invariant amplitudes. We investigated in details the contribution of the continuum (non-resonant mechanism, see see Fig.\,1) to different double and single differential distributions over invariant variables in the case of unpolarized particles. The key moment in these investigation is the choice of the electromagnetic form factors. 

In the present work this formalism is extended to polarization observables. 

The earlier study of the phase space in terms of invariant variables  and the 
derived analytical expressions for the corresponding invariant amplitudes allow to calculate also different polarization observables. The importance of polarization observables can not be overestimated, as shown by the proton form factor measurements in the spacelike region. For the  elastic electron proton scattering  the method earlier suggested in Refs. \cite{Akhiezer:1968ek,Akhiezer:1974em}, that requires a longitudinally polarized electron beam and the measurement of the transverse polarization of the few GeV recoil proton (or the asymmetry with a transversely polarized target) could be systematically applied only with the advent of high duty cycle, highly polarized electrons beams at Jefferson Laboratory \cite{Puckett:2017flj}. The precise determination of the electric to magnetic form factor ratio, achieved by the JLab-GEP collaboration showed that the electric and magnetic distributions in the proton are different, contrary to what was previously assumed and gave rise to a deep revision of nucleon models.

In the timelike region, form factors which are of complex nature  and the study of   polarization phenomena is necessary for their full determination. 

\begin{figure}
\centering
\includegraphics[width=0.4\textwidth]{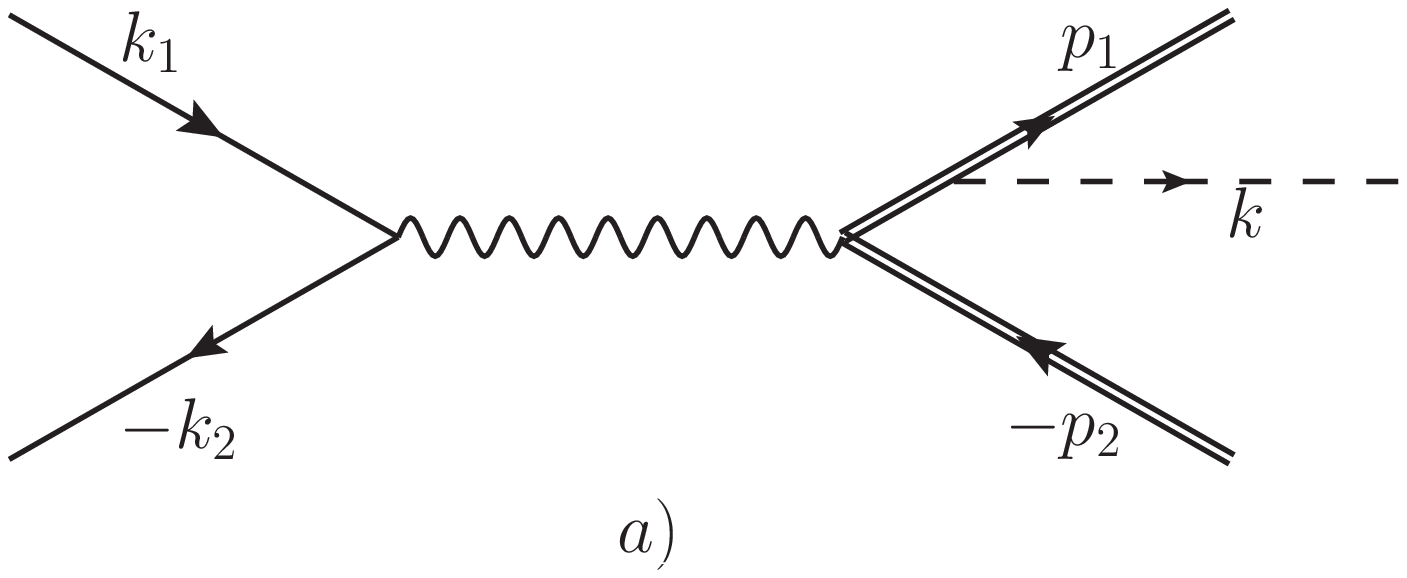}
\includegraphics[width=0.4\textwidth]{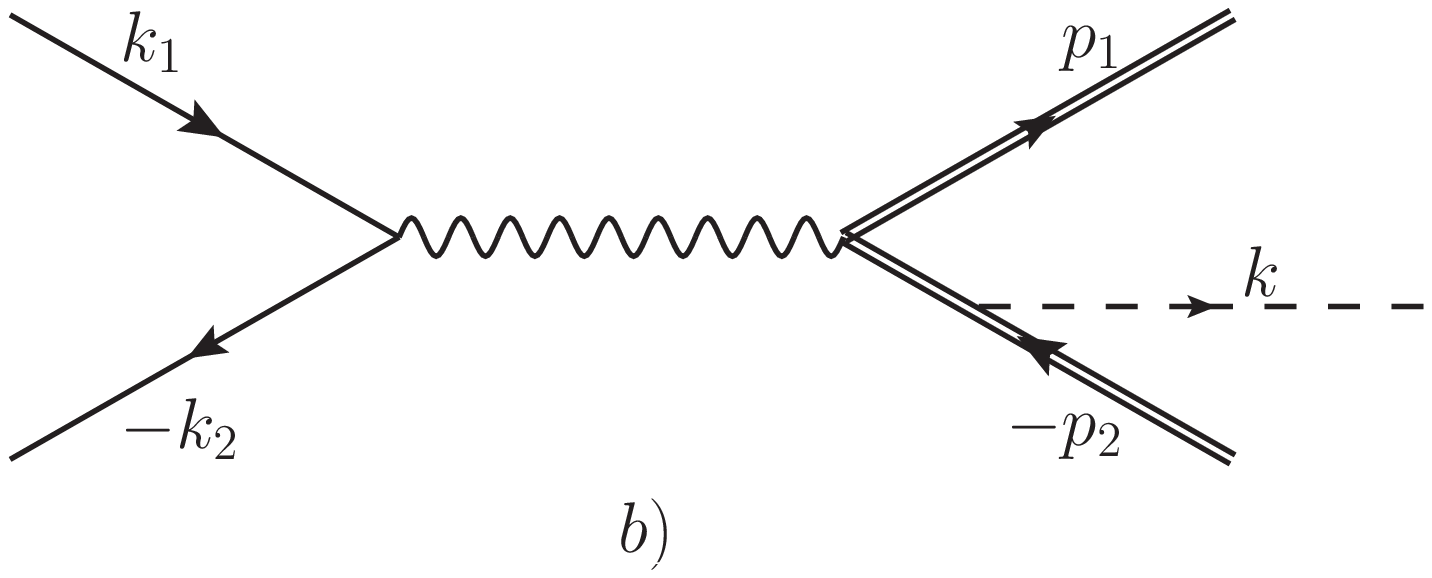}
%\hspace{1cm}
%\includegraphics[width=0.3\textwidth]{gena2019c.eps}
 \parbox[t]{0.9\textwidth}{\caption{The simplest Feynman diagrams which describe the continuum (non-resonant) contribution to the process (\ref{eq:1}); (a)$-$ with intermediate nucleon,
 (b)$-$ with intermediate antinucleon.}\label{fig.1}}
\end{figure}

In the present paper we consider polarization phenomena in the approximation of the non-resonant mechanism (see Fig. \ref{fig.1}) widely using the results of \cite{Gakh:2022fad} concerning the final particles phase space.  

The hadron tensor for a longitudinally polarized electron beam and polarized nucleon  is explicitely derived. In these conditions we can investigate the single-spin effects due to the polarization of the electron beam (single-spin beam asymmetry) or to the polarization of the nucleon, as well the double-spin observables, i.e.,  the correlation between electron and nucleon polarizations.

\section{Formalism}

In our analysis we consider three different possible independent polarization states of the nucleon which described in \cite{Gakh:2022fad} as longitudinal $S^L,$ transverse $S^T$ and normal
$S^N$. Our calculation follows different steps. First we obtain the full differential  cross section over four invariant variables
$$s_1=(k+p_1)^2,\,s_2=(k+p_2)^2,\, t_1=(k_1-p_1)^2,\,t_2=(k_2-p_2)^2,$$
 accounting for also the terms depending on the nucleon spin states. Then we perform the analytical integration over two variables using the sets of limits defined in \cite{Gakh:2022fad} and find the spin-dependent double differential distributions over the pairs $(s_1,\,s_2),\,(t_2,\,s_1)$ and $(s_{12},\,s_1),\,\, s_{12}=(p_1+p_2)^2$. These distributions are derived not neglecting any particle mass (even the electron one).

The corresponding polarization of the nucleon (longitudinal, transverse or normal)
is defined by the ratio of the spin-dependent part of the cross section to the spin-independent one. Performing one more integration we obtain the differential cross sections  over one invariant variable
$(s_1,\, s_{12}, t_2)$ and then the corresponding single- and double-spin distributions for the nucleon polarizations.

The full differential cross section of the process (\ref{eq:1}) can be written, in terms of the convolution of the leptonic and hadronic tensors
and the final particle phase space, as follows
\begin{equation}\label{eq:6difsec4}
d\,\sigma =\frac{\alpha^2}{8\pi^3\,q^6}L^{\mu\nu}\,H_{\mu\nu}\,d\,R_3, \ \ d\,R_3
=\frac{d^3p_1}{2\,E_1}\,\frac{d^3p_2}{2\,E_2}\,\frac{d^3k}{2\,E}\,\delta(k_1+k_2-p_1-p_2-k),
\end{equation}
where $E_1\,(E_2)$ is the nucleon (antinucleon) energy and $E$ is the pion one. We suggest that there is the possibility to choose the coordinate system in such a way that one of the final 3-momentum will belong to definite plane, for example to the $zx$ one. Such choice, in fact, corresponds to the integration over one azimuthal angle.
In this case, we can  express the phase space in terms of the invariant variables, namely \cite{Byckling:1971vca}:
\begin{equation}\label{eq:phasespace5}
d\,R_3=\frac{\pi}{16(s-2m_e^2)}
\frac{dt_1\,dt_2\,ds_1\,ds_2}{\sqrt{-\Delta}},
\end{equation}
where $\Delta$ is the Gramian determinant (see \cite{Gakh:2022fad} for the details). Both, the convolution of the tensors and the Gramian determinant can be expressed through the $q^2= s$ and chosen invariants for description of the phase space. Moreover, the condition $-\Delta >0$ alone sets the limits of the variation for all the chosen invariant variables.

The spin-dependent part of the hadronic tensor, in the general case, has been obtained in \cite{Gakh:2022fad} in terms of the invariant amplitudes. We need to calculate its convolution with the leptonic tensor for the non-resonant mechanism, which contributes to  single-spin effects. Using the connection between the Dirac and Pauli form factors, $F_1$ and $F_2$  and the  corresponding invariant amplitudes (see Eqs.\,(26) in \cite{Gakh:2022fad}) we have

\begin{eqnarray}
L^{\mu\nu}\,H^{(s)}_{\mu\nu}(S) &=& \frac{4\,g^2_{\pi^0 N\bar{N}}\,Im[F_1\,F_2^*]}{M (k\cdot q - p\cdot q - q^2)[(k\cdot q)^2 - (p\cdot q)^2]^2}\Big[ 4k\cdot q( k\cdot q k_1\cdot p - k\cdot k_1 p\cdot q)
(S_1(k_1kqS)+\nn\\
&& +S_2(k_1pqS))+S_3(kpqS)\Big], \ p=p_1-p_2, \ (abcd) =  \varepsilon^{\mu\nu\lambda\rho}\, a_\mu b_\nu c_\lambda d_\rho,
\label{eq:singlespin6}
\end{eqnarray}
where $S$ is the nucleon spin 4-vector, $M\,(m)$ is the nucleon (neutral pion) mass and

\begin{eqnarray}
S_1&=&q^2(q^2-p^2-4M^2)-2k\cdot q(q^2-2M^2) +(k\cdot q)^2 -(p\cdot q)^2 -4M^2\,p\cdot q, \nn\\
S_2&=&(p\cdot q-k\cdot q)^2 +m^2q^2,\nn\\
S_3 &=&(k\cdot q-p\cdot q)\Bigl \{ p\cdot q(q^2+p\cdot q)\left [ 4(k_1\cdot k)^2+m^2 q^2\right ] + \nn\\
&&
+(k\cdot q)^2\left [ 4 (k_1\cdot p)^2+4k_1\cdot p k_1\cdot q +p^2 q^2\right ]-\nn\\
&&-k\cdot q \left [q^2 k\cdot p (q^2 +2p\cdot q) + 4k_1\cdot k [k_1\cdot q p\cdot q +k_1\cdot p(q^2 +2p\cdot q) ] \right ]\Bigr\}-\nn\\
&&
-q^2 k\cdot q \Bigl \{( k\cdot q)^3 + (p\cdot q)^3+(k\cdot q)^2(4M^2-2q^2- p\cdot q)+m^2q^2p\cdot q +
\nn\\
&&
+k\cdot q \left [q^2(q^2+p^2-4M^2)-(p\cdot q)^2-4M^2 p\cdot q\right ]\Bigr \}.
\label{singlespin7}
\end{eqnarray}
All the scalar products in Eq. (\ref{eq:singlespin6}) can be expressed via the invariant variables.
Note also that
\begin{equation}\label{eq:FG7}
Im[F_1\,F_2^*] = \frac{Im[G_EG_M^*]}{\tau-1}, \ \ \tau=\frac{q^2}{4M^2}.
\end{equation}
Eq. ( \ref{eq:FG7})  shows  that the  nucleon polarization due to the non-resonant mechanism gives information about the phase difference of the electric and magnetic form factors ($G_E$ and $G_M$ are the commonly used Sachs form factors \cite{PhysRev.126.2256}, linearly related to $F_1$ and $F_2$).

Let us consider the effect originated by the longitudinal polarization of the nucleon when the direction of its three-vector polarization (in the nucleon rest frame) is  along ${\bf{n}}=-{\bf{q}}/|{\bf{q}}|$. In this case \cite{Gakh:2022fad}
\begin{equation}\label{eq:SL8}
S_\mu=S^L_\mu = \frac{p_1\cdot q \,p_{1\mu}-M^2\,q_\mu}{M\,K}, \ \ K= \sqrt{(p_1\cdot q)^2-M^2q^2},
\end{equation}
and the r.h.s. of Eq.\,(\ref{eq:singlespin6}) becomes very simple, namely
\begin{equation}\label{eq:singlLong9}
L^{\mu\nu}\,H^{(s)}_{\mu\nu}(S^L) = -\frac{4\,g^2_{\pi^0 N\bar{N}}Im[F_1\,F_2^*](s_1+s_2-2M^2)(k_1\,k_2\,p_1\,p_2)}{K(s_1-M^2)^2(s_2-M^2)}\,I(s_1,s_2,t_1,t_2),
\end{equation}
with 
$$I(s_1,s_2,t_1,t_2)=2M^4+(s-2m_e^2)(s_1+s_2)-2s_1s_2+2(s_1t_1+s_2t_2)-2M^2(s+t_1+t_2-2m_e^2).$$

If the nucleon is polarized in such a way that the direction of the three-vector polarization (in the nucleon rest frame) is along 
$[{\bf q}\times[{\bf k}\times{\bf{q}}]]/{\left|[{\bf q}\times[{\bf k}\times{\bf{q}}]]\right|}$, we have
\begin{equation}\label{eq:ST10}
S_\mu =S^T_\mu = \frac{(q^2\,k\cdot p_1-q\cdot p_1\,k\cdot q)\,\tilde{p}_1^{\mu} + [(q\cdot p_1)^2-q^2M^2]\,\tilde{k}^\mu}{K\, N}\,, \ \tilde{a}^{\mu}= a^\mu -\frac{a\cdot q\,q^\mu}{q^2},
\end{equation}
with 
$$N=\sqrt{-(\mu k p_1 q)(\mu k p_1 q)}\,, \ N^2=2 k\cdot q\,k\cdot p_1\,q\cdot p_1-q^2(k\cdot p_1)^2-M^2(k\cdot q)^2-m^2(q\cdot p_1)^2+q^2 M^2 m^2$$ \,.
In this case
\begin{equation}\label{eq:singlTrans11}
L^{\mu\nu}\,H^{(s)}_{\mu\nu}(S^T) = \frac{g^2_{\pi^0 N\bar{N}}Im[F_1\,F_2^*](s_1+s_2-2M^2)(k_1\,k_2\,p_1\,p_2)}{M K N\,(s_1-M^2)^2(s_2-M^2)^2}\,J(s_1,s_2)\,I(s_1,s_2,t_1,t_2),
\end{equation}
and 
$$J(s_1,s_2)=3M^6-M^4(s+s_1+4s_2)+M^2[s(s_1+s_2)+s_2^2-3m^2 s]+(s-s_2)(m^2s-s_1s_2).$$
In terms of invariant variables the following relations hold 
\begin{eqnarray}
K^2&=&\frac{1}{4}\big[(s-s_2)^2-2M^2(s+s_2)+M^4\big],\nn\\
N^2&=&\frac{1}{4}\Bigl \{ -2M^6+M^4(s+s_1+s_2+m^2)-M^2[s(s_1+s_2)-2s_1s_2-m^2(2s-s_1-s_2)]+\nn\\
&&\ \ \ \ \ + m^2[s(s_1+s_2)+s_1s_2-s^2]-m^4s +s_1s_2(s-s_1-s_2) \Bigr \}.
\label{eq:K2N2}
\end{eqnarray}
In both cases the convolution of the leptonic and hadronic tensors contains the  product $(k_1\,k_2\,p_1\,p_2)\,I(s_1,s_2,t_1,t_2).$ Therefore,  all the dependence on the  variables $t_1$ and $t_2$ of the spin-dependent part of the full differential cross section is contained in the factor $I(s_1,s_2,t_1,t_2)$ because the factor $(k_1\,k_2\,p_1\,p_2)$ just cancels the Gramian determinant in the phase space.

To calculate the corresponding double differential $(s_1,\,s_2)-$distribution one needs to integrate with respect $t_1$ and $t_2$. This results in: 
\begin{equation}\label{eq:Intt1t212}
\int\limits_{t_{1-}}^{t_{1+}} d\,t_1\int\limits_{t_{2-}}^{t_{2+}}\,dt_2\,I(s_1,s_2,t_1,t_2) =0.
\end{equation}
For $t_{1\pm}$ and $t_{2\pm}$ see Eqs.\,(19) and (20) in \cite{Gakh:2022fad}.
The $(s_1,\,s_2)-$distribution for the longitudinal and transverse nucleon polarizations vanishes,  but this is not the  case for the double $(t_2,\,s_1)-$distribution. Of course, after the integration over $t_2$, this last distribution also vanishes.

Here we need to note that the factor $(k_1\,k_2\,p_1\,p_2)$ may be expressed in terms  of the used invariant variables up to the sign only. To understand this problem let us consider the 
c.m.s. of the the initial particles with  the $z$ axis along the direction ${\bf{k}_1}$  and ${\bf{p}_1}$ in the plane $(x,\,z)$. In this system
$$(k_1\,k_2\,p_1\,p_2) = \frac{s}{2}|\bf{p}_1|\cdot|\bf{p}_2|\sin{\theta_1}\,\sin{\theta_2}\,\sin{\phi},$$
where $\theta_1\,(\theta_2)$ is the polar angle of the nucleon (antinucleon) and $\phi$ is the azimuthal angle of the antinucleon. We can express explicitly $\cos{\phi}$ in terms of invariant variables, giving the quantity $\sin{\phi}$ up to the sign only.

Let us consider the normal nucleon polarization
\begin{equation}\label{eq:SN13}
S_\mu=S_\mu^N = \frac{(\mu k p_1 q)}{N}.
\end{equation}
In this case, the convolution of the tensors is more complicated and we report its expression in the limit $m_e\to 0$:
\begin{eqnarray}
L^{\mu\nu}\,H^{(s)}_{\mu\nu}(S^N) &=& \frac{g^2_{\pi^0 N\bar{N}}Im[F_1\,F_2^*]}{Z}\Big\{(2M^2-s_1-s_2)I(s_1,s_2,t_1,t_2)\big[(3M^4+M^2(s_1-3s_2)-  \label{eq:LH}\\
 &&-s_1s_2+m^2s)C_1+(m^2s-s_1s_2+M^2(s_1+s_2)-M^4)C_2\big] +4(s_2-M^2)\,C_3\,C_4\Big\},\nn
 \end{eqnarray}
 with 
 \begin{eqnarray}
&Z=&4 M\,N(s_1-M^2)^2(s_2-M^2)^2(s+M^2-s_2),\nn\\
&C_1=&(s_1 + s_2) [s_1 (s_2 - t_1) - s_2 t_2] + s [s_1 ( t_1 - t_2 -2 s_2) + s_2 (t_2-t_1)] +
\nn\\
 &&
+m^2s (2 s - s_1 - s_2 + 2 t_1 + 2 t_2) +2 M^6  -M^4 (2 s + s_1 + s_2 + 2 t_1 + 2 t_2) + \nn\\
 &&
+ M^2 [-2 m^2 s - 2 s_1 s_2 + 2 s (s_1 + s_2) + 3 s_1 t_1 + s_2 t_1 + s_1 t_2 + 3 s_2 t_2],\nn\\
 &C_2=&s_1^2 (t_1-s_2) + (2 s - s_2) [s (t_1 - t_2) + s_2 t_2] +  s_1 [s_2^2 + s (t_2-3 t_1) + s_2 (t_2-t_1)] +
 \nn\\
 &&
 +m^2 s (s_1 - s_2 + 2 t_1 - 2 t_2)+M^2 [-2 s (s_1 - s_2 + 2 t_1 - 2 t_2) - (s_1 - s_2) (t_1 + t_2)] +\nn\\
 &&
 +M^4 (s_1 - s_2),\nn\\
 &C_3=&s_1s_2(s_1+s_2-s) +m^2[m^2s-M^4 +M^2(s_1+s_2-2s)+s^2-s_1s_2-s(s_1+s_2)]+ \nn\\
 &&+2M^6-M^4(s+s_1+s_2)+M^2[s(s_1+s_2)-2s_1s_2],\nn\\
 &&C_4=m^2[2 s (s_1-M^2) (M^2 + s - s_2)]+ 2M^8 -2 M^6\, [3 s + 2 (t_1 + t_2)]+\nn\\
 && 2 M^4 [-2 s_1 s_2 + 2 s_1 t_1 + t_1^2 + 2 s_2 t_2 + 2 t_1 t_2 + t_2^2 + s (3 s_1 + 2 s_2 + 3 t_1 + t_2)]-\nn\\
 &&-M^2 [(2 s^2 (t_1 - t_2) + 4 (t_1 + t_2) (s_1 (t_1-s_2) + s_2 t_2) +   s (s_1^2 + s_2^2 + 7 s_1 t_1 + s_2 t_1 + 2 t_1^2 + \nn\\
 &&
  +3 s_1 t_2 +5 s_2 t_2 - 2 t_2^2)]+s^2 (s_1 + s_2) (t_1 - t_2) + 2 [s_1 (t_1-s_2 ) + s_2 t_2]^2 +\nn\\
 &&s [s_1^2 (3t_1-2 s_2) - s_1 s_2 (t_1 - 5 t_2) + 2 s_1 t_1 (t_1 - t_2)+ s_2 t_2(s_2 + 2 t_1 - 2 t_2)].
\label{eq:singlNorm14}
\end{eqnarray}
Note that in this case one has not problem with the ambiguity, so further we focus on the normal polarization.
%%%%%%%%%%%%%%%%%%
\section{Single spin asymmetry}
%%%%%%%%%%%%%%%%%%
The single-spin beam asymmetry is defined by the convolution of the spin-dependent antisymmetrical part of the leptonic tensor:
\begin{equation} \label{eq:LT}
L_{\mu\nu}=L_{\mu\nu}^{(s)} + L_{\mu\nu}^{(a)}, \ \ L_{\mu\nu}^{(s)} = -q^2\,g_{\mu\nu} +2(k_{1\mu}k_{2\nu}+k_{1\nu}k_{2\mu}), \ \  L_{\mu\nu}^{(a)} = 2\,i\,m_e(\mu\nu \eta q),
\end{equation}
where $m_e$ is the electron mass and $\eta$ is the four-vector of its longitudinal polarization,
$$q=k_1+k_2, \ \ (\mu\nu \eta q)= \varepsilon^{\mu\nu\lambda\rho}\,\eta_\lambda q_\rho, \ \ \varepsilon^{0123} = +1,$$
and the antisymmetrical spin-independent part of the hadronic tensor:
\begin{equation} \label{eq:HT}
H_{\mu\nu} = \frac{1}{2}\bigg(H_{\mu\nu}^{(s)}(0)+H_{\mu\nu}^{(a)}(0)\bigg) + H_{\mu\nu}^{(s)}(S) + H_{\mu\nu}^{(a)}(S).
\end{equation}
where we use the same notations as in \cite{Gakh:2022fad}.

Both spin-independent and spin-dependent parts of the hadronic tensor are defined in \cite{Gakh:2022fad} in general case in terms of bilinear combinations of invariant amplitudes and corresponding independent tensor structures. Note, that for non-resonant mechanism the spin-independent antisymmetrical part vanish: $H_{\mu\nu}^{(a)}(0)=0$.  That is why
in present paper we do not consider the single-spin beam asymmetry. However  such situation is not general. For example, taking into  account the decay of  the virtual photon into a real $\pi^0$ and a virtual vector meson $V^0$ which then interacts with the real nucleon-antinucleon pair (see Fig.\,2), leads to a nonzero value of $H_{\mu\nu}^{(a)}(0)$ and, consequently, to nonzero single-spin beam asymmetry.

\begin{figure}
\centering
%\includegraphics[width=0.4\textwidth]{gena2019a.eps}
%\includegraphics[width=0.4\textwidth]{gena2019b.eps}
%\hspace{1cm}
\includegraphics[width=0.6\textwidth]{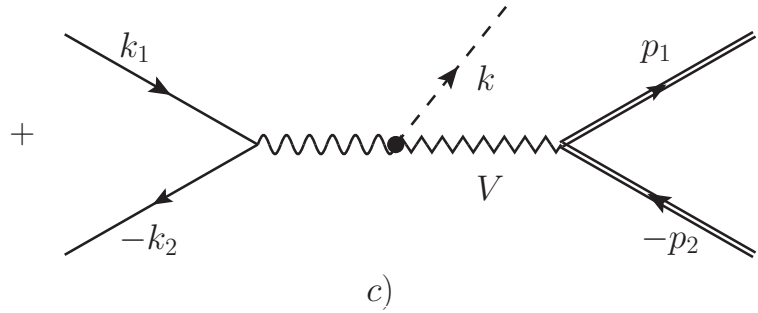}
 \parbox[t]{0.9\textwidth}{\caption{The Feynman diagrams which describe the decay $\gamma^*\to \pi^0 +V^0$ with subsequent transition $V^0\to N + \bar{N}$.}\label{fig.2}}
\end{figure}

Thus, the considered single-spin effect arises due to the nucleon polarization and is defined by convolution of the symmetrical spin-independent part of the leptonic tensor and symmetrical spin-dependent part of the hadronic one
$$L^{(s)\mu\nu}\,H^{(s)}_{\mu\nu}(S).$$ The double-spin effect is defined by
$$L^{(a)\mu\nu}\,H^{(a)}_{\mu\nu}(S).$$
In this paper we study only the single-spin effects due to the nucleon polarization.
%%%%%%%%%%%%%%%%%%%%%%%
\section{Double differential distributions}
%%%%%%%%%%%%%%%%%%%%%%
We consider the double differential distributions $(t_2,\,s_1),\, (s_1,\,s_2)$ and $(s_1,\,s_{12})$. We analyze the spin-dependent part of the cross section $d\,\sigma^N$ and the nucleon normal polarization defined as
\begin{equation}\label{eq:defP15}
P^{N} = \frac{d\sigma^{N}}{d\sigma},
\end{equation}
where $d\sigma$ is the unpolarized differential cross section. We obtained analytical expressions for these distributions but we list below only the $(s_1,\,s_2)$ one
(see Eq.\,(\ref{eq:s1s2norm16})). The $(s_1,\,s_{12})-$distribution can be derived from it by simple algebraic exercize and the expression for the $(t_2,\,s_1)-$distribution is very lengthy to be given in this paper.

The corresponding numerical results are plotted in Figs.\,4,\,5 and 6 for both, $\pi^0\,p\,\bar{p}-$ and $\pi^0\,n\,\bar{n}-$channels using the dimensionless invariant variables
$x_4=t_2/s,\,x_{1,2} =s_{1,2}/s,\, x_{12}=s_{12}/s.$

Note that nucleon polarization depends on the single combination of electromagnetic form factors $Im[G_E\,G_M^*]$ that is proportional to
$\sin{(Arg[G_E\,G_M^*])}$. It means that the corresponding measurements
have to probe, in principle, the phase difference between electric and magnetic form factors ($Arg[G_E\,G_M^*]$) for both, proton and neutron. This phase difference depends strongly  on parametrization of the form factors. In Fig.\,3 we show the dependence of the  $Arg[G_E\,G_M^*]$ on $q^2$ for two different choices of form factors, the one used in this paper and the one labeled in  \cite{Gakh:2022fad} as the "new" version \cite{Bijker:2004yu}. Looking at these plots one can conclude that the predictions for the nucleon polarizations in the processes (\ref{eq:1}) depend radically on the form factor choice, what  increases the  interest of their measurements.

\begin{figure}
\centering
\includegraphics[width=0.4\textwidth]{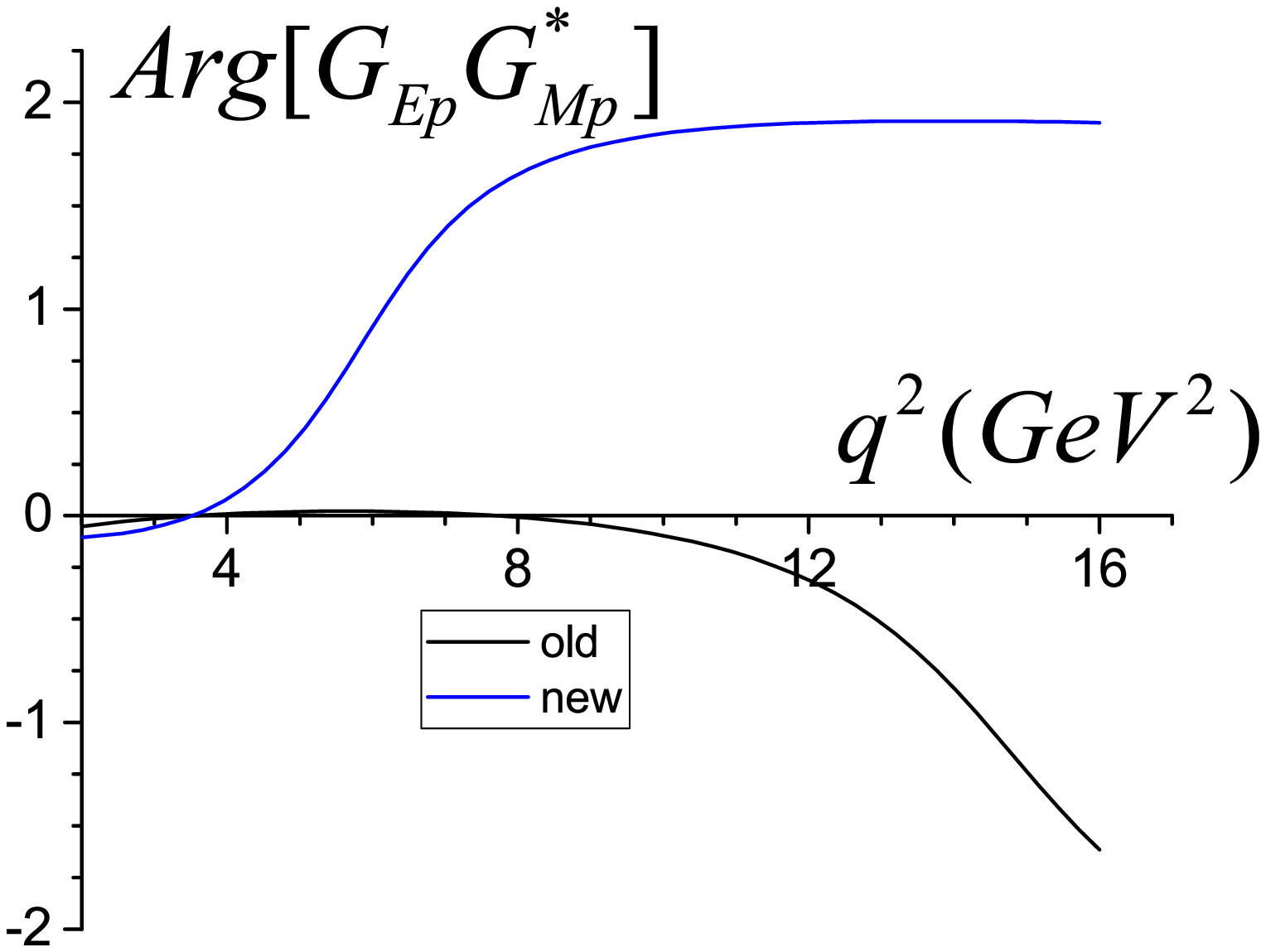}
\includegraphics[width=0.4\textwidth]{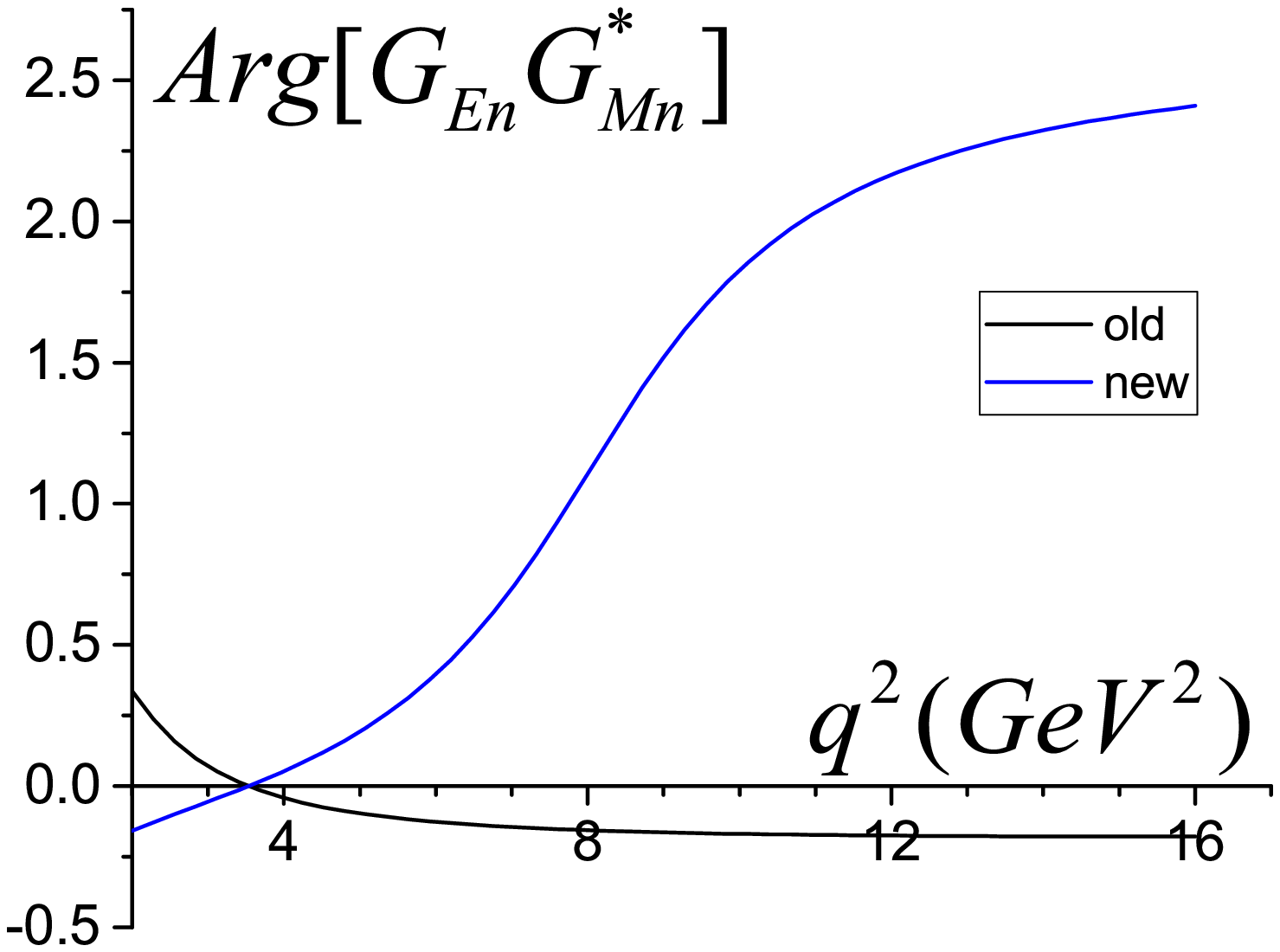}

 \parbox[t]{0.9\textwidth}{\caption{Phase difference $Arg[G_E\,G_M^*]$ (in radian) for two different parameterizations of the electric and magnetic nucleon form factors.}\label{fig.3}}
\end{figure}
To decrease the number of curves in Figures,  we make the choice of using the nucleon electromagnetic form factors labeled in \cite{Gakh:2022fad} as the "old" version
 (only in the next section we show plots of the polarization $P^N$ for single differential distributions over $x_4,\,x_1$ and $x_{12}$ calculated by the "new" version). Let us remind that we consider the non-resonant contribution.

\begin{figure}
\centering
\includegraphics[width=0.22\textwidth]{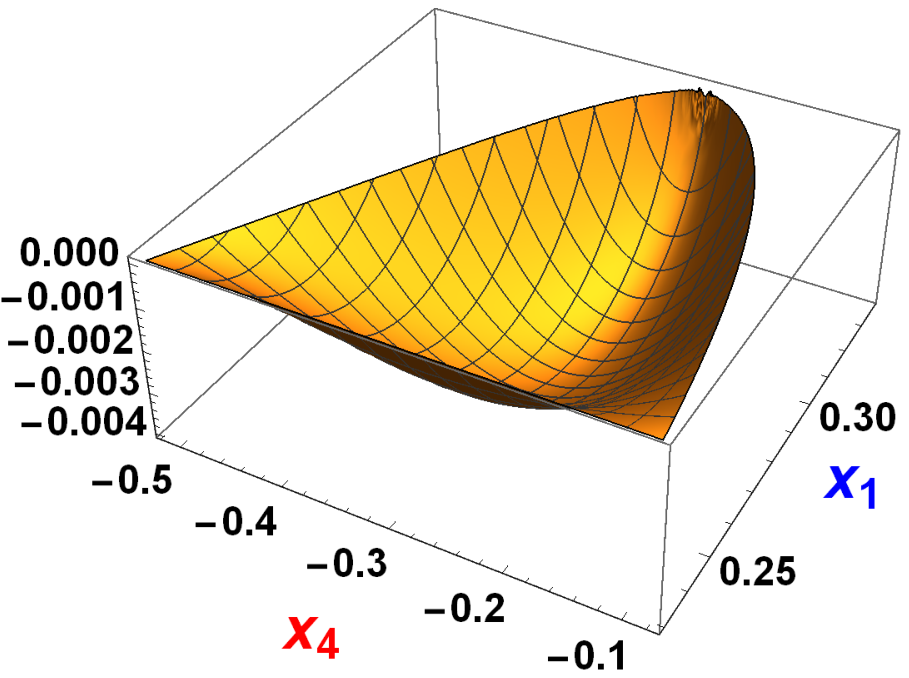}
\includegraphics[width=0.22\textwidth]{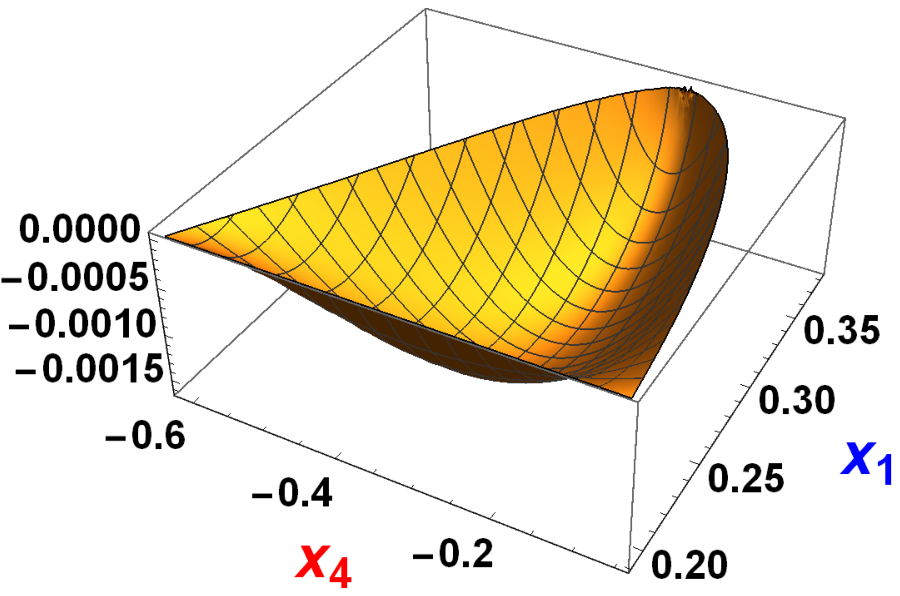}
\includegraphics[width=0.22\textwidth]{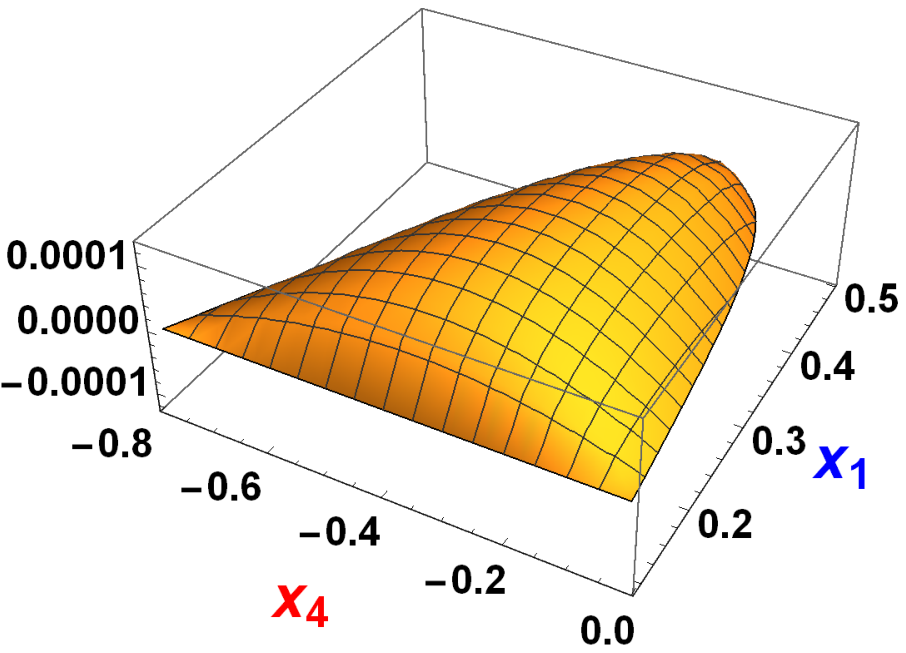}
\includegraphics[width=0.22\textwidth]{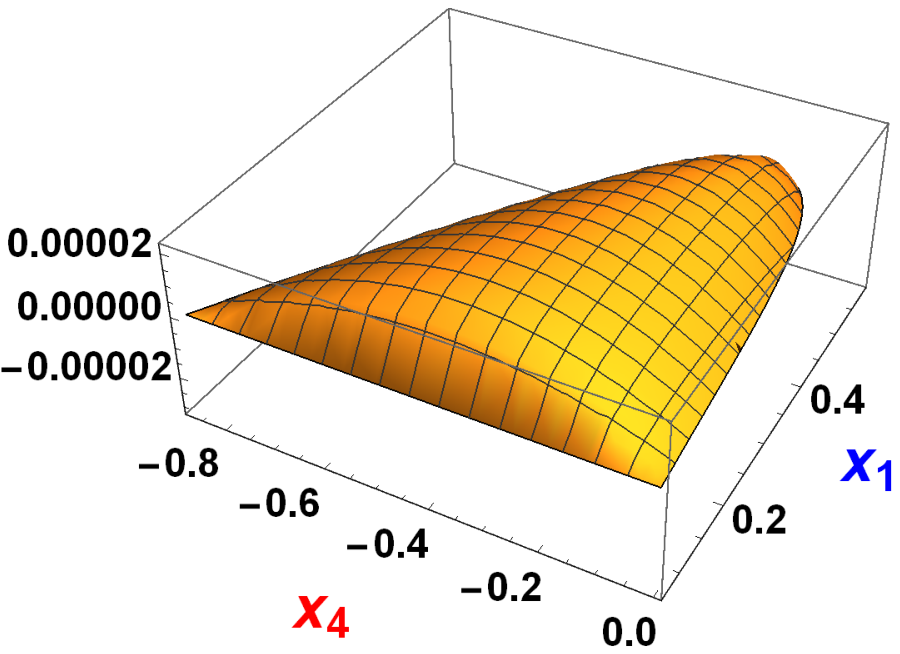}

\vspace{0.3cm}
\includegraphics[width=0.22\textwidth]{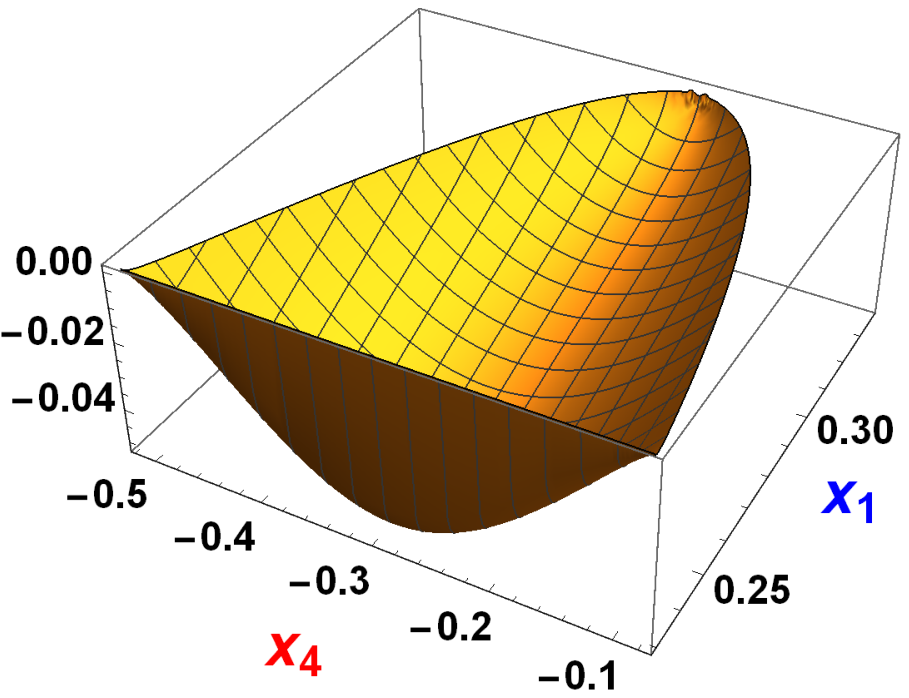}
\includegraphics[width=0.22\textwidth]{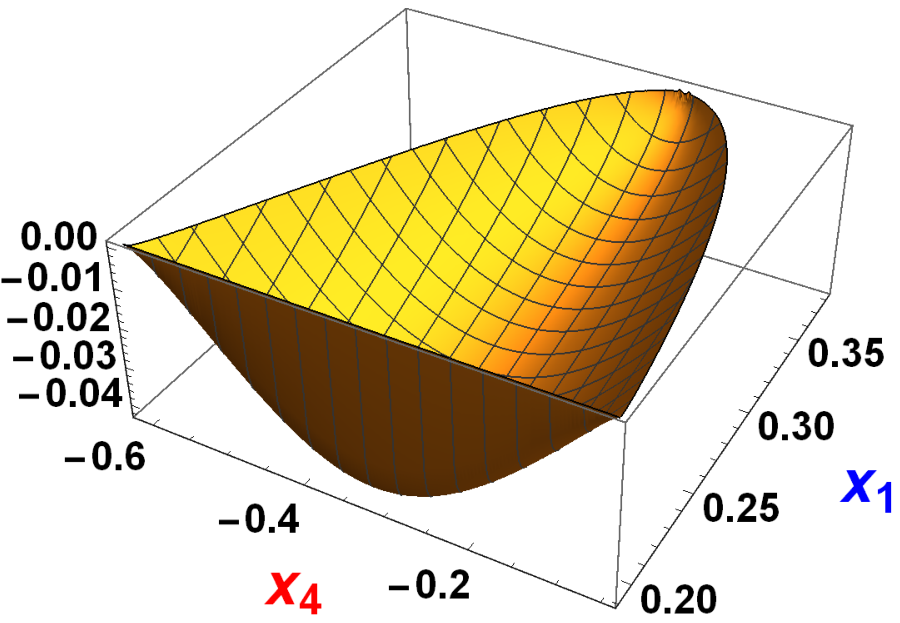}
\includegraphics[width=0.22\textwidth]{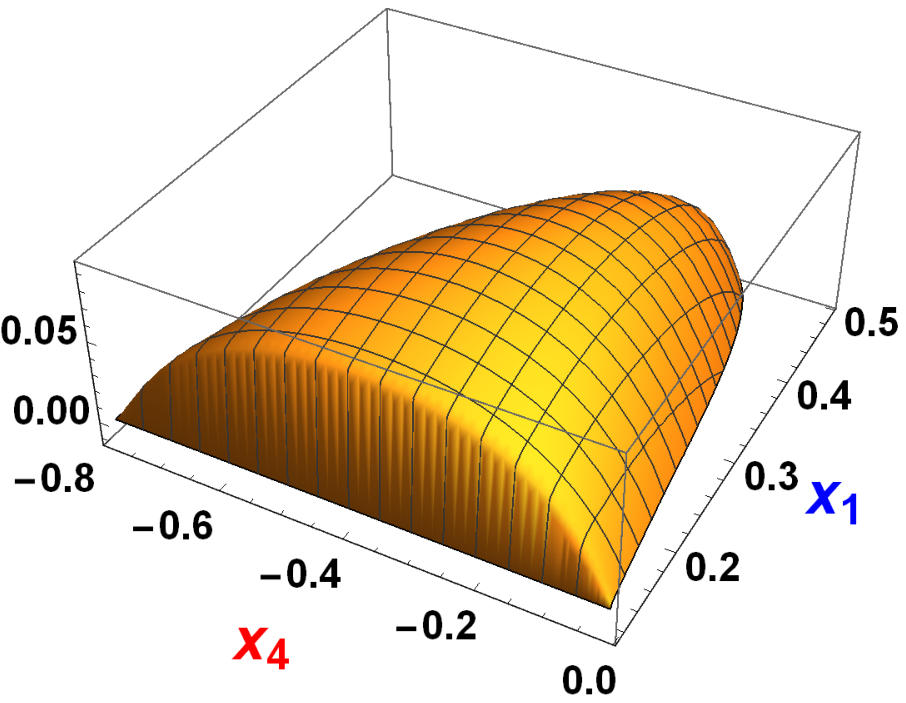}
\includegraphics[width=0.22\textwidth]{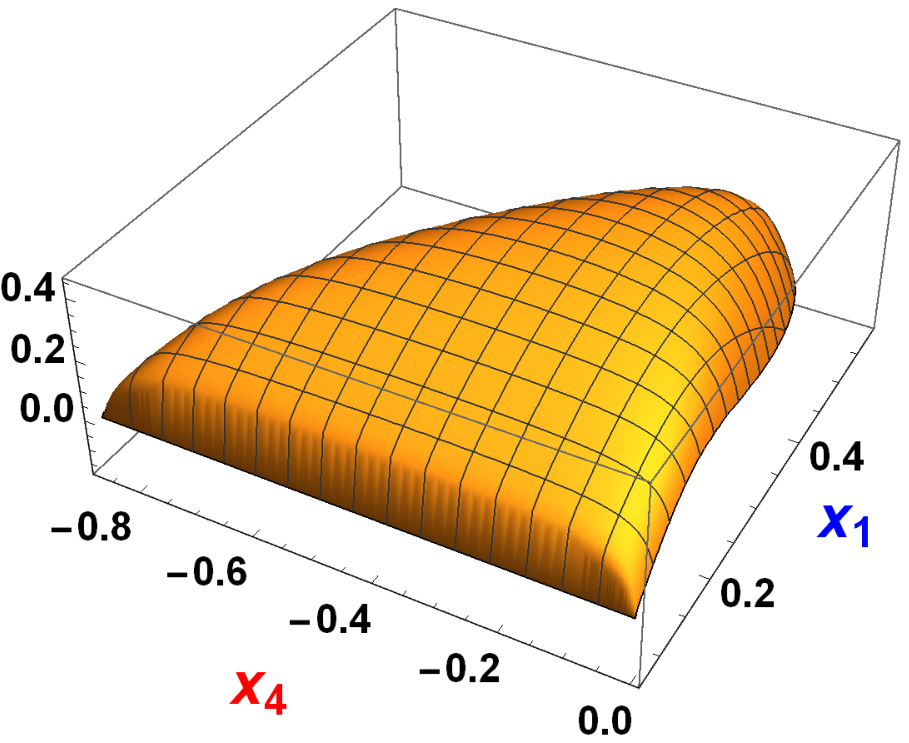}

\vspace{0.3cm}
\includegraphics[width=0.22\textwidth]{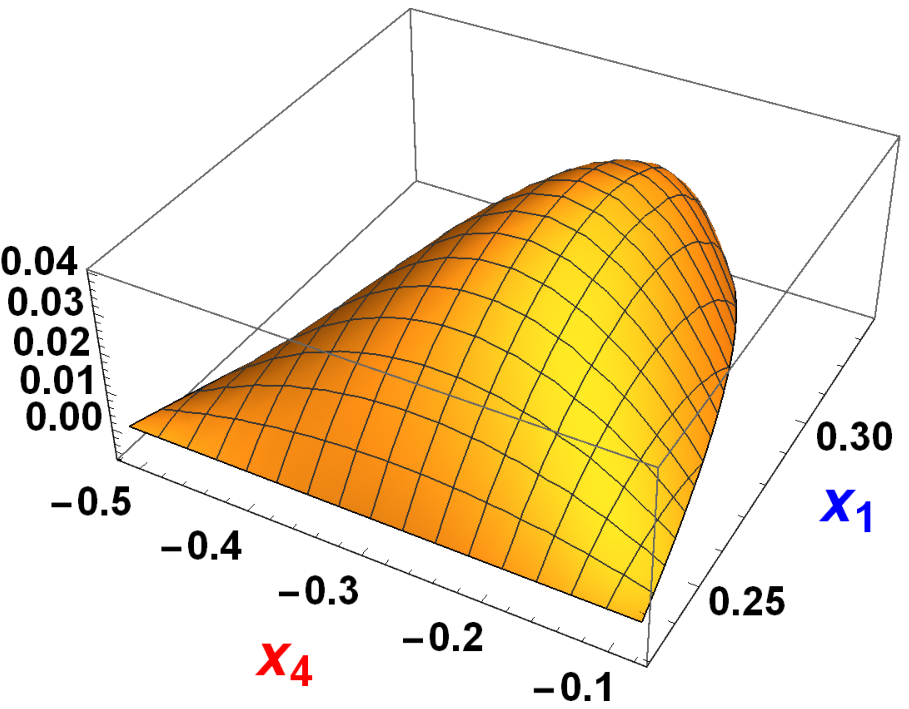}
\includegraphics[width=0.22\textwidth]{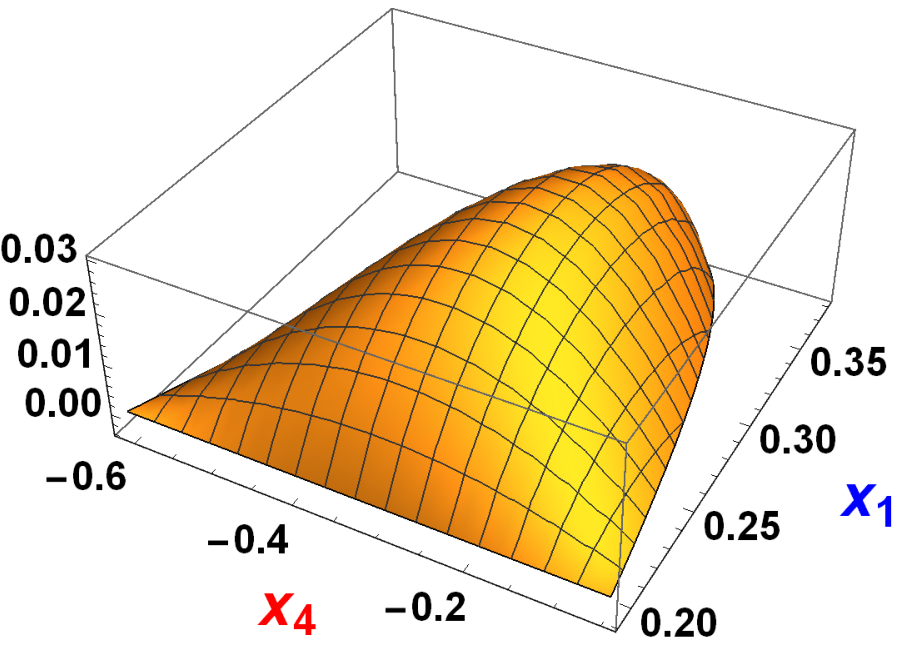}
\includegraphics[width=0.22\textwidth]{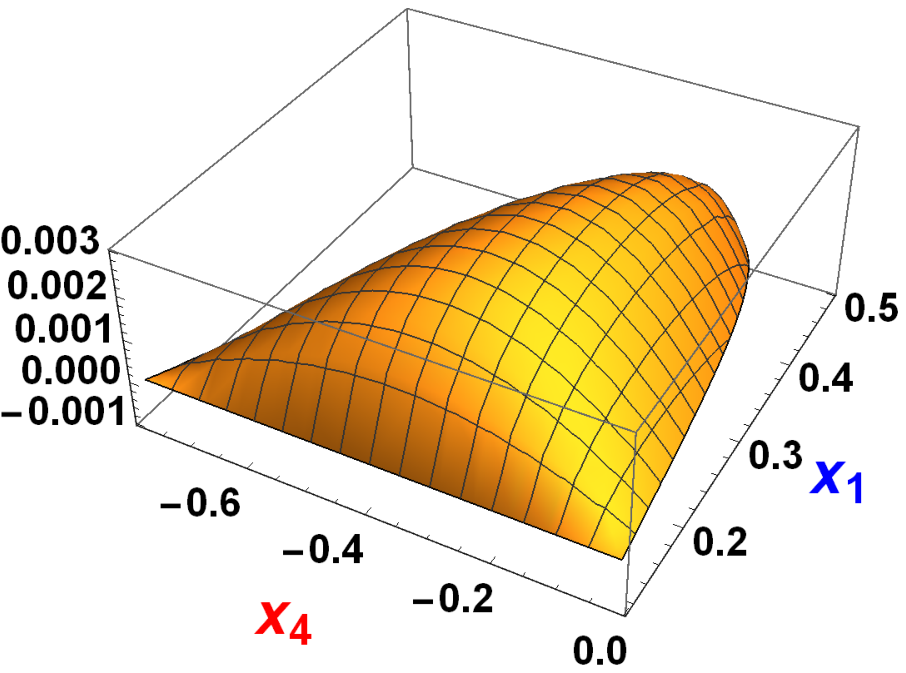}
\includegraphics[width=0.22\textwidth]{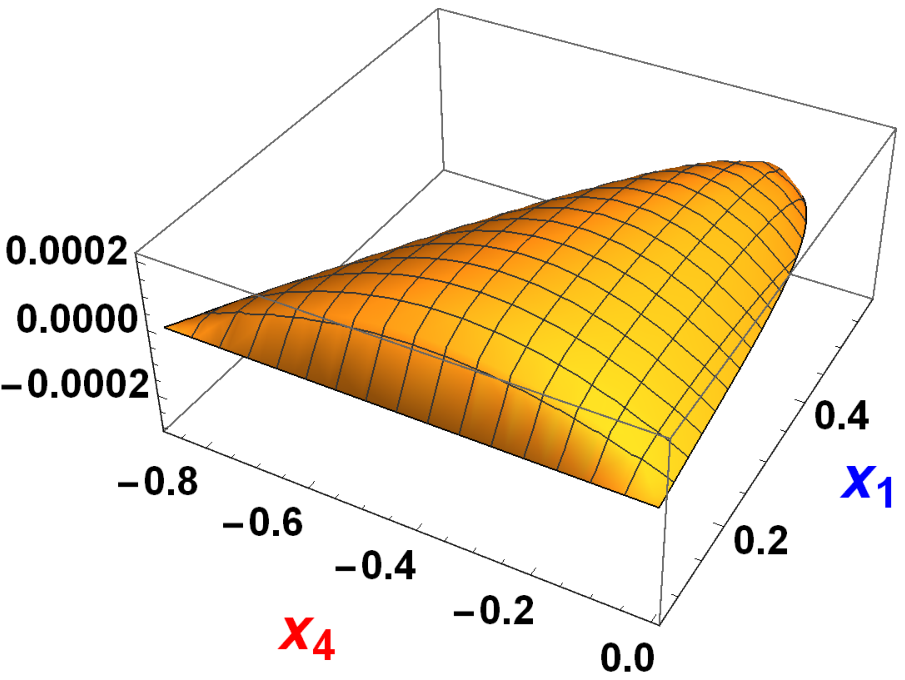}

\vspace{0.3cm}
\includegraphics[width=0.22\textwidth]{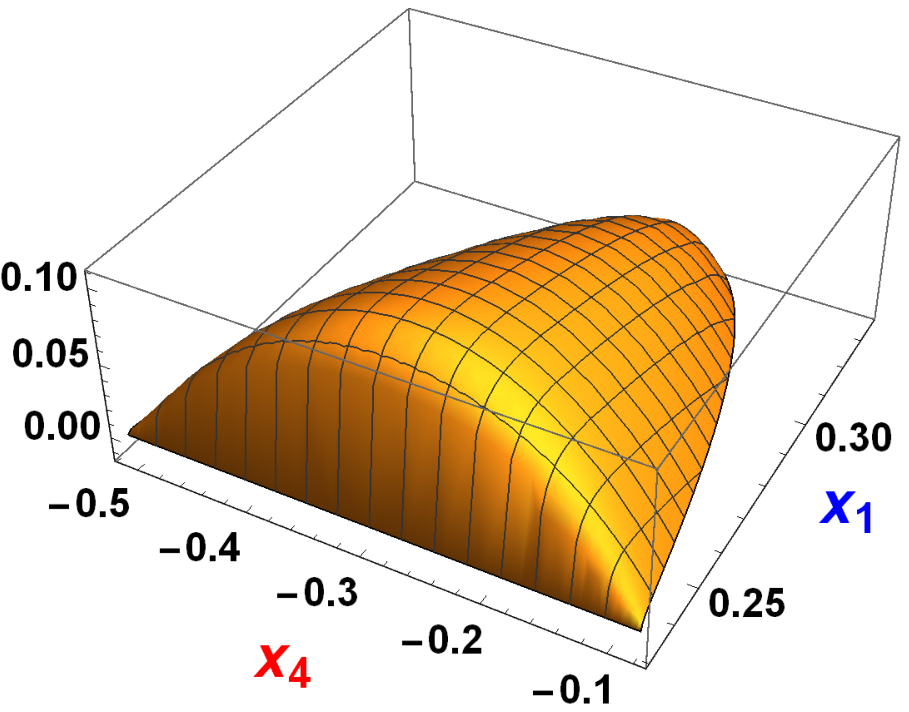}
\includegraphics[width=0.22\textwidth]{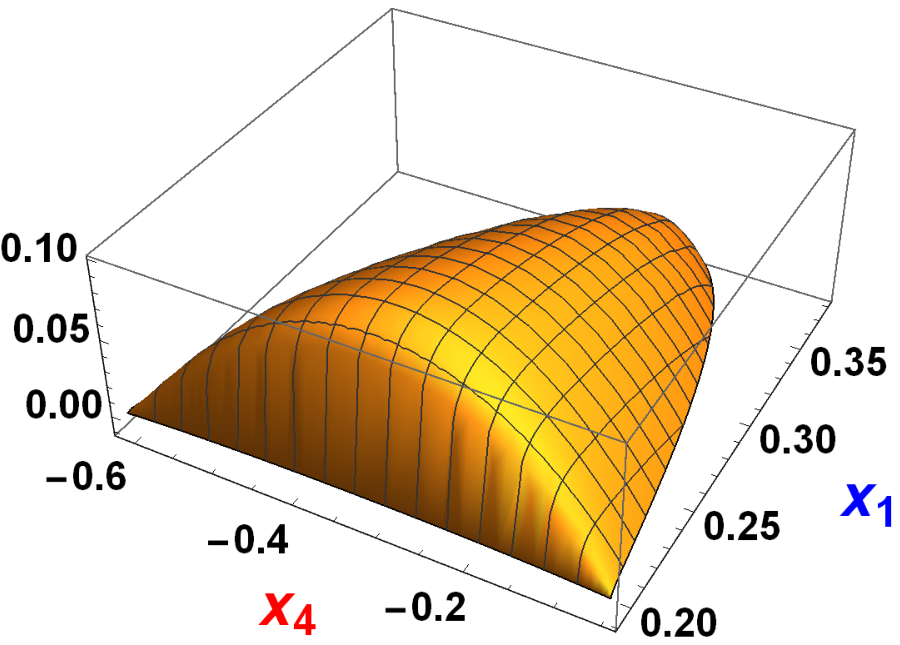}
\includegraphics[width=0.22\textwidth]{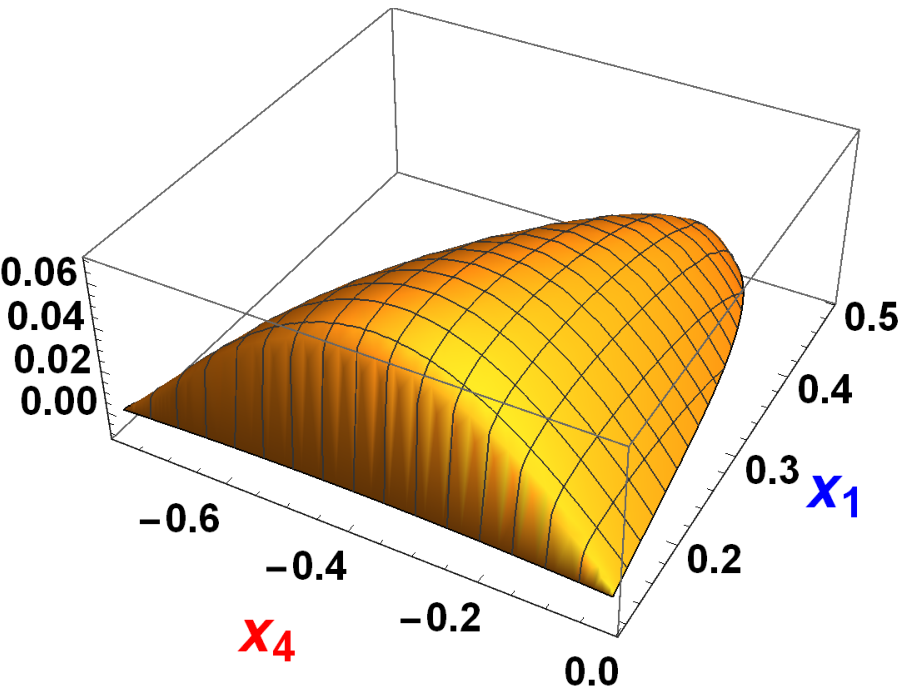}
\includegraphics[width=0.22\textwidth]{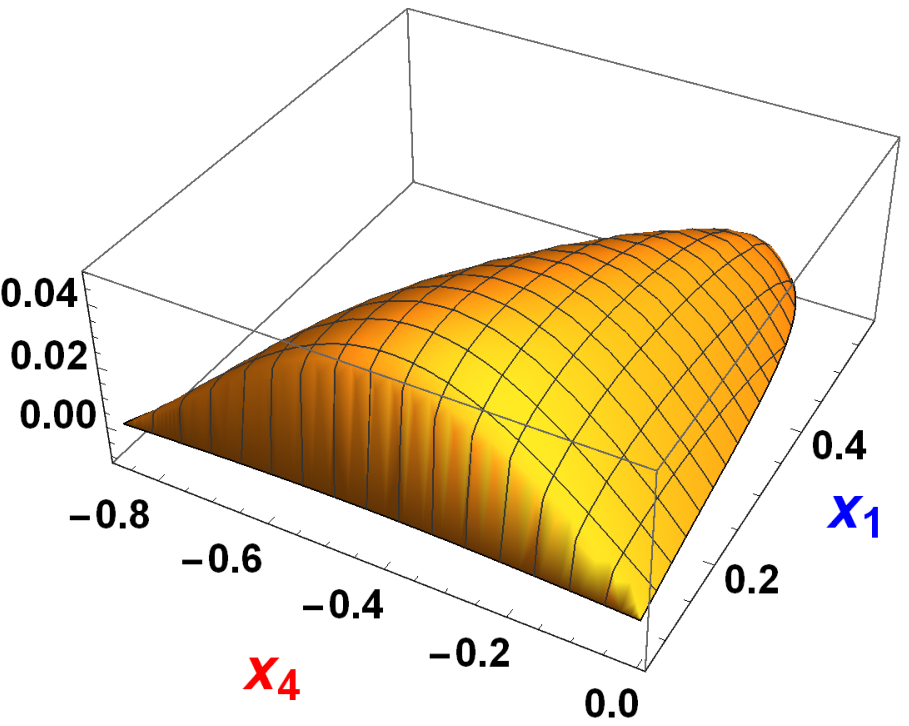}

 \parbox[t]{0.9\textwidth}{\caption{First row: double differential distribution for the spin-dependent part of the cross section $d\sigma^N$  in nb units over the dimensionless invariant variables  $x_1=s_1/s,\, x_4=t_2/s,$. Second row: corresponding proton normal polarization $P^N$ in the process $e^++e^- \to \pi^0+p+\bar{p}$ for $s=5,\,6,\,10,\,16$ GeV$^2$ from left to right, respectively. Third and fourth row: same as the first and second rows but for the process $e^++e^- \to \pi^0+n+\bar{n}.$}\label{fig.4}}
\end{figure}

%\begin{figure}
%\centering
%\includegraphics[width=0.22\textwidth]{1spint2s1trans_5.eps}
%\includegraphics[width=0.22\textwidth]{1spint2s1trans_6.eps}
%\includegraphics[width=0.22\textwidth]{1spint2s1trans_10.eps}
%\includegraphics[width=0.22\textwidth]{1spint2s1trans_16.eps}
%
%\vspace{0.3cm}
%\includegraphics[width=0.22\textwidth]{p1spint2s1trans_5.eps}
%\includegraphics[width=0.22\textwidth]{p1spint2s1trans_6.eps}
%\includegraphics[width=0.22\textwidth]{p1spint2s1trans_10.eps}
%\includegraphics[width=0.22\textwidth]{p1spint2s1trans_16.eps}

%\includegraphics[width=0.3\textwidth]{gena2019c.eps}
 %\parbox[t]{0.9\textwidth}{\caption{The same as in Fig.\,4 but in the case of the transverse polarization.}\label{fig.5}}
%\end{figure}

The double differential $(t_2,\,s_1)-$distribution of the spin-independent part of the cross section is symmetrical under the substitution $t_2\to t_1,\,s_1 \to s_2$ but the spin-dependent part loses this symmetry.

Consider now the $(s_1,\,s_2)-$ and $(s_1,\,s_{12})-$distributions. After integration over $t_1$ and $t_2$ the spin-dependent part of the cross section looks very simple, namely
\begin{equation}\label{eq:s1s2norm16}
\frac{d\,\sigma^N}{d\,s_1\,ds_2} = \frac{g^2_{\pi^0 N \bar{N}}\alpha^2\,M\,N(s_1+s_2-2M^2-2m^2)Im[G_EG^*_M]}{3\,\pi\,s^3(4M^2-s)(M^2-s_1)(M^2-s_2)},
\end{equation}
where we use the limit $m_e\to 0$ and the quantity  $N$ is defined in  Eq. (\ref{eq:ST10}). This distribution is symmetrical under the change $s_1\rightleftarrows s_2.$ The corresponding $(s_1,\,s_{12})-$distribution can be obtained from (\ref{eq:s1s2norm16}) by simple substitution
$s_{12} = s+2M^2+m^2-s_1-s_2.$

In Fig.\,5  and  Fig. \,6 the   $(x_1,\,x_2)-$ and the  $(x_1,\,x_{12})-$  distributions are plotted for both nucleon channels.

\begin{figure}
\centering
\includegraphics[width=0.22\textwidth]{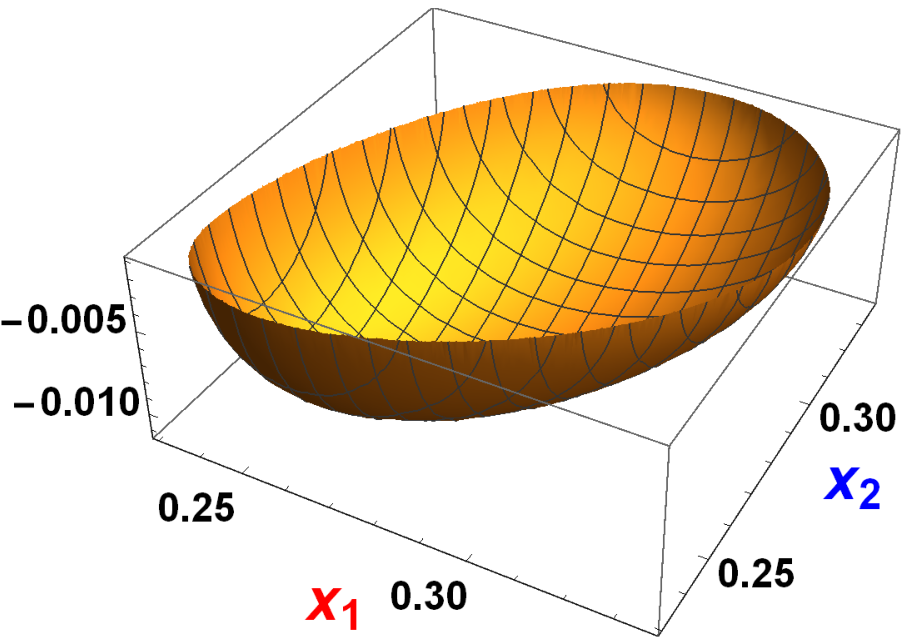}
\includegraphics[width=0.22\textwidth]{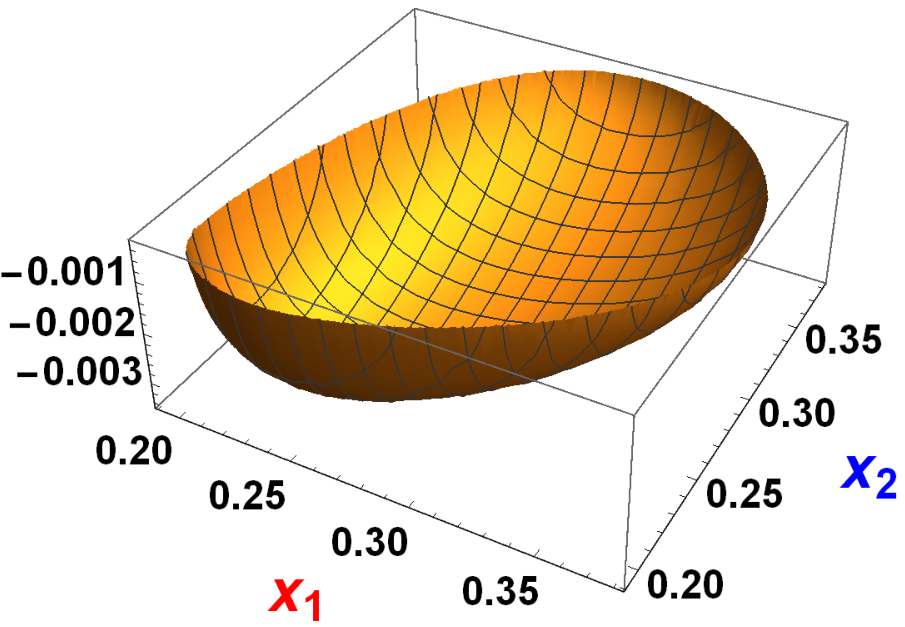}
\includegraphics[width=0.22\textwidth]{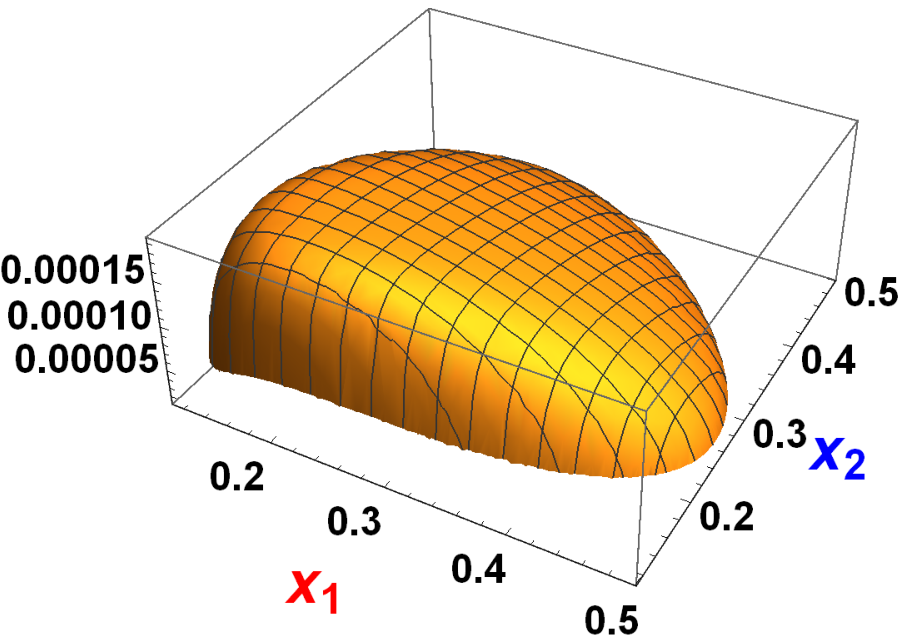}
\includegraphics[width=0.22\textwidth]{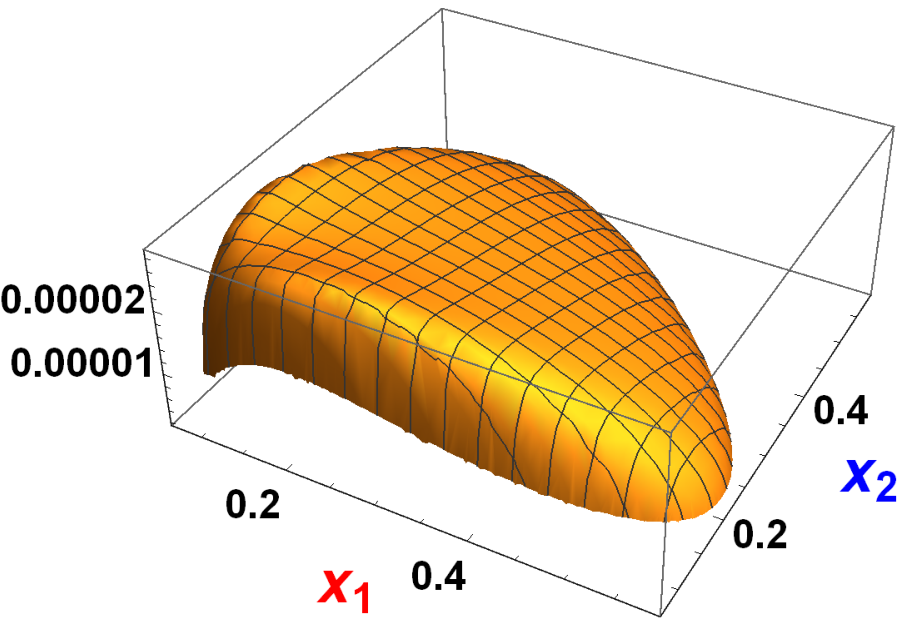}

\vspace{0.3cm}
\includegraphics[width=0.22\textwidth]{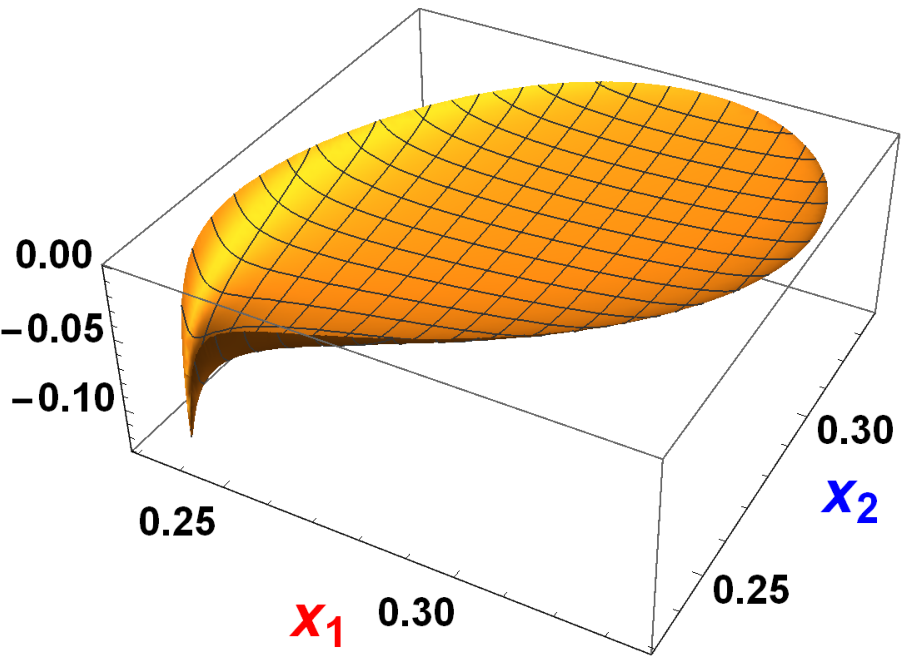}
\includegraphics[width=0.22\textwidth]{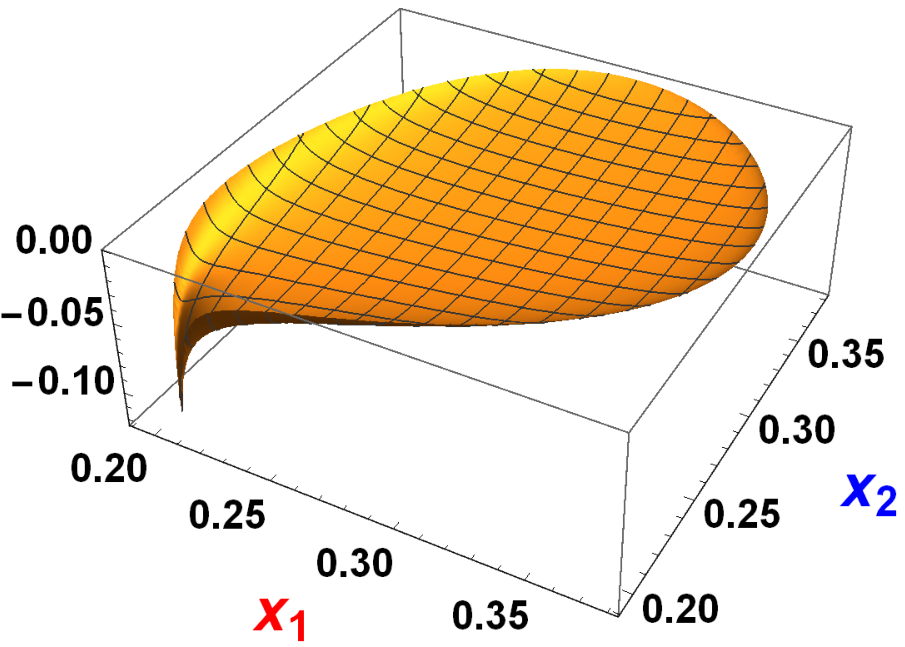}
\includegraphics[width=0.22\textwidth]{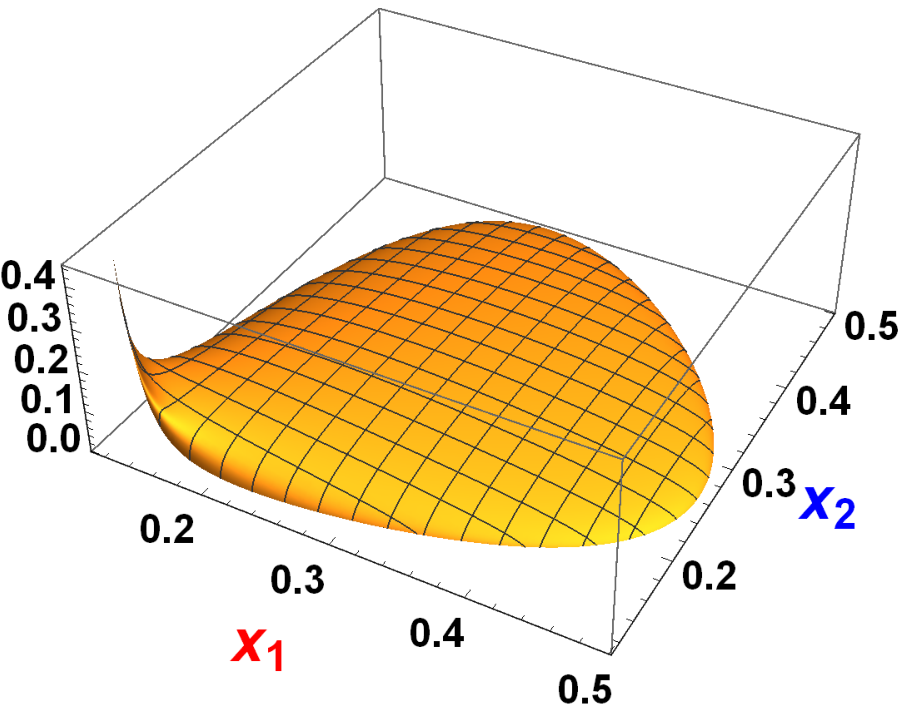}
\includegraphics[width=0.22\textwidth]{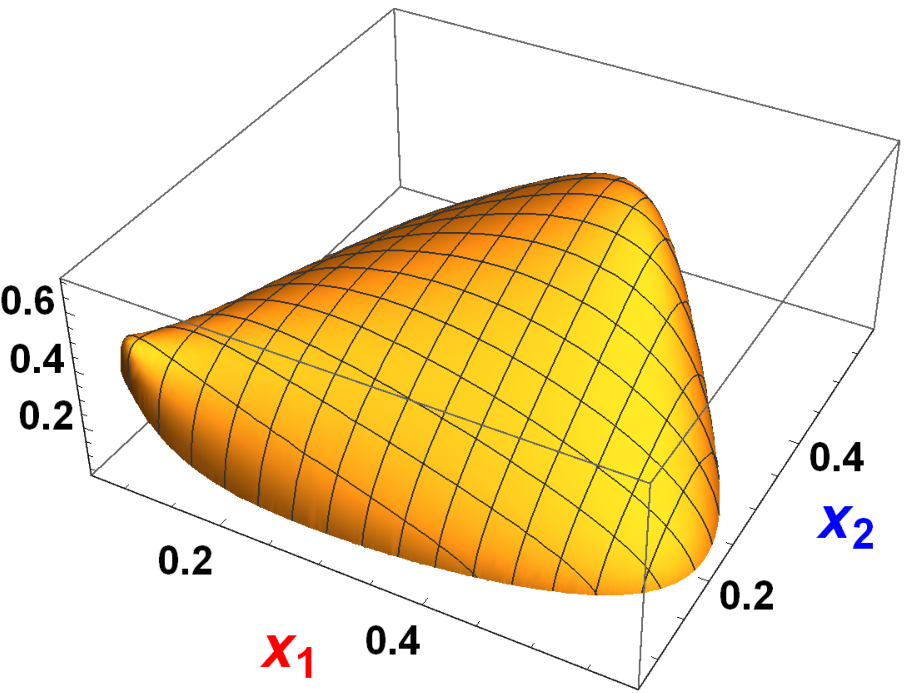}
%\vspace{0.3cm}
%\includegraphics[width=0.22\textwidth]{1spins1s12norm_5.eps}
%\includegraphics[width=0.22\textwidth]{1spins1s12norm_6.eps}
%\includegraphics[width=0.22\textwidth]{1spins1s12norm_10.eps}
%\includegraphics[width=0.22\textwidth]{1spins1s12norm_16.eps}

\vspace{0.3cm}
\includegraphics[width=0.22\textwidth]{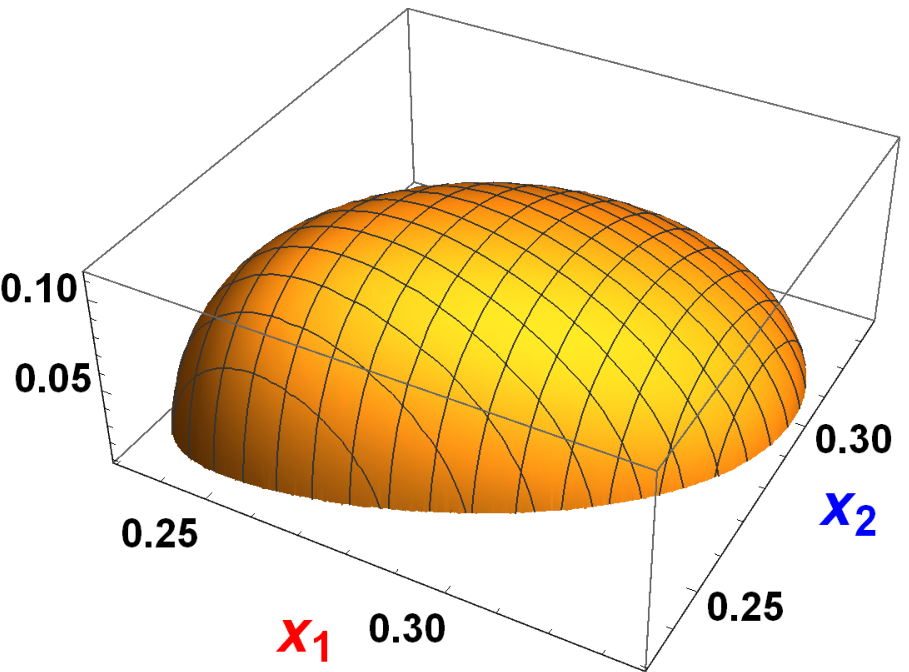}
\includegraphics[width=0.22\textwidth]{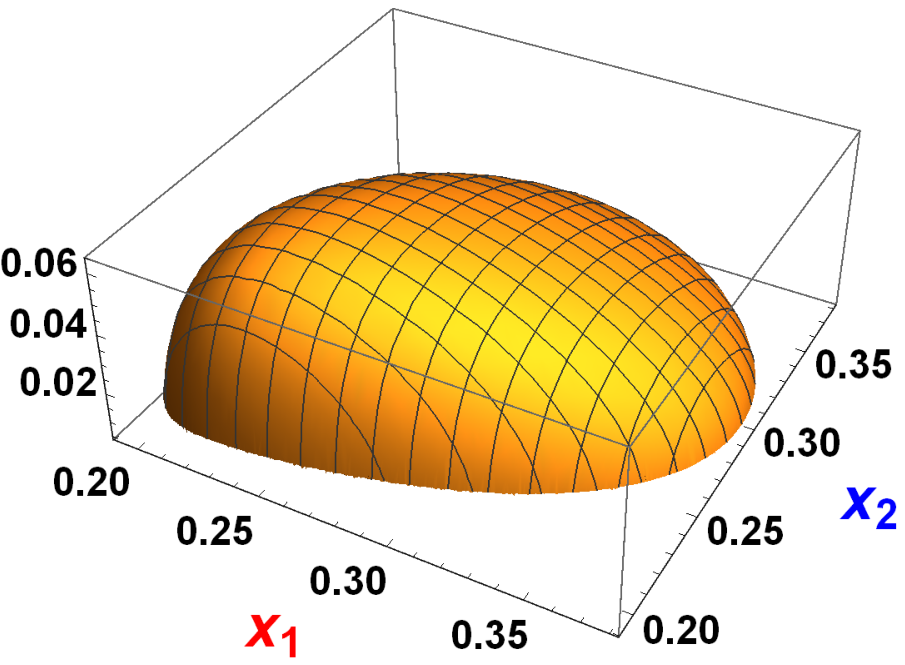}
\includegraphics[width=0.22\textwidth]{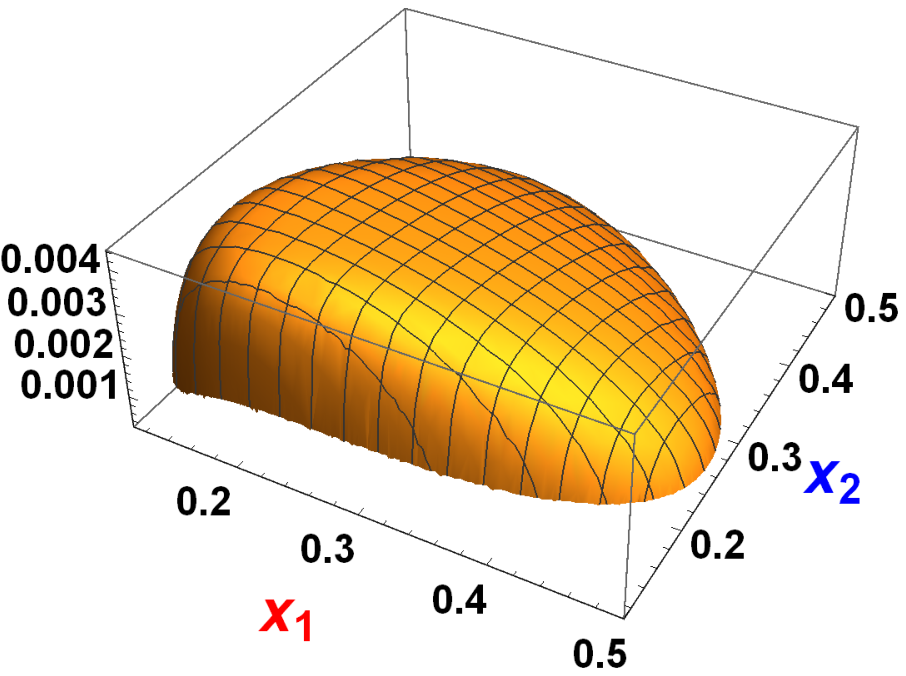}
\includegraphics[width=0.22\textwidth]{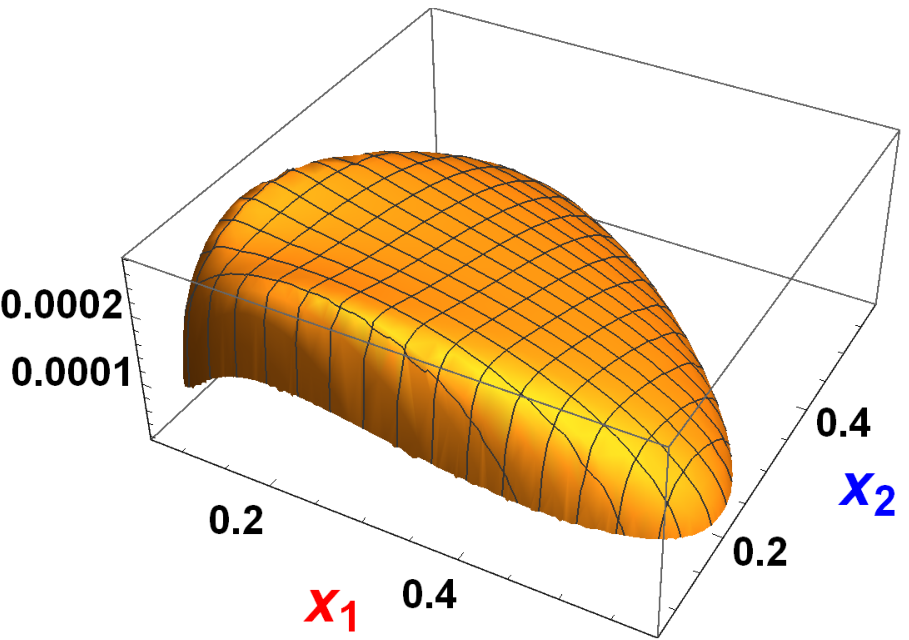}

\vspace{0.3cm}
\includegraphics[width=0.22\textwidth]{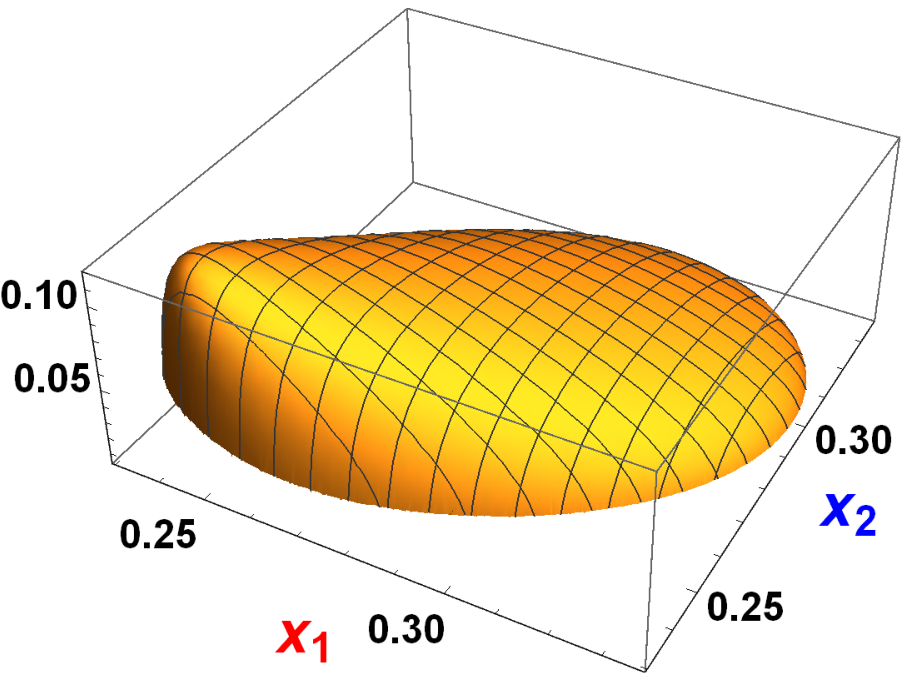}
\includegraphics[width=0.22\textwidth]{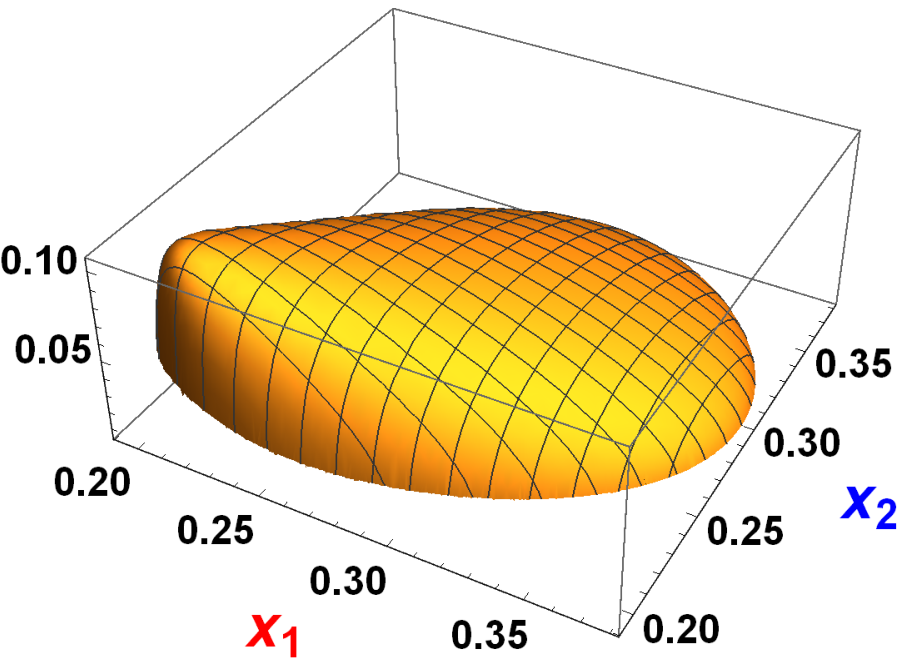}
\includegraphics[width=0.22\textwidth]{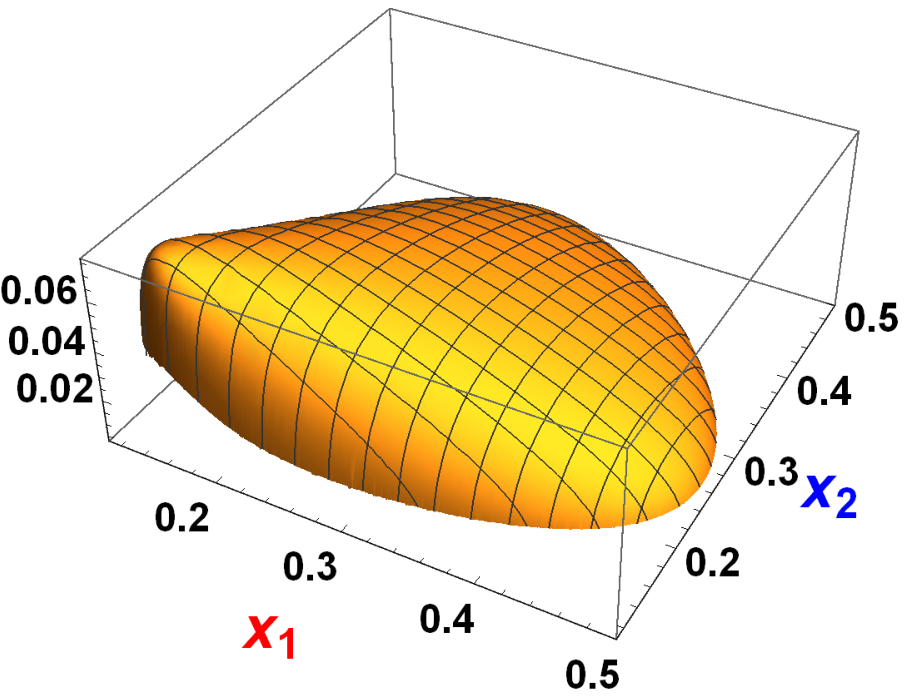}
\includegraphics[width=0.22\textwidth]{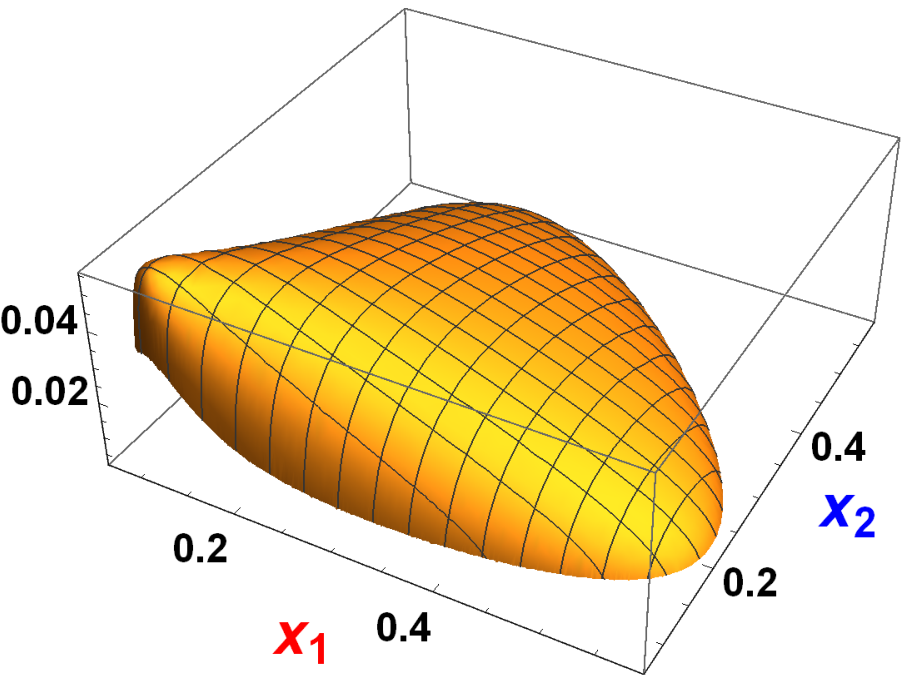}

 \parbox[t]{0.9\textwidth}{\caption{The same as in Fig.\,4 but for the double $(x_1,\,x_2)$-distribution.}\label{fig.5}}
 % same as in Fig.\,4 double $(x_1,\,x_2)$-distribution of the $d\sigma^N$ (the first row) and corresponding polarization $P^N$ (the second row)
% in the case of the $\pi^0 p \bar{p}$ channel for $ s=5,\,6,\,10,\,16$ GeV$^2$ from left  to right and the same for the $(x_1,\,x_{12})$-distribution (the third and fourth %rows); $x_2=s_2/s,\, x_{12}=s_{12}/s.$}\label{fig.5}}
\end{figure}

\begin{figure}
\centering
\includegraphics[width=0.22\textwidth]{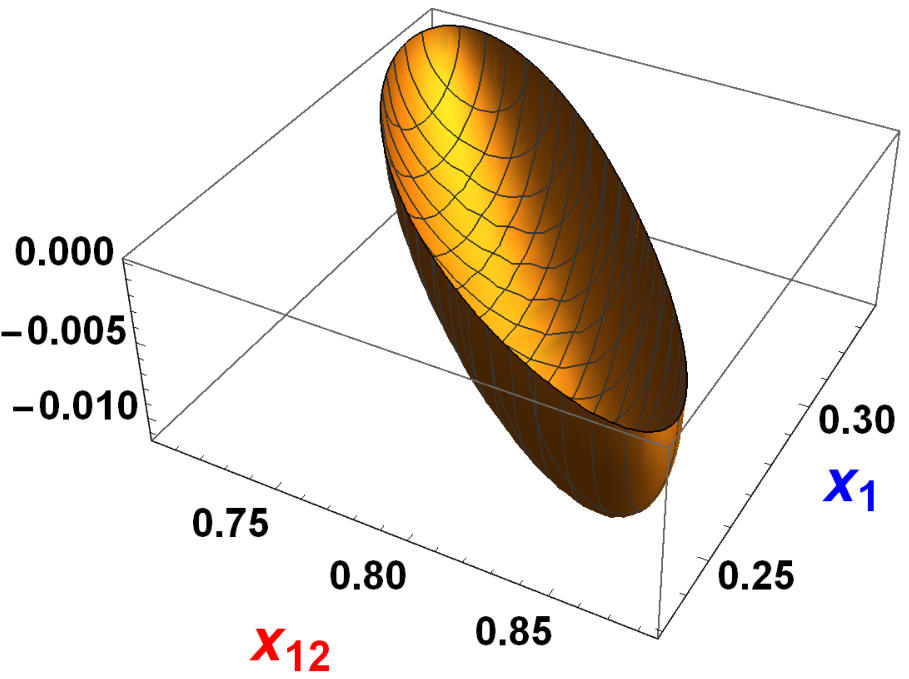}
\includegraphics[width=0.22\textwidth]{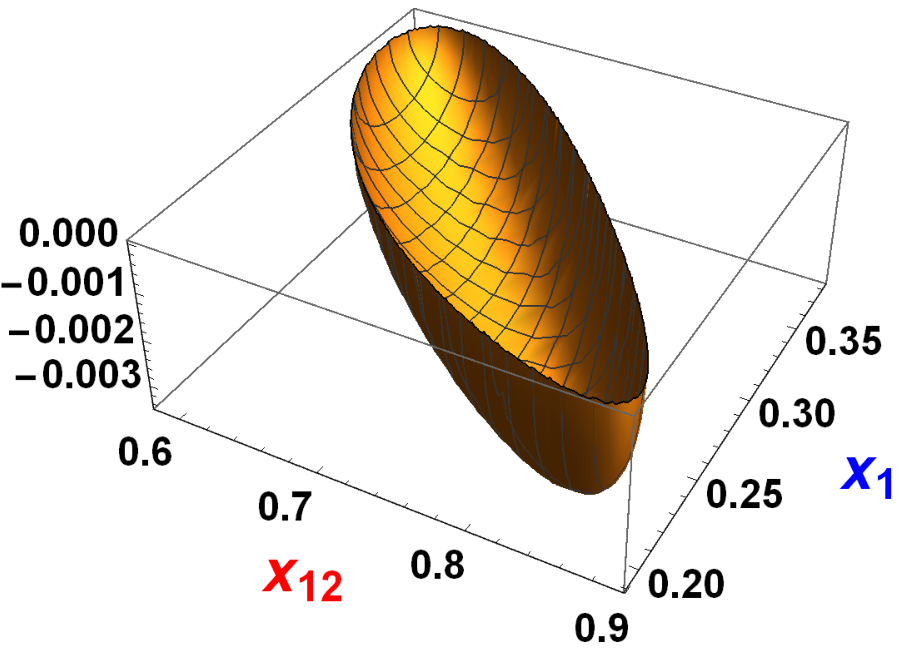}
\includegraphics[width=0.22\textwidth]{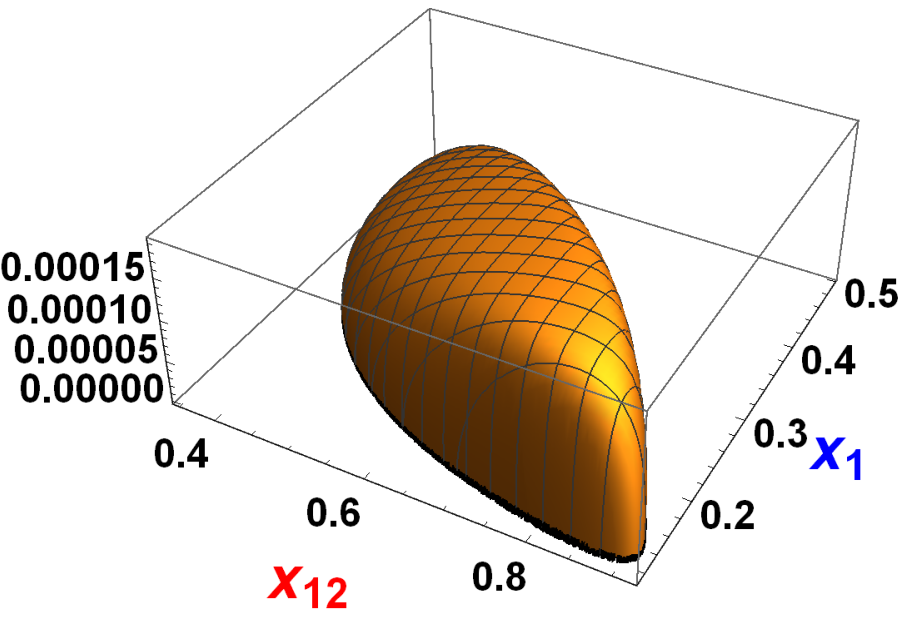}
\includegraphics[width=0.22\textwidth]{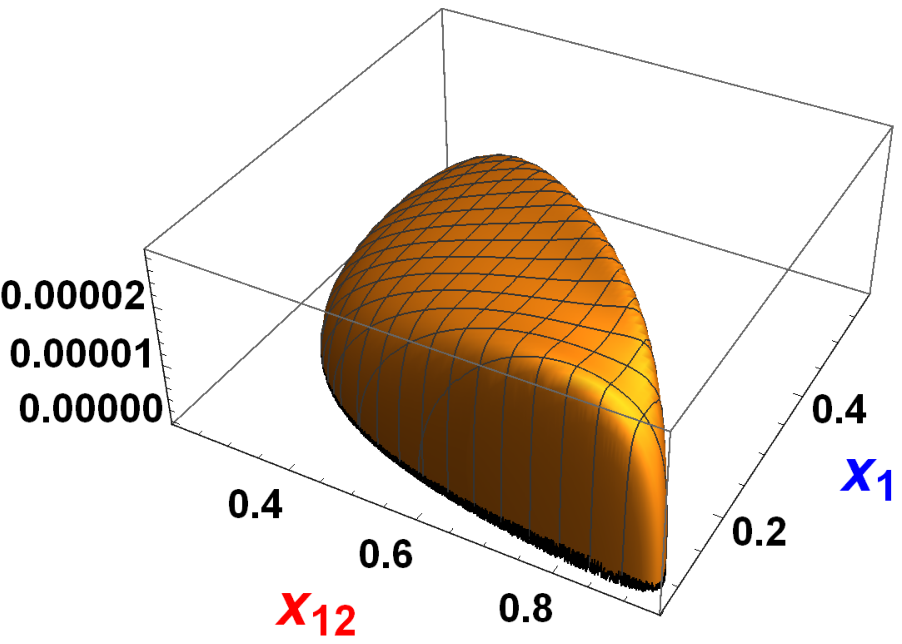}

\vspace{0.3cm}
\includegraphics[width=0.22\textwidth]{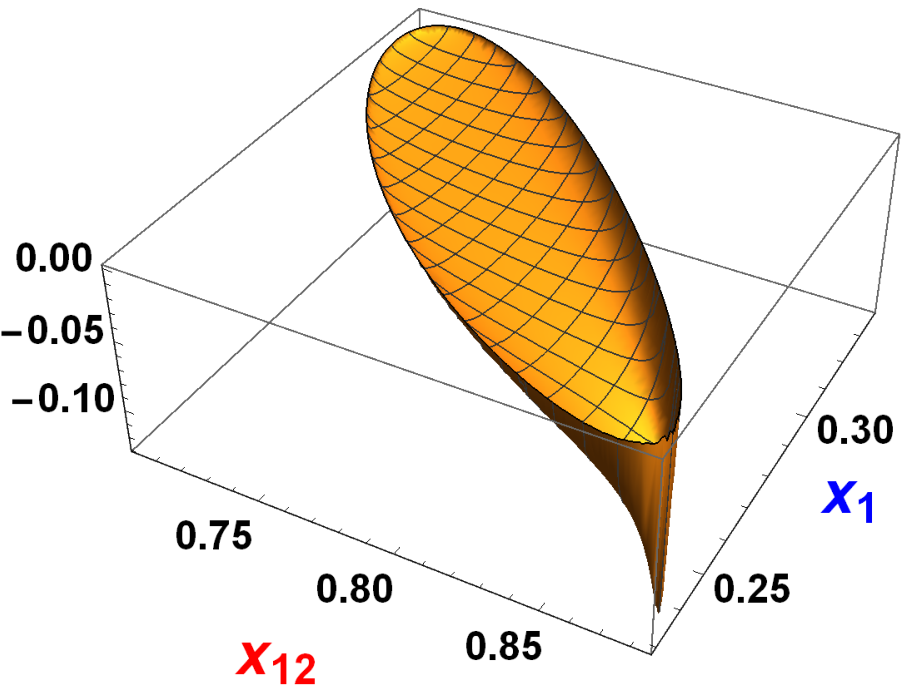}
\includegraphics[width=0.22\textwidth]{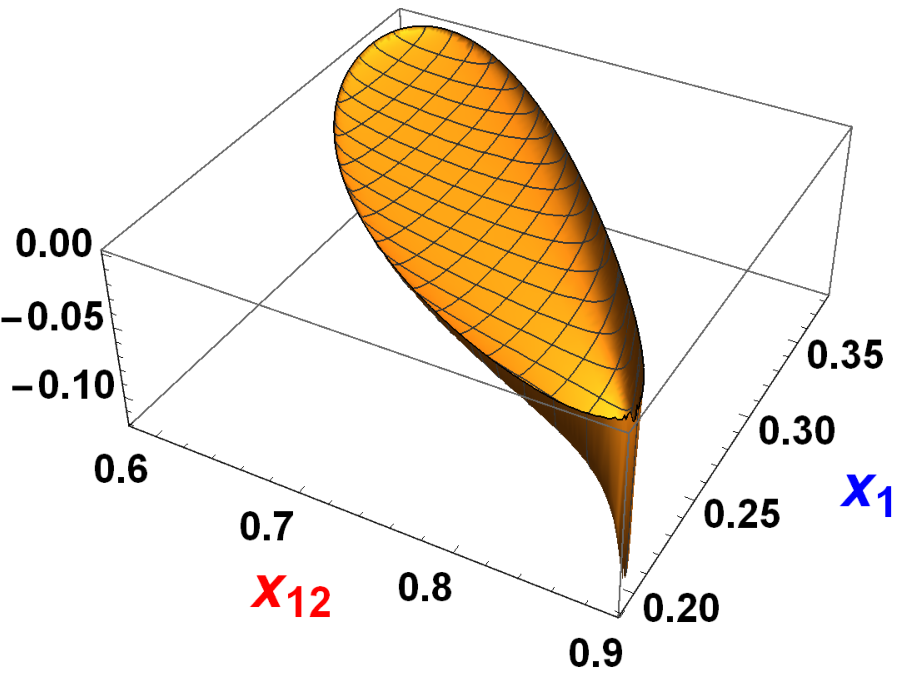}
\includegraphics[width=0.22\textwidth]{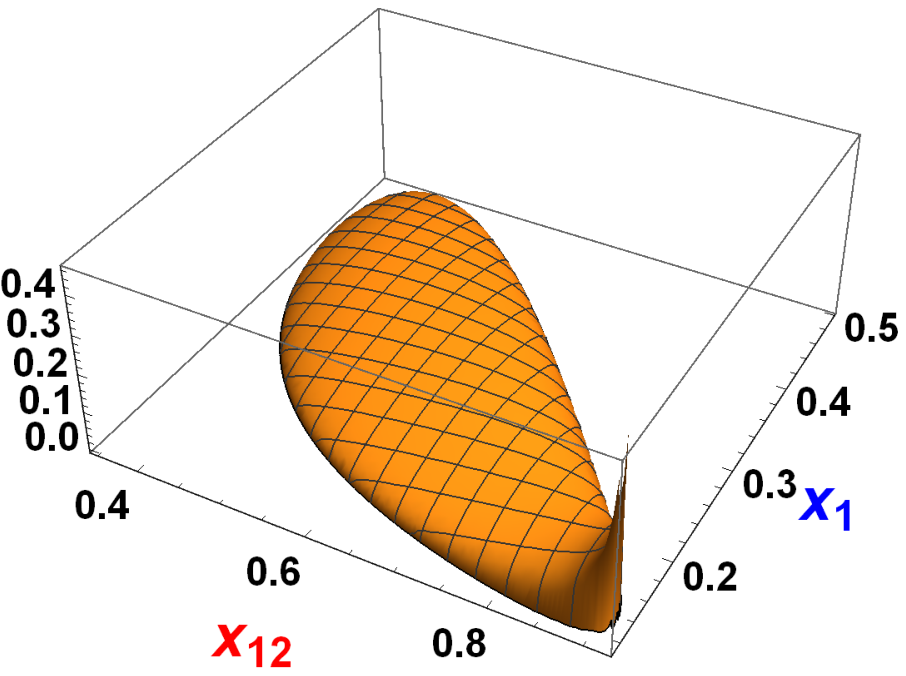}
\includegraphics[width=0.22\textwidth]{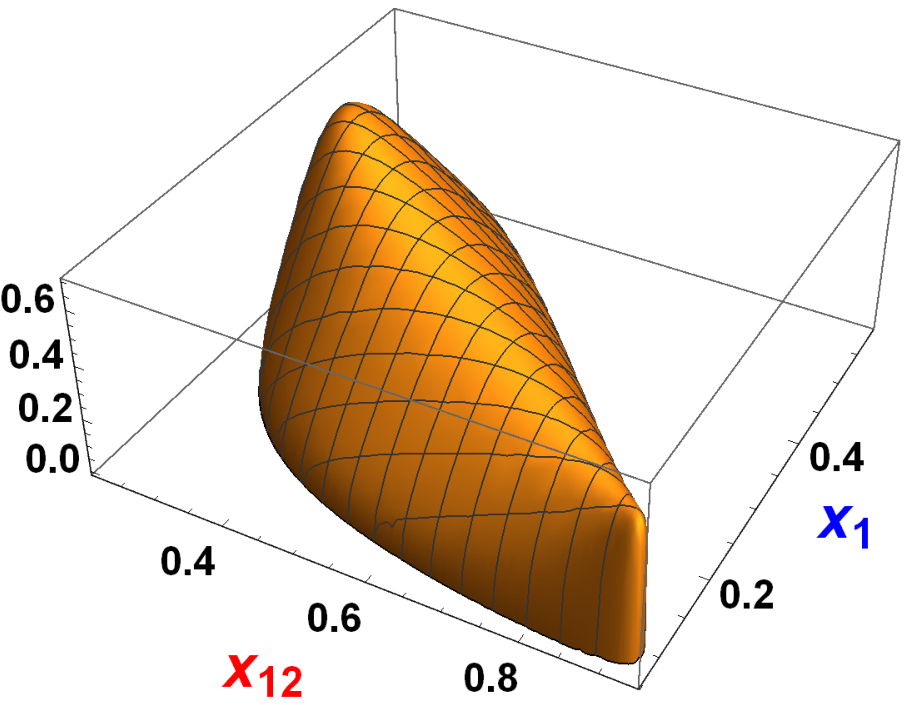}

\vspace{0.3cm}
\includegraphics[width=0.22\textwidth]{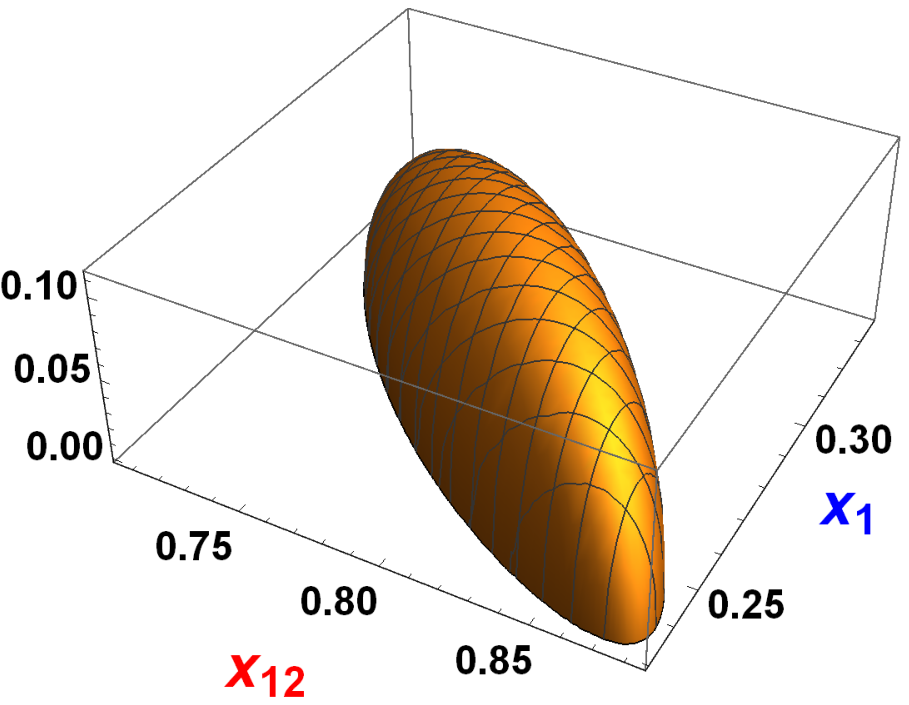}
\includegraphics[width=0.22\textwidth]{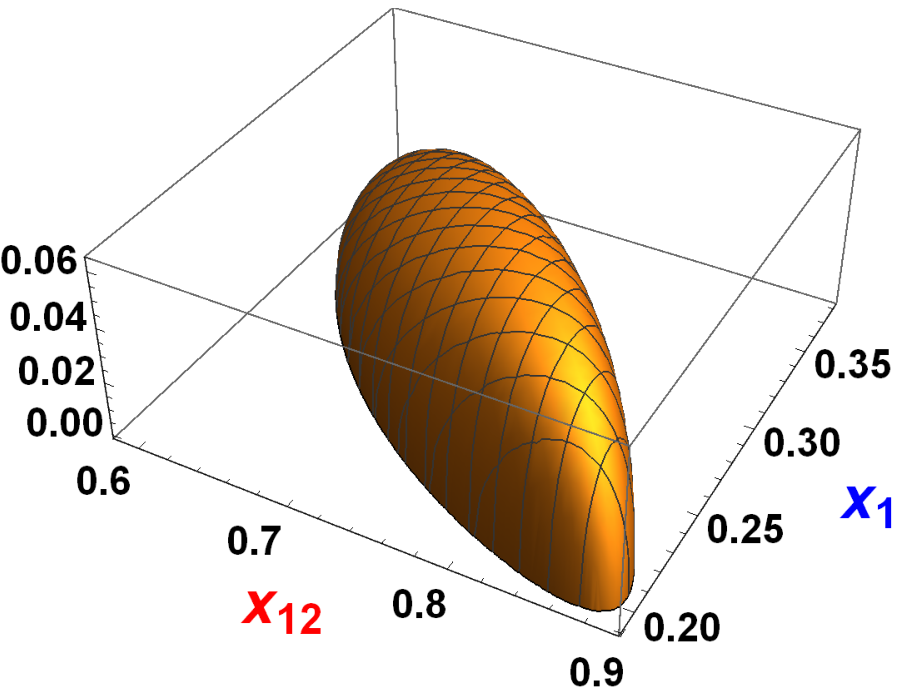}
\includegraphics[width=0.22\textwidth]{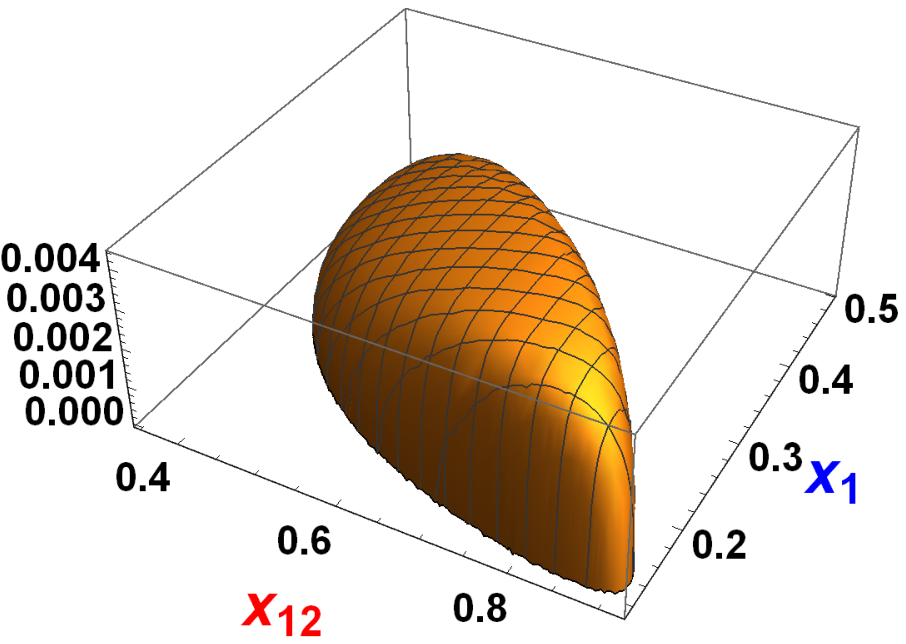}
\includegraphics[width=0.22\textwidth]{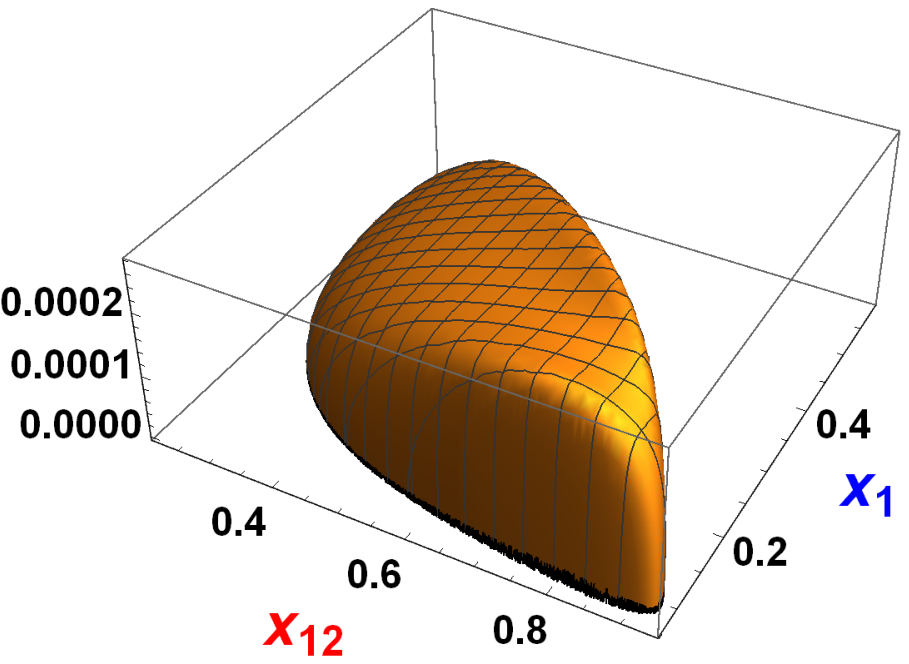}

\vspace{0.3cm}
\includegraphics[width=0.22\textwidth]{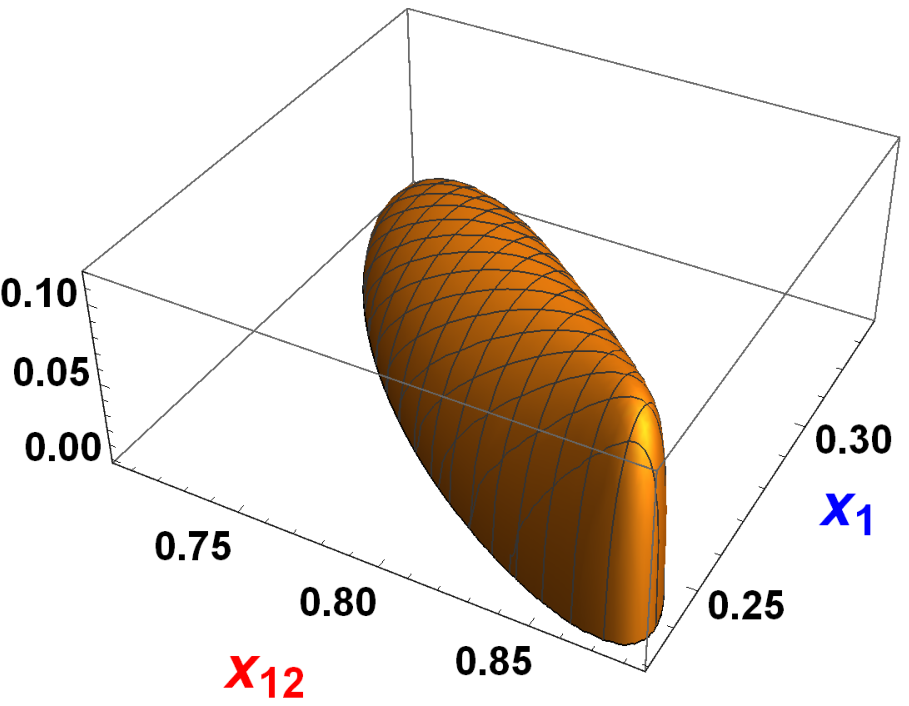}
\includegraphics[width=0.22\textwidth]{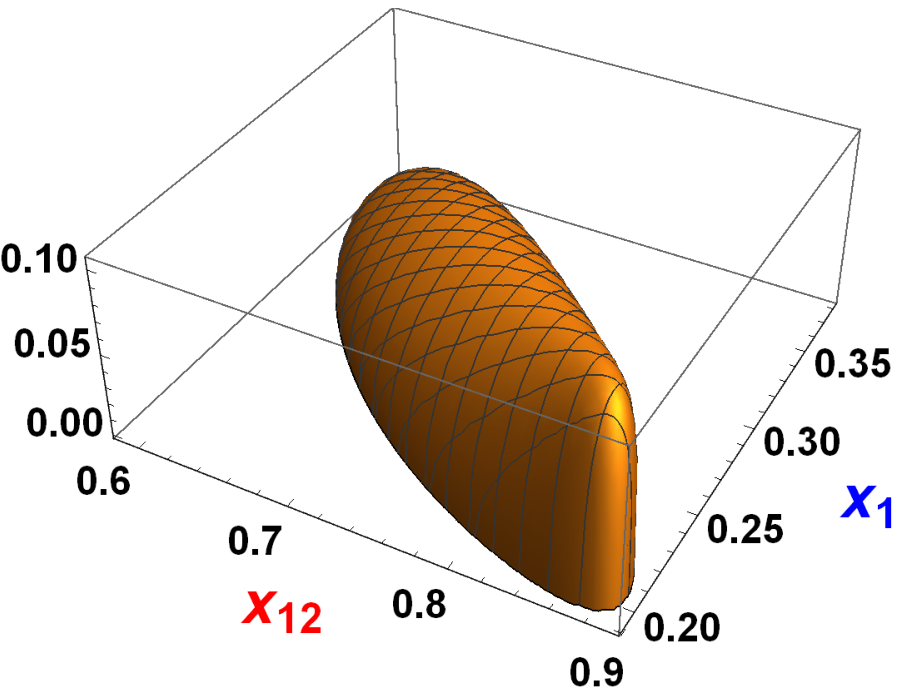}
\includegraphics[width=0.22\textwidth]{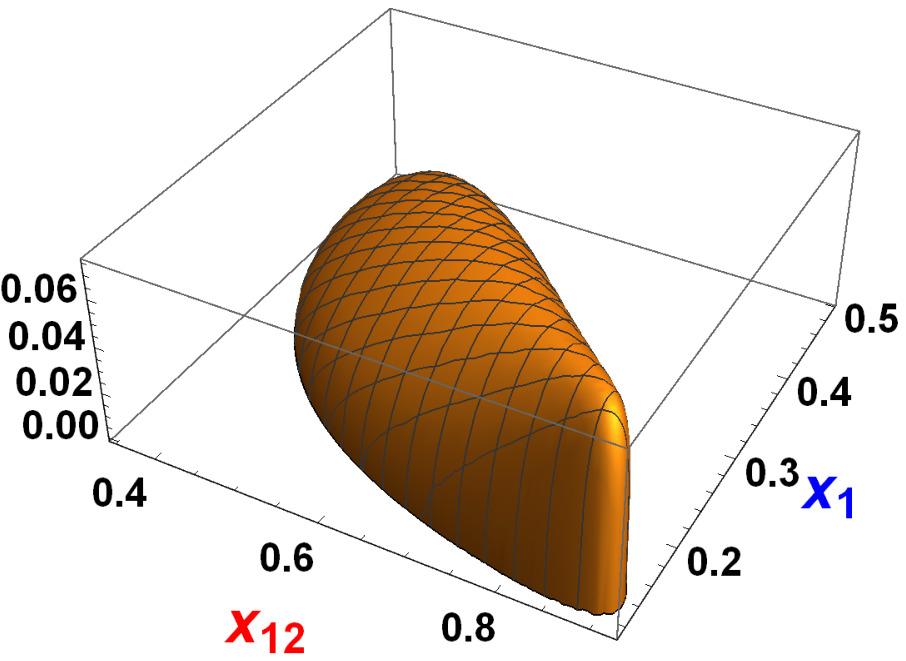}
\includegraphics[width=0.22\textwidth]{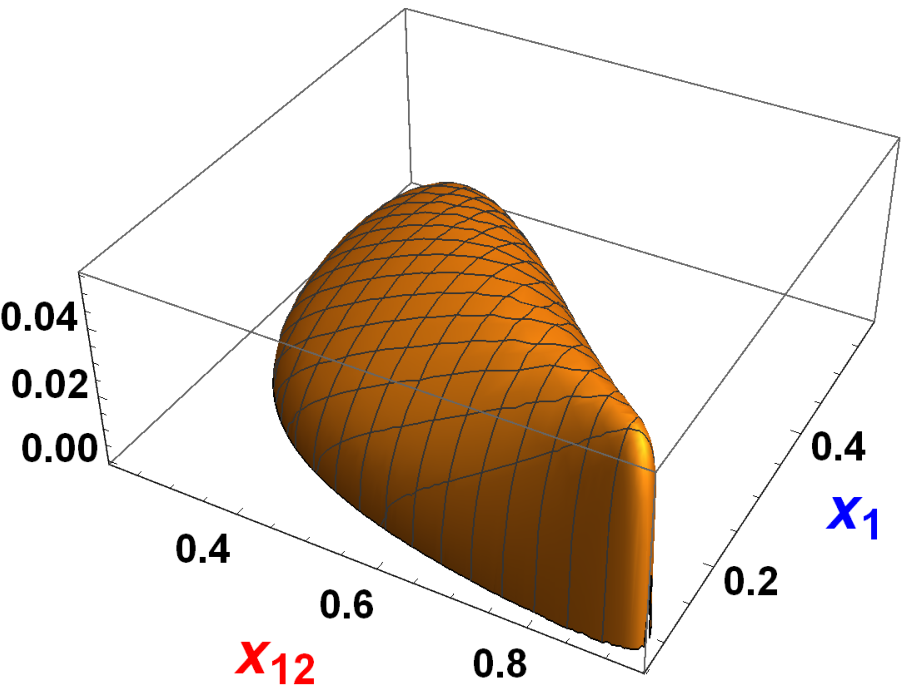}

 \parbox[t]{0.9\textwidth}{\caption{The same as in Fig.\,4 but for the $(x_1,\,x_{12})-$ double distribution.}\label{fig.6}}
\end{figure}

\section{Single differential distributions}

Let us consider the single differential distributions. In the case of the $s_1$ and the $s_{12}$ distributions the expressions of the spin-dependent parts of the cross section are
\begin{eqnarray}
\frac{d\sigma^N}{d\,s_{12}}&=&\frac{g^2_{\pi^0 N \bar{N}}\alpha^2\,M\,(s-s_{12}-m^2)Im[G_E\,G^*_M]}{6\,s^3\,(s-4M^2)(s-s_{12} +m^2)}Z_1,\label{eq:s12norm17}\\
Z_1&=&-(s+m^2-s_{12})\sqrt{s_{12}} +2 \sqrt{M^2[(s-s_{12})^2+m^4-2m^2(s+s_{12})]+m^2s\,s_{12}},\nonumber
\end{eqnarray}

\begin{eqnarray}
\frac{d\sigma^N}{d\,s_{1}}&=&\frac{g^2_{\pi^0 N \bar{N}}\alpha^2\,M\,Im[G_E\,G^*_M]}{48\,s^3\,(s-4M^2)(M^2-s_1)s_1^{3/2}}Z_2,
\label{eq:s1norm18}\\
Z_2 &=&M^8-2M^6(s+4s_1+m^2)+M^4[s^2+6ss_1+10s_1^2+m^2(4s-2s_1)+m^4] +\nn\\
 &&+ s_1^2(s^2+2ss_1-3s_1^2)-2M^2[ss_1(s+3s_1)+m^2(s^2-6ss_1+3s_1^2+m^4(s-3s_1)] +\nn\\
&&+m^4(s^2-10ss_1-7s_1^2)-2m^2s_1(s^2-5s_1^2)+
\nn\\
&&
+8(M^2+2m^2-s_1)s_1^{3/2}\sqrt{M^6-2M^4s_1+M^2(s_1^2-3m^2s)+m^2s(s-s_1+m^2)}.\nn
\end{eqnarray}
In Figs.\,7 and 8 we plot the spin-dependent part of the cross section and the nucleon polarization for both, $\pi^0 p \bar{p}$  and $\pi^0 n \bar{n},$ channels as
function of dimensionless variable $x_{12}$ and $x_1,$ correspondingly. The differential distributions over $x_4$ are plotted in Fig.\,9.
\begin{figure}
\centering
\includegraphics[width=0.35\textwidth]{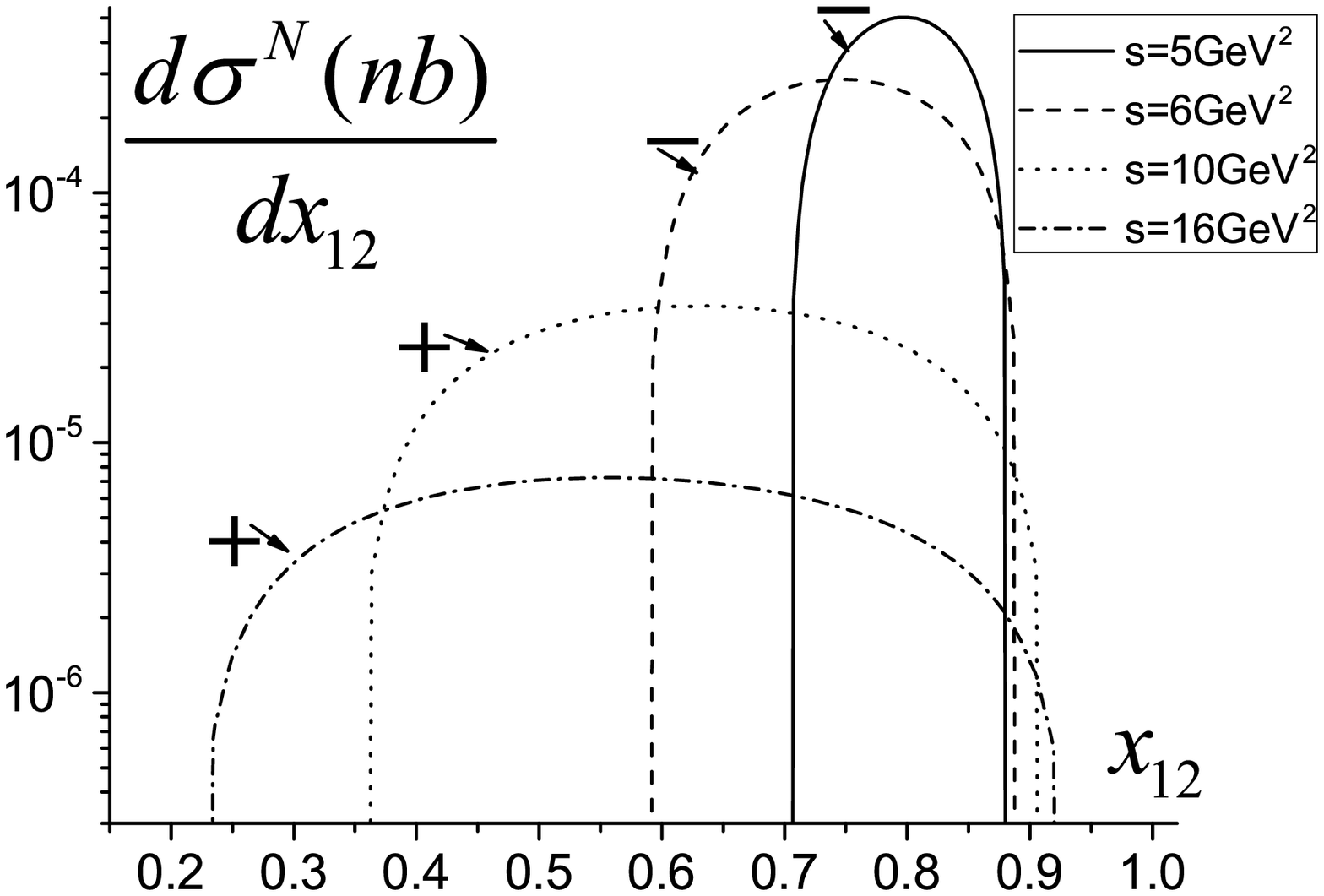}
\includegraphics[width=0.32\textwidth]{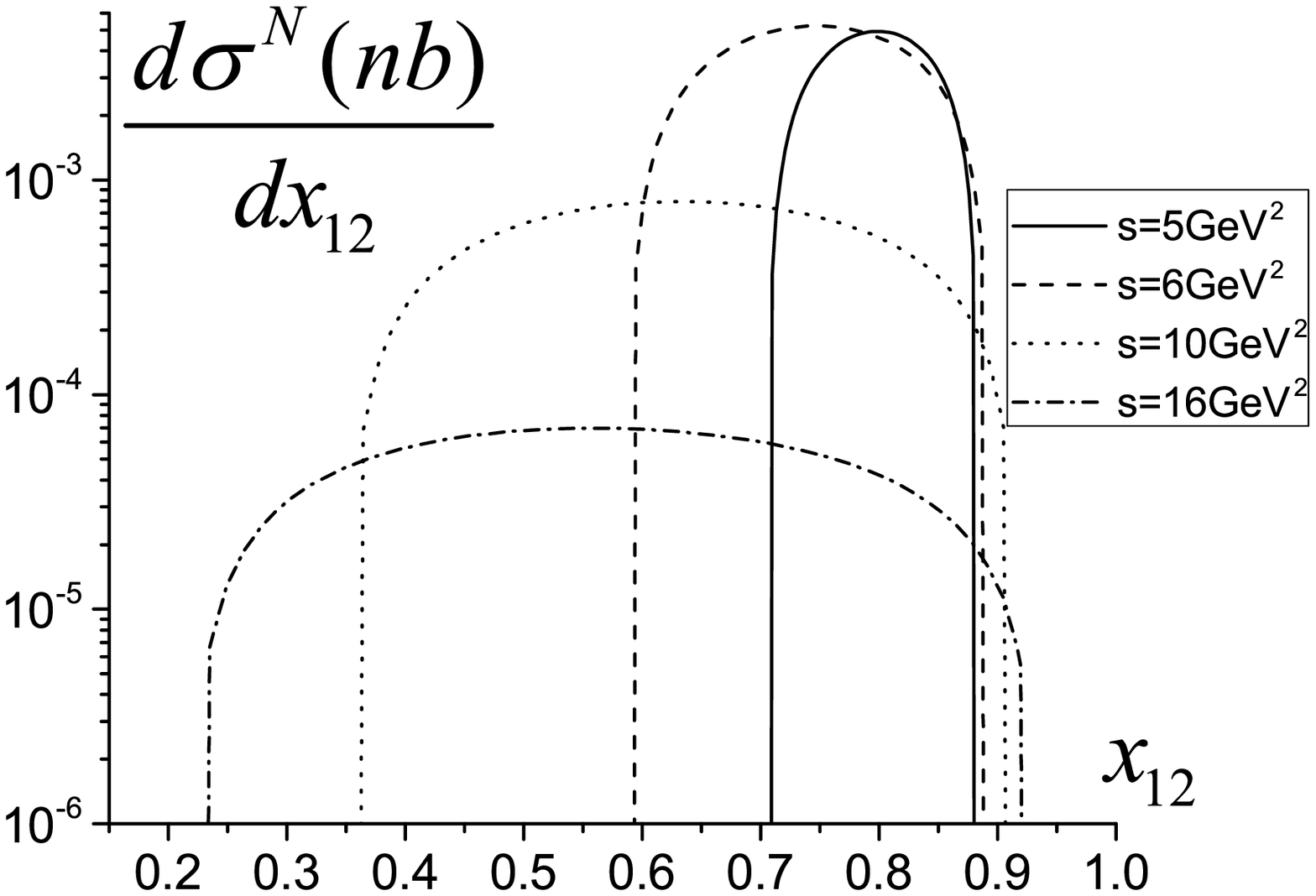}
\includegraphics[width=0.32\textwidth]{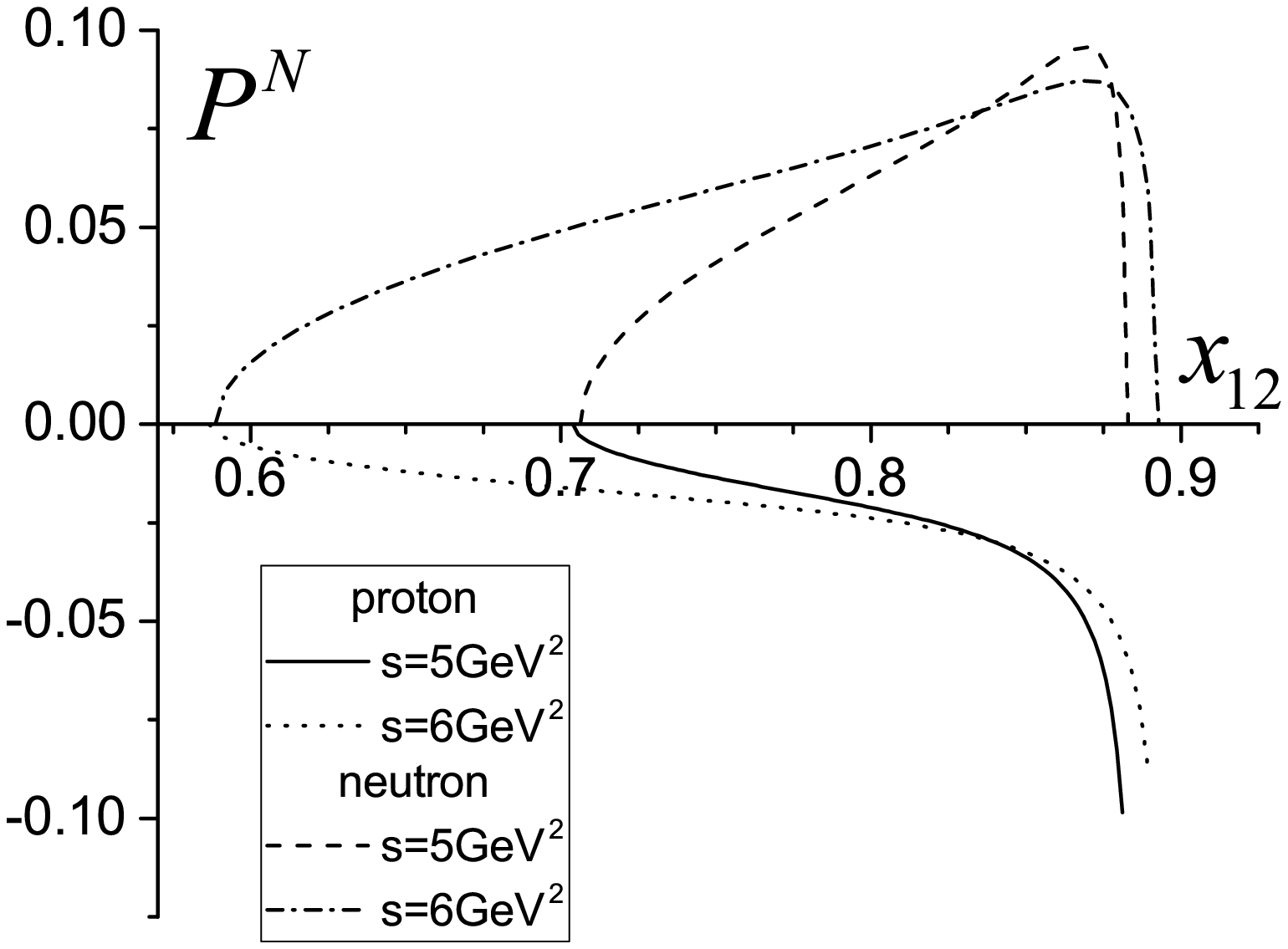}
\includegraphics[width=0.32\textwidth]{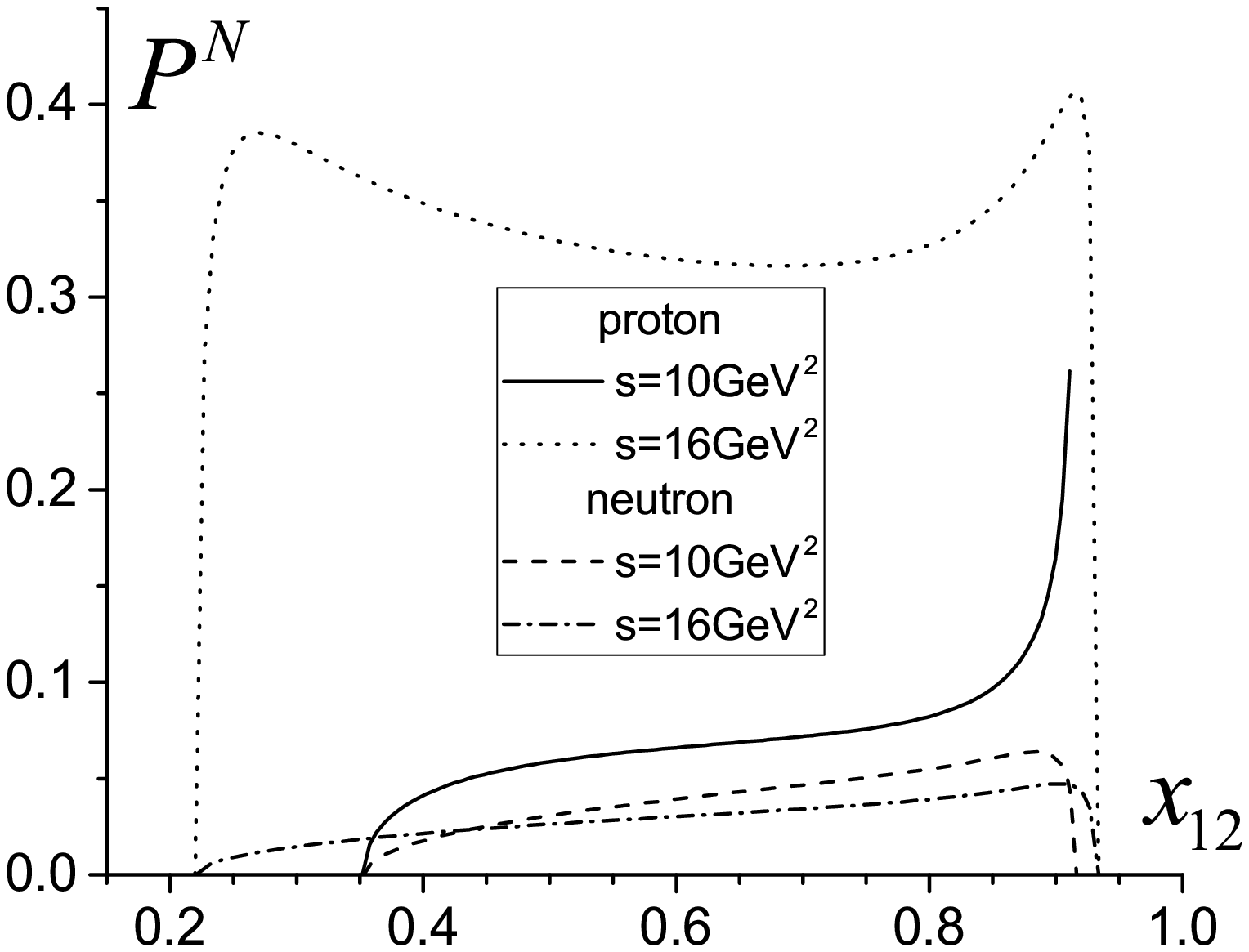}
 \parbox[t]{0.9\textwidth}{\caption{First row: $x_{12}$-distribution of the spin-dependent part of the cross section for  the
 $\pi^0 p \bar{p}$-channel (left panel) and $\pi^0 n \bar{n}$ channel (right panel) as defined by Eq.(\ref{eq:s12norm17}); the sign +\,(-) indicates that the corresponding quantity is positive (negative). Second row:  the corresponding nucleon normal polarization is plotted.}\label{fig.7}}
\end{figure}

\begin{figure}
\centering
\includegraphics[width=0.35\textwidth]{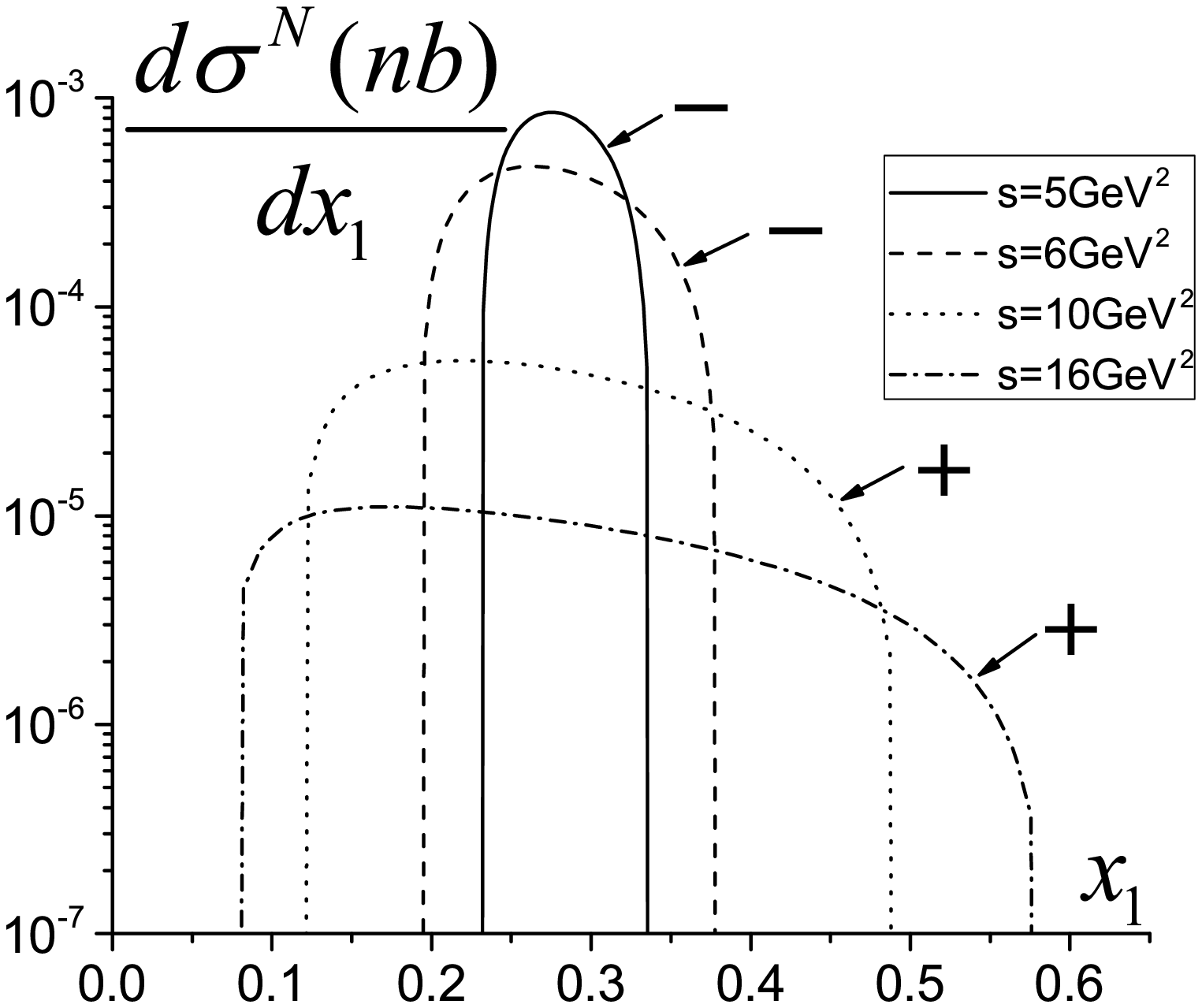}
\includegraphics[width=0.32\textwidth]{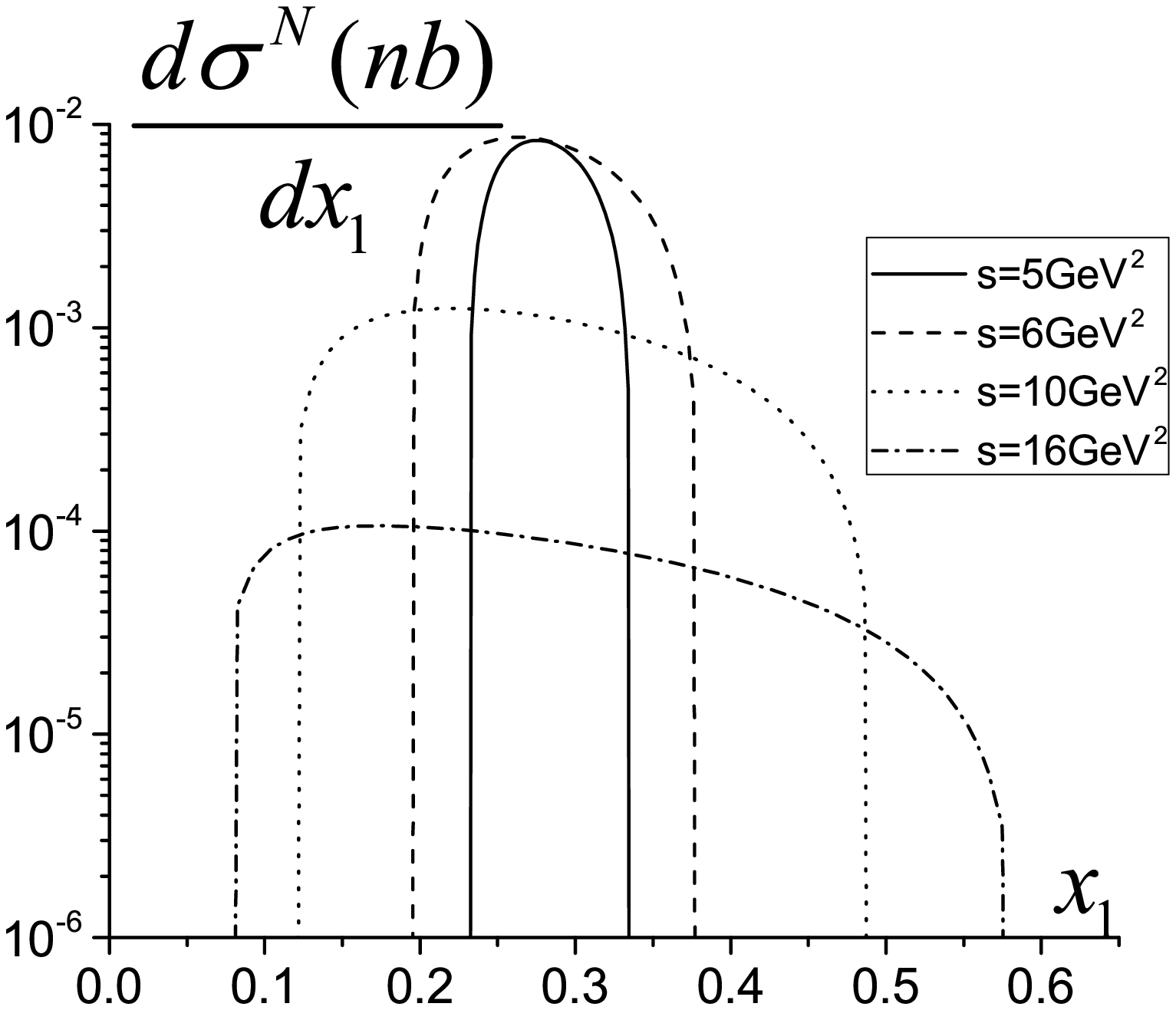}

\includegraphics[width=0.32\textwidth]{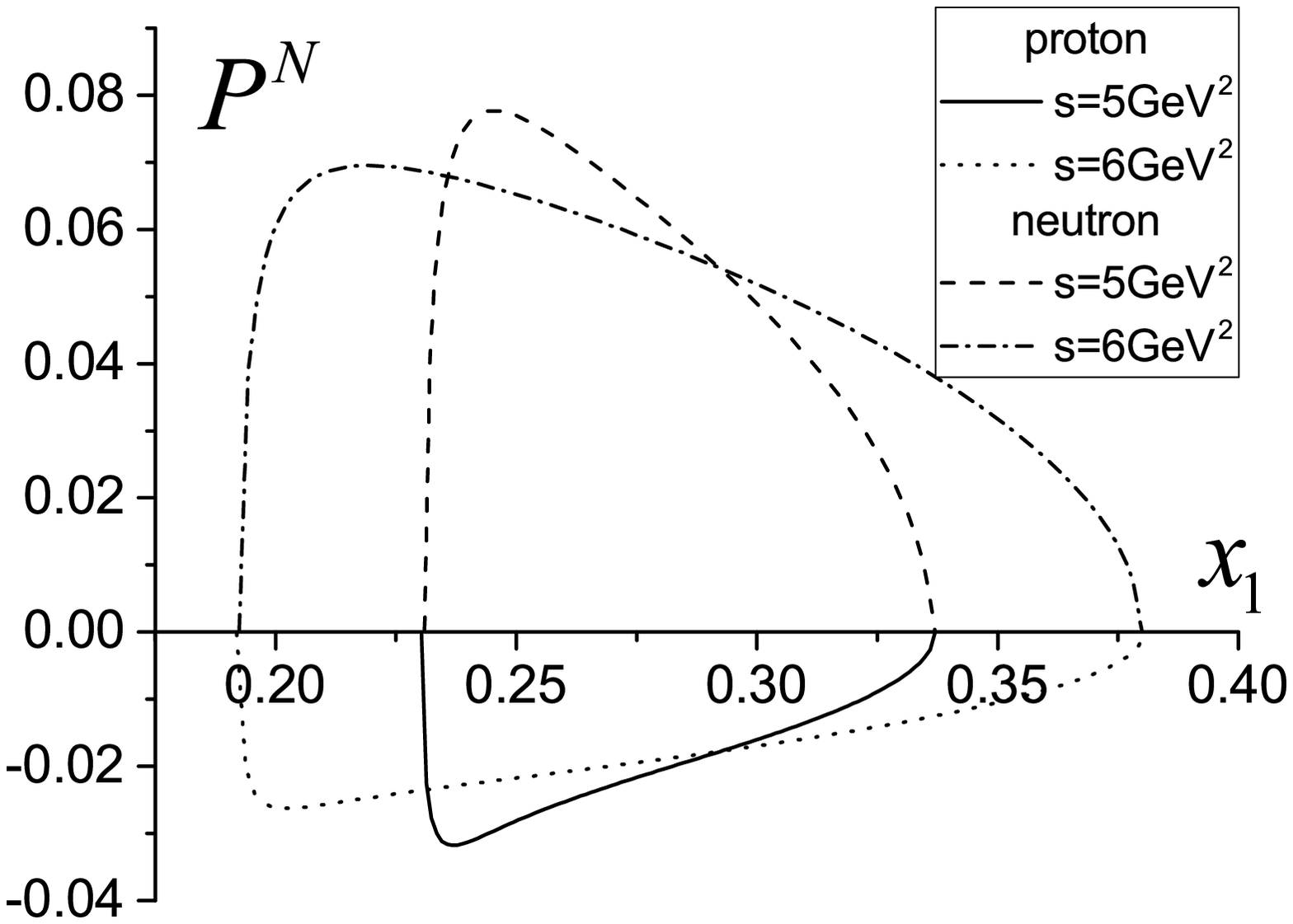}
\includegraphics[width=0.32\textwidth]{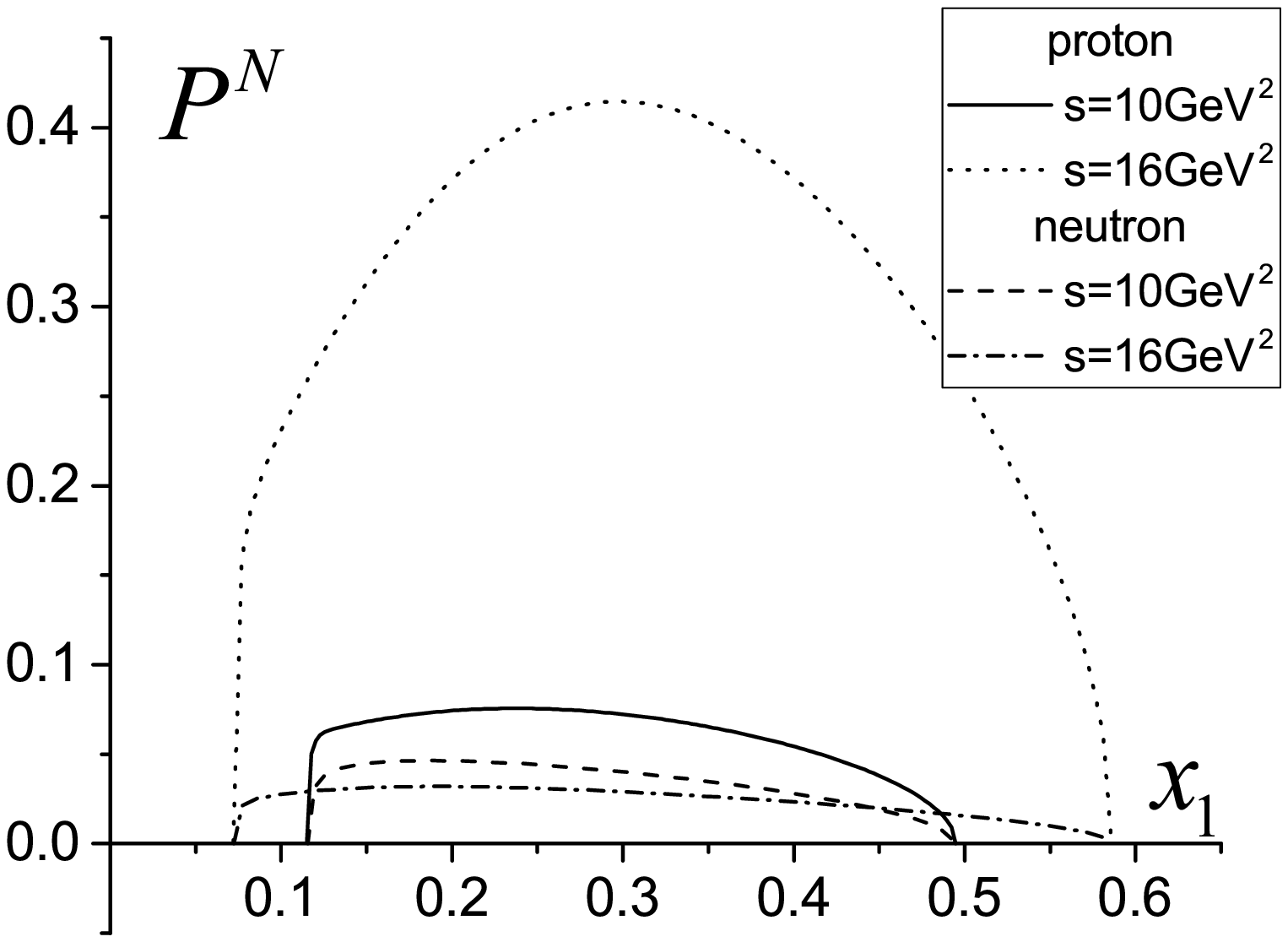}

 \parbox[t]{0.9\textwidth}{\caption{The same as in Fig.\,8 but for  the $x_1-$distribution as defined by Eq.(\ref{eq:s1norm18}).}\label{fig.8}}
\end{figure}

\begin{figure}
\centering
\includegraphics[width=0.35\textwidth]{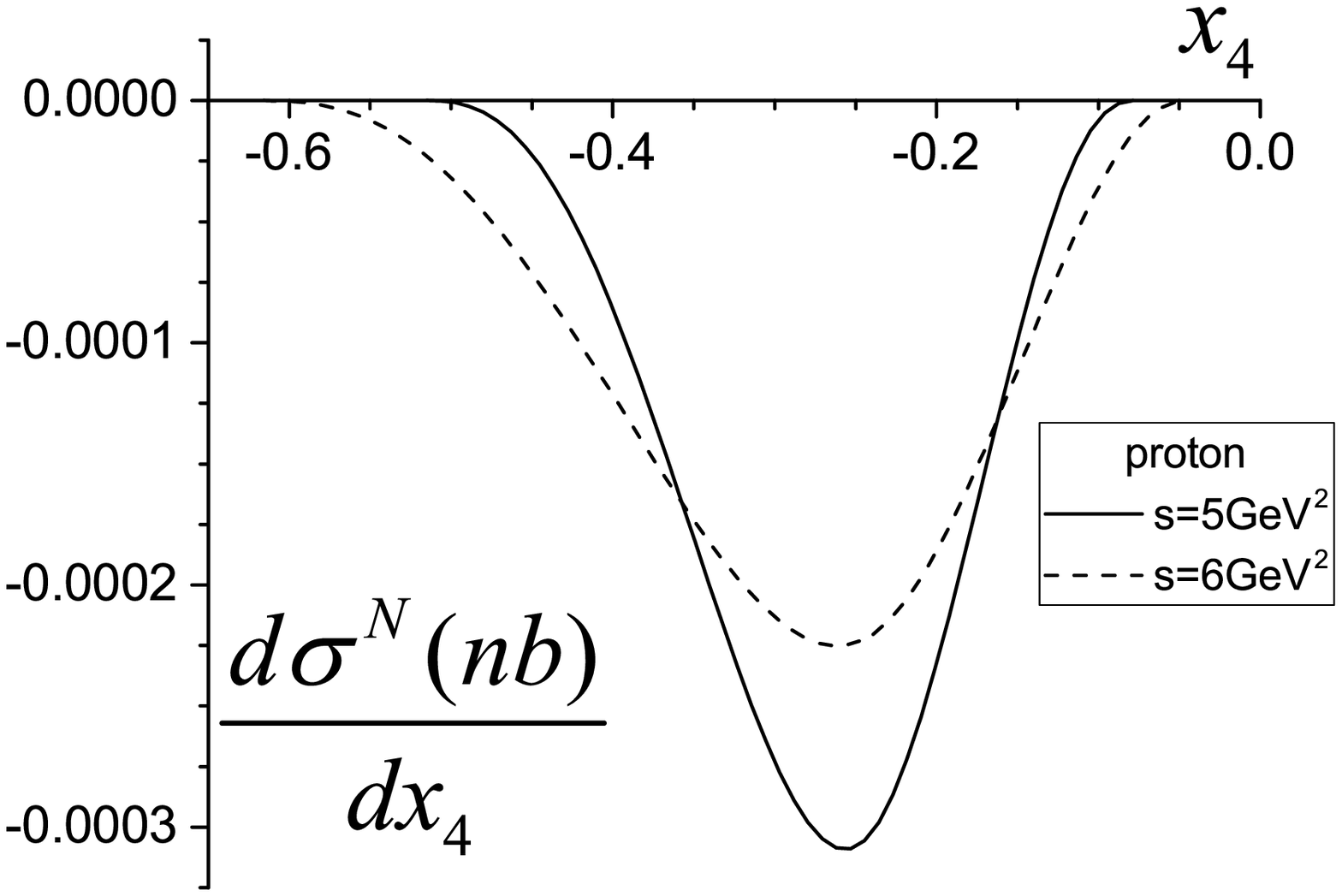}
\includegraphics[width=0.32\textwidth]{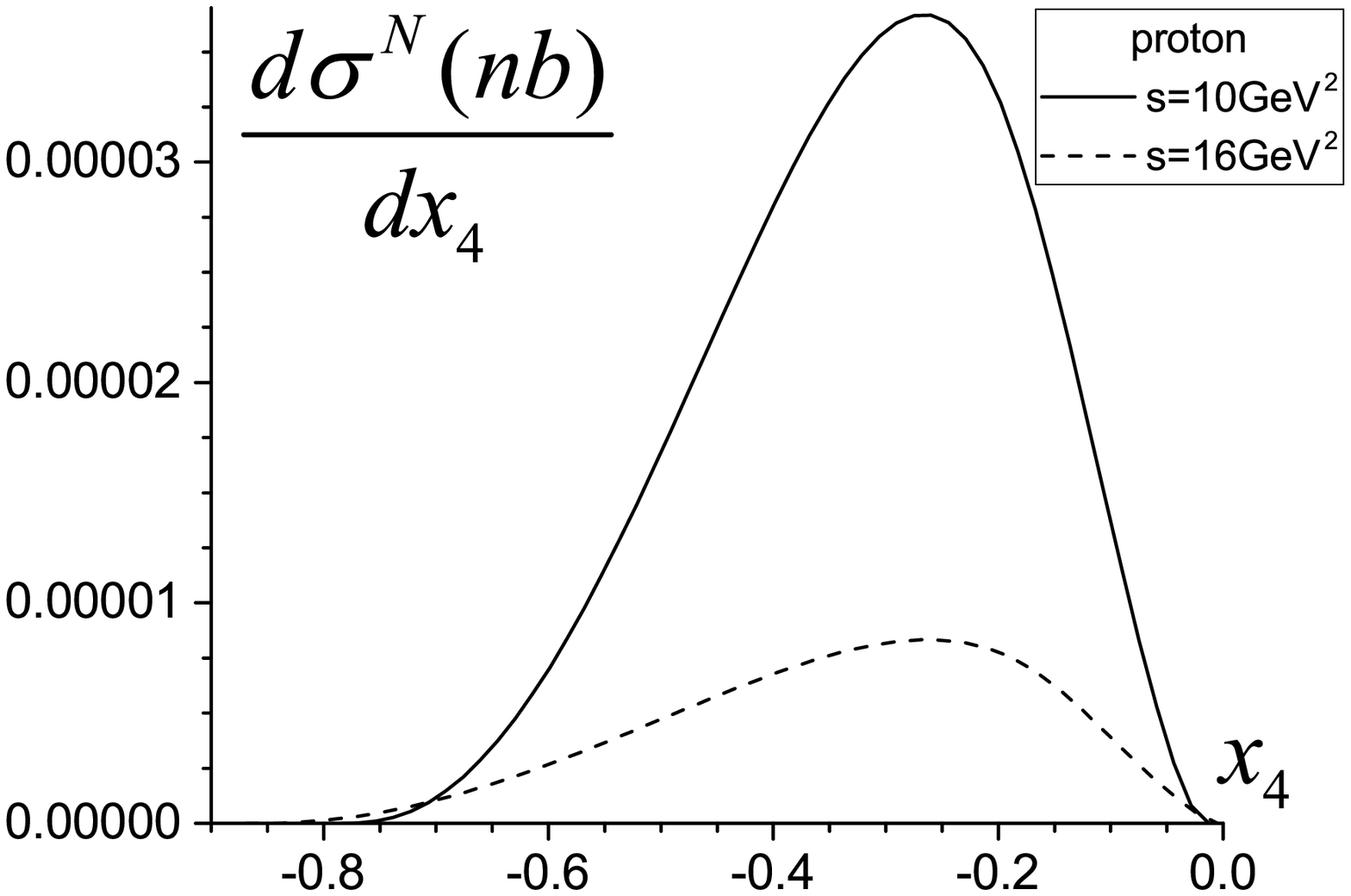}
\includegraphics[width=0.35\textwidth]{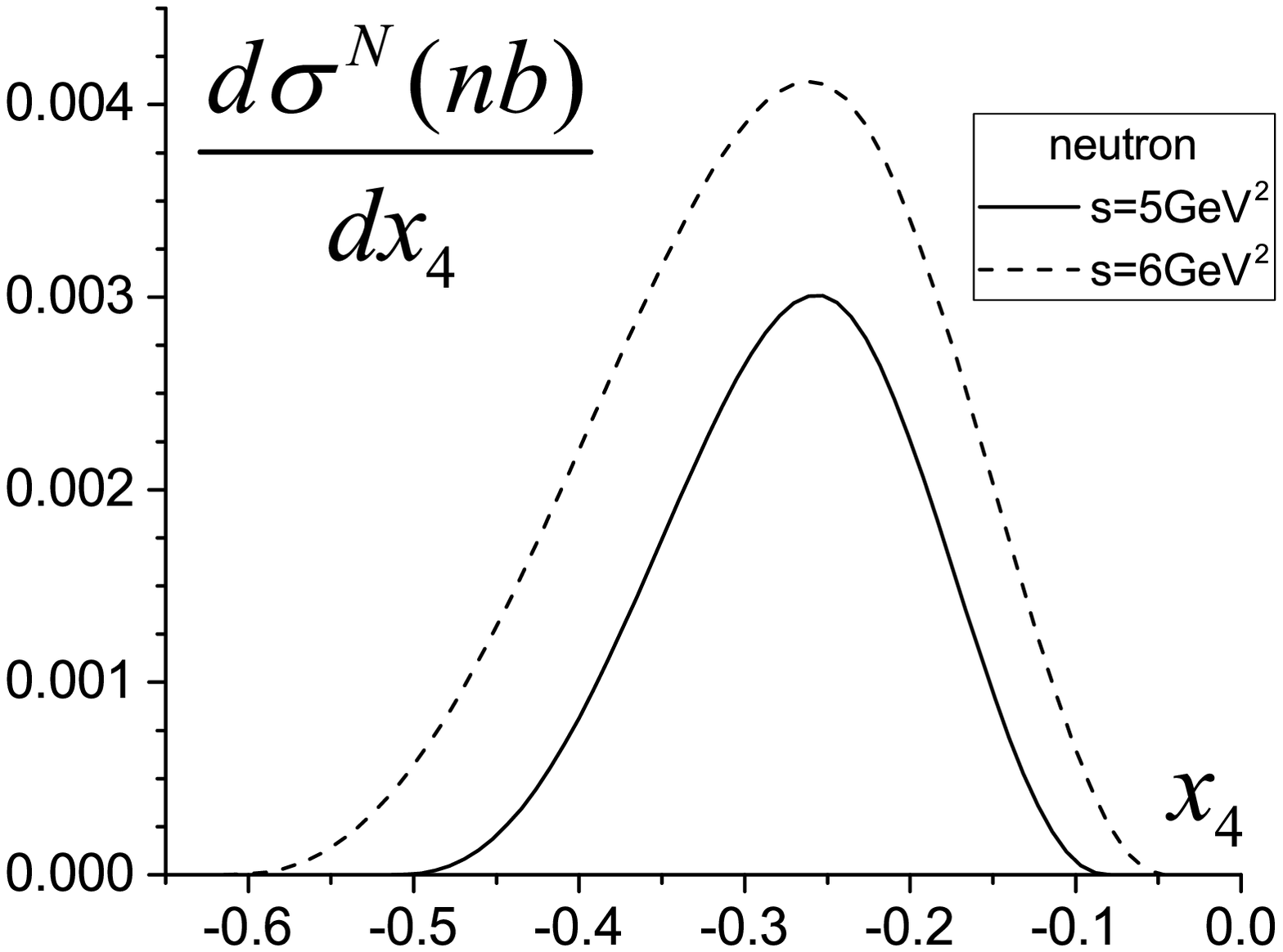}
\includegraphics[width=0.32\textwidth]{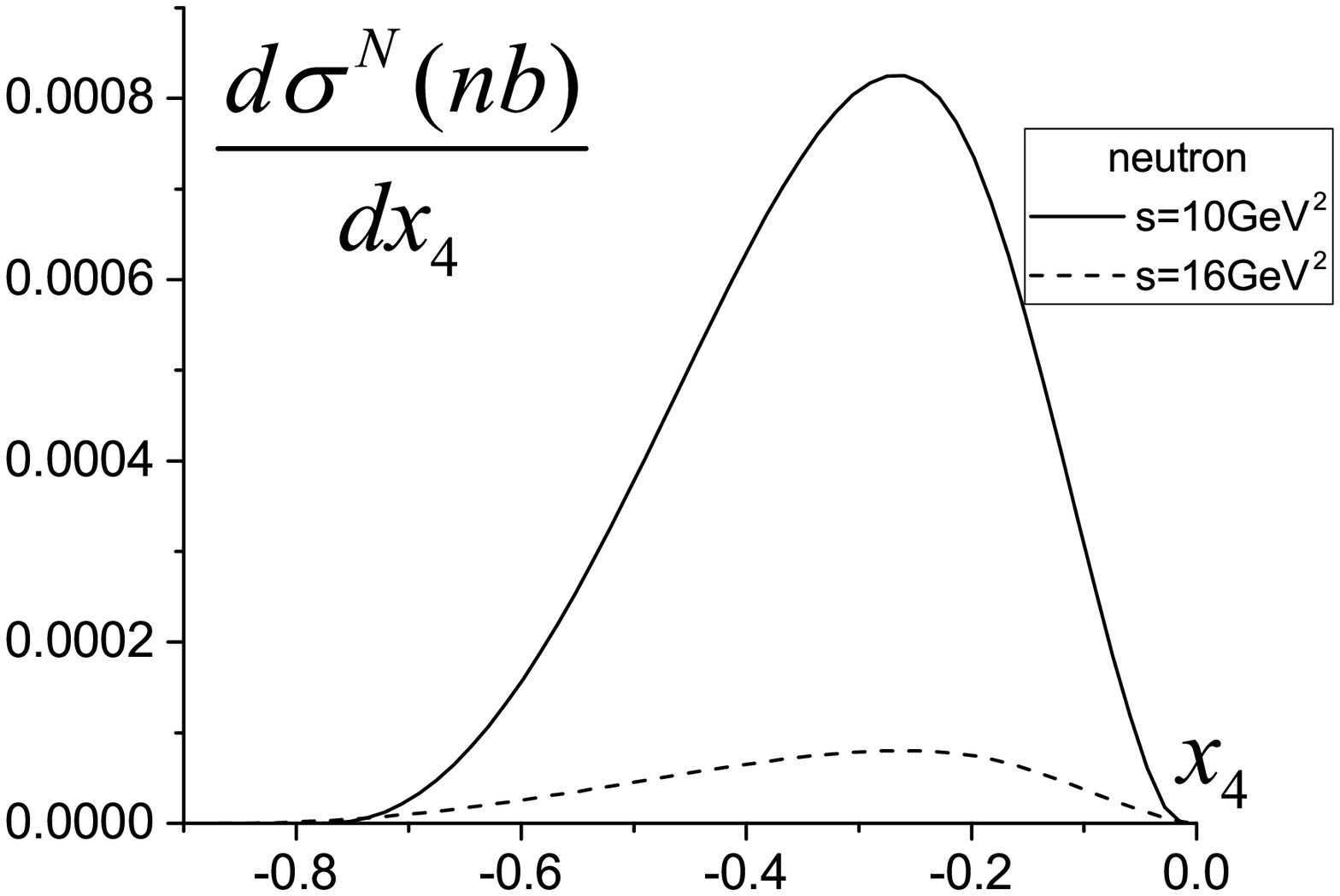}

\includegraphics[width=0.32\textwidth]{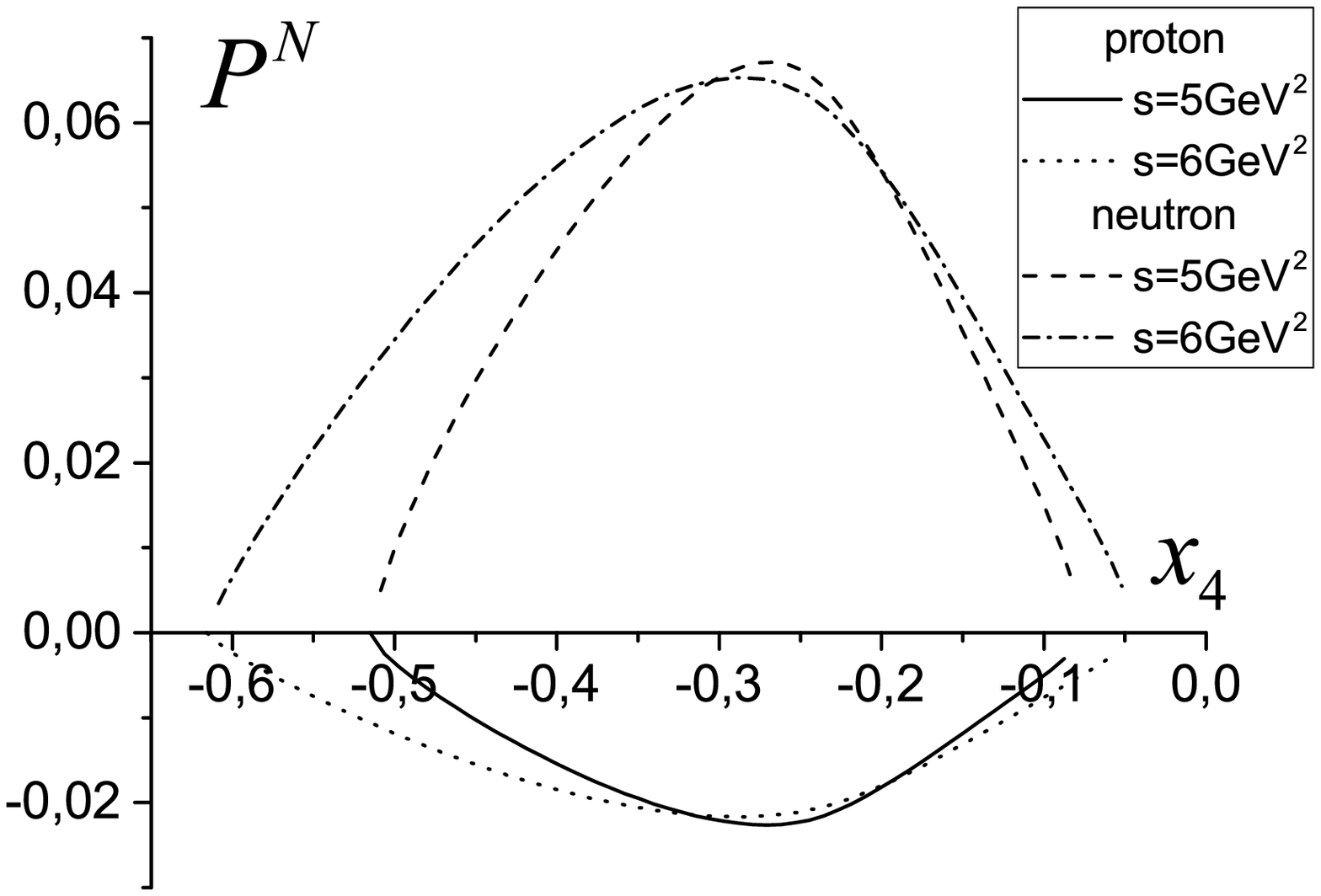}
\includegraphics[width=0.32\textwidth]{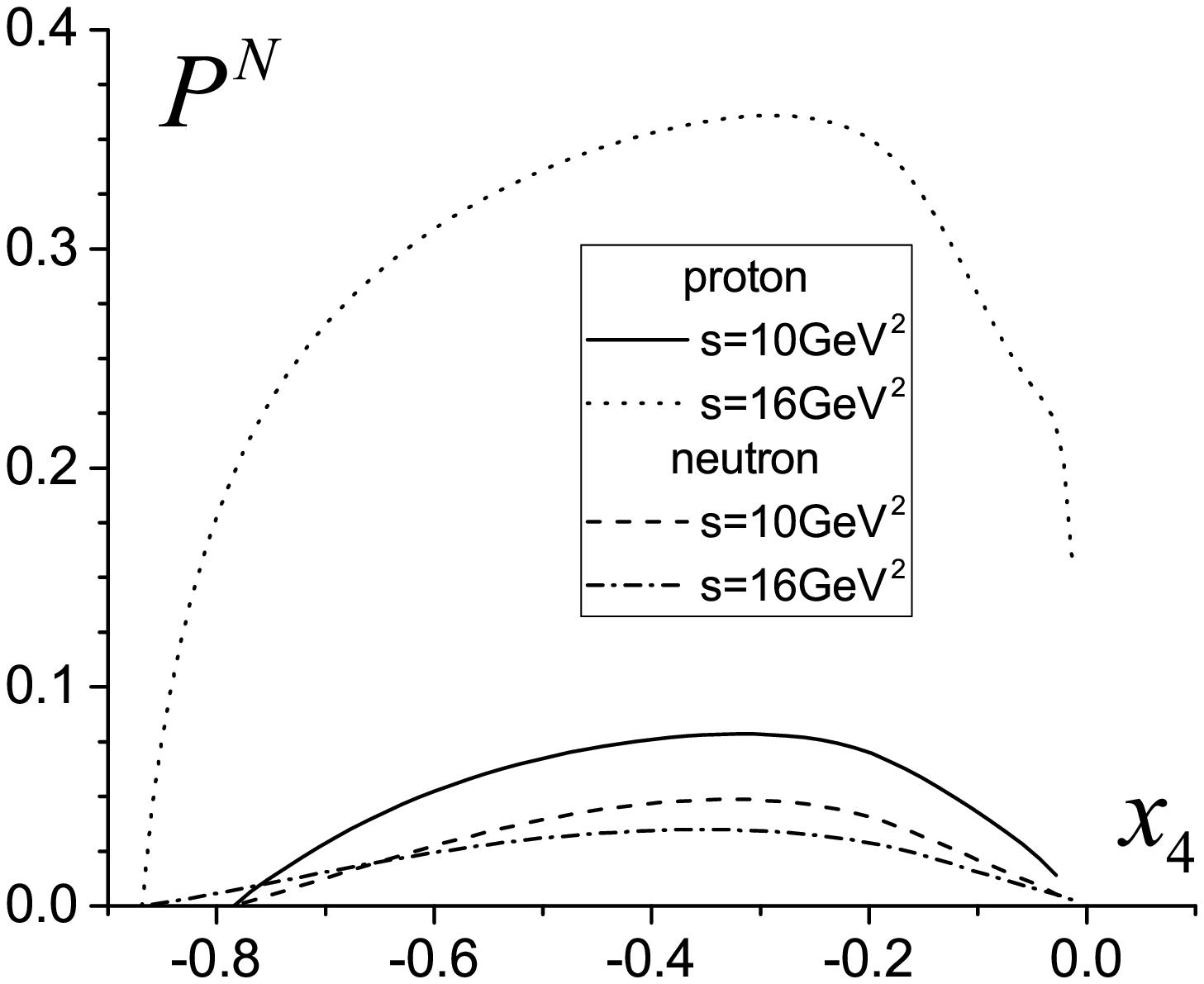}

 \parbox[t]{0.9\textwidth}{\caption{Differential distributions over $x_4$ of the $d\sigma ^N$ and  corresponding nucleon polarization $d\sigma ^N/d\sigma$.}\label{fig.9}}
\end{figure}

In Fig.\,10 we plot the single distributions of the polarization $P^N$ with respect to dimensionless variables $x_1,\,x_{12}$ and $x_4$ calculated with the other choice of form factors ("new" version in \cite{Gakh:2022fad}). By comparison with the corresponding curves in Figs.\,7,\,8,\,9 one can see that the predictions for $P^N$ depend essentially on form factors, therefore the measurement can give additional information about  proton and neutron electromagnetic form factors.

\begin{figure}
\centering
\includegraphics[width=0.35\textwidth]{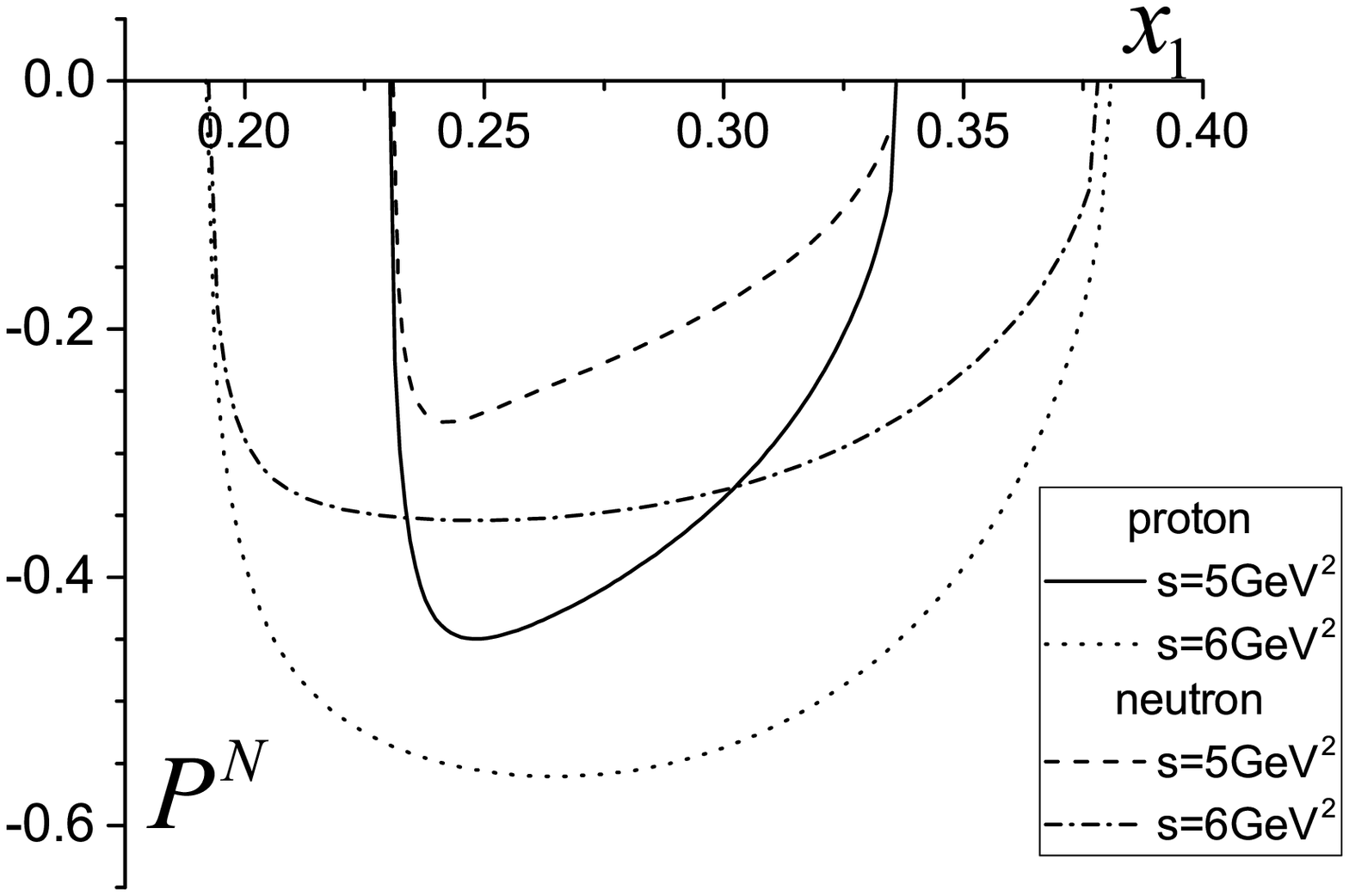}
\includegraphics[width=0.32\textwidth]{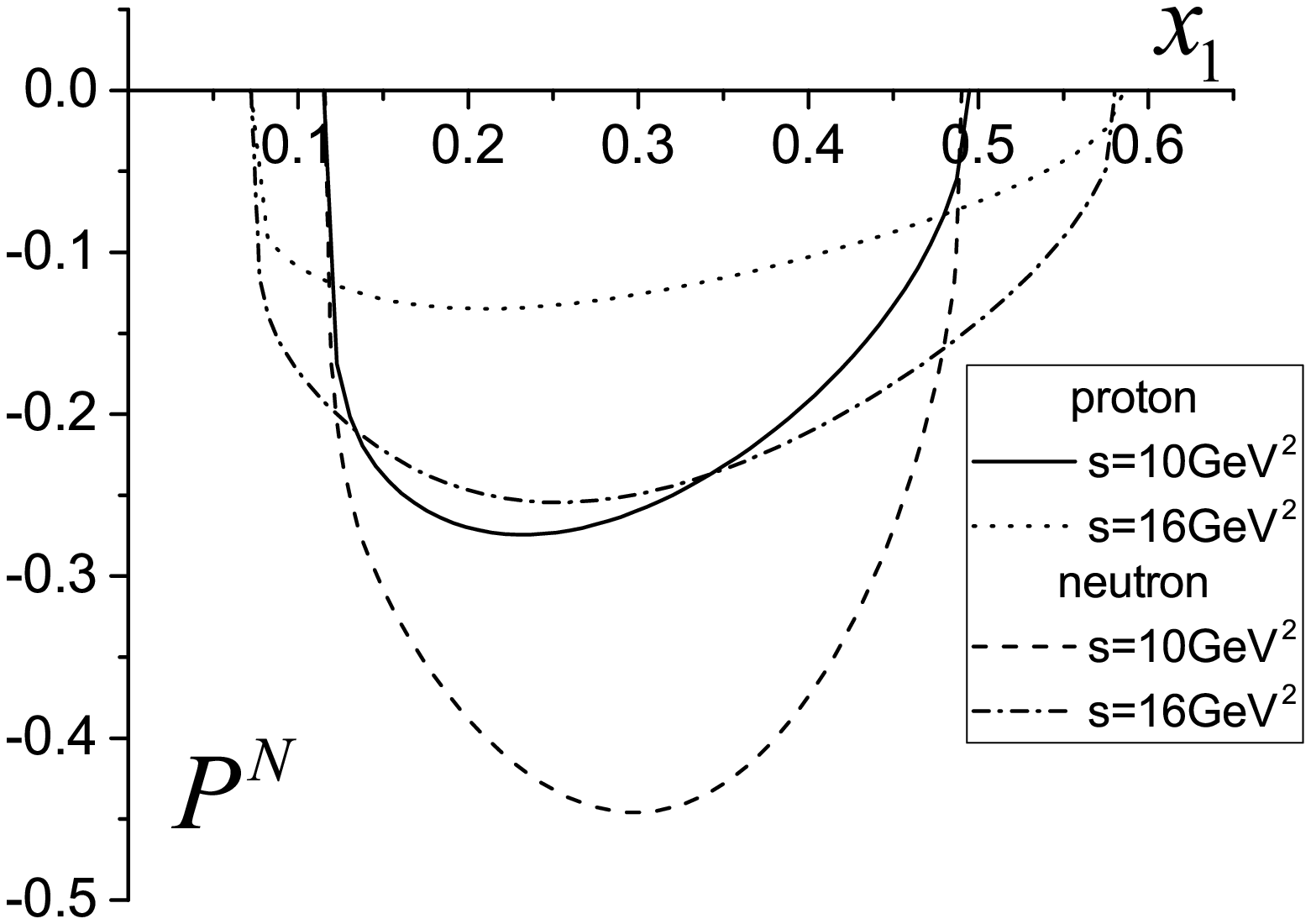}
\includegraphics[width=0.35\textwidth]{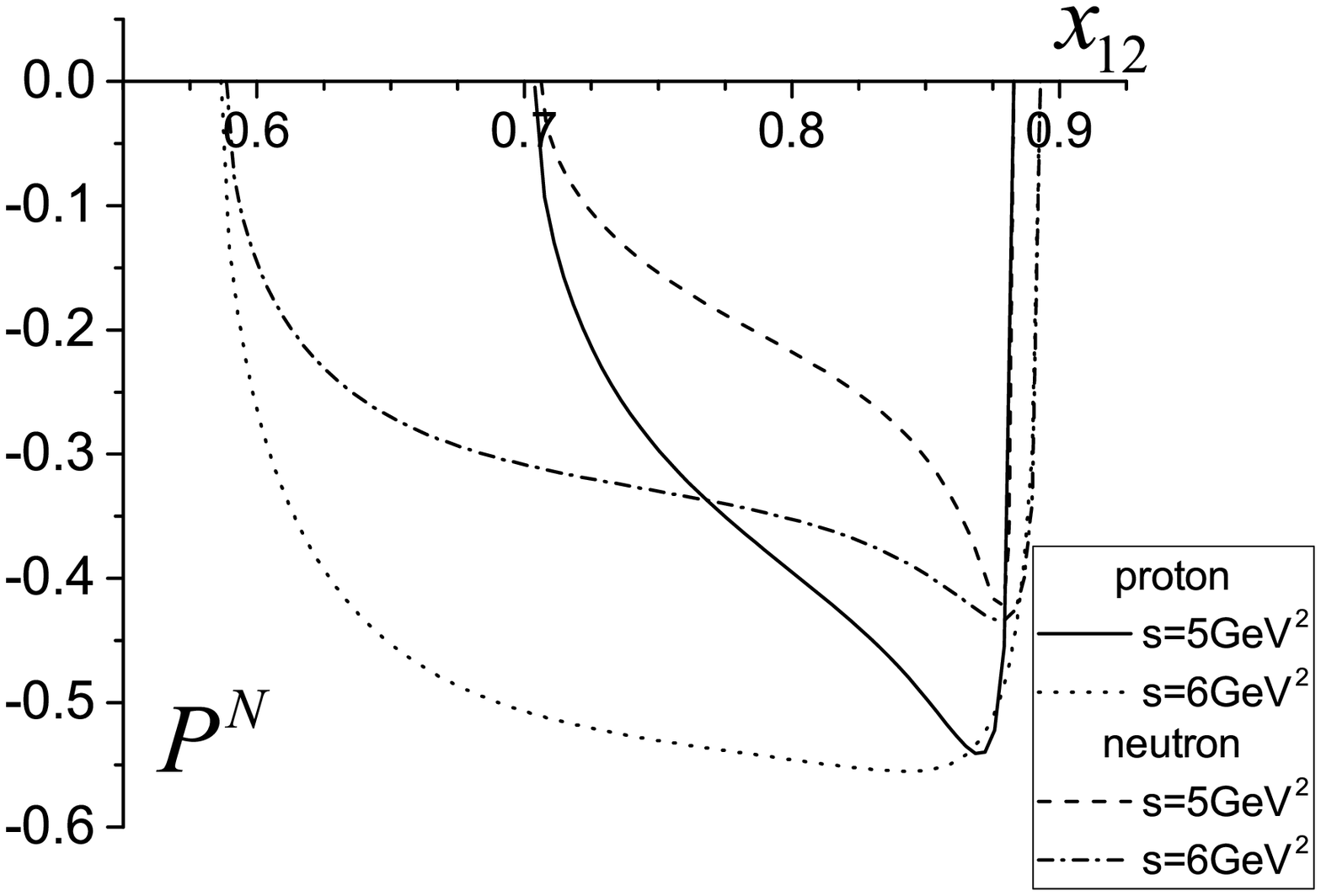}
\includegraphics[width=0.32\textwidth]{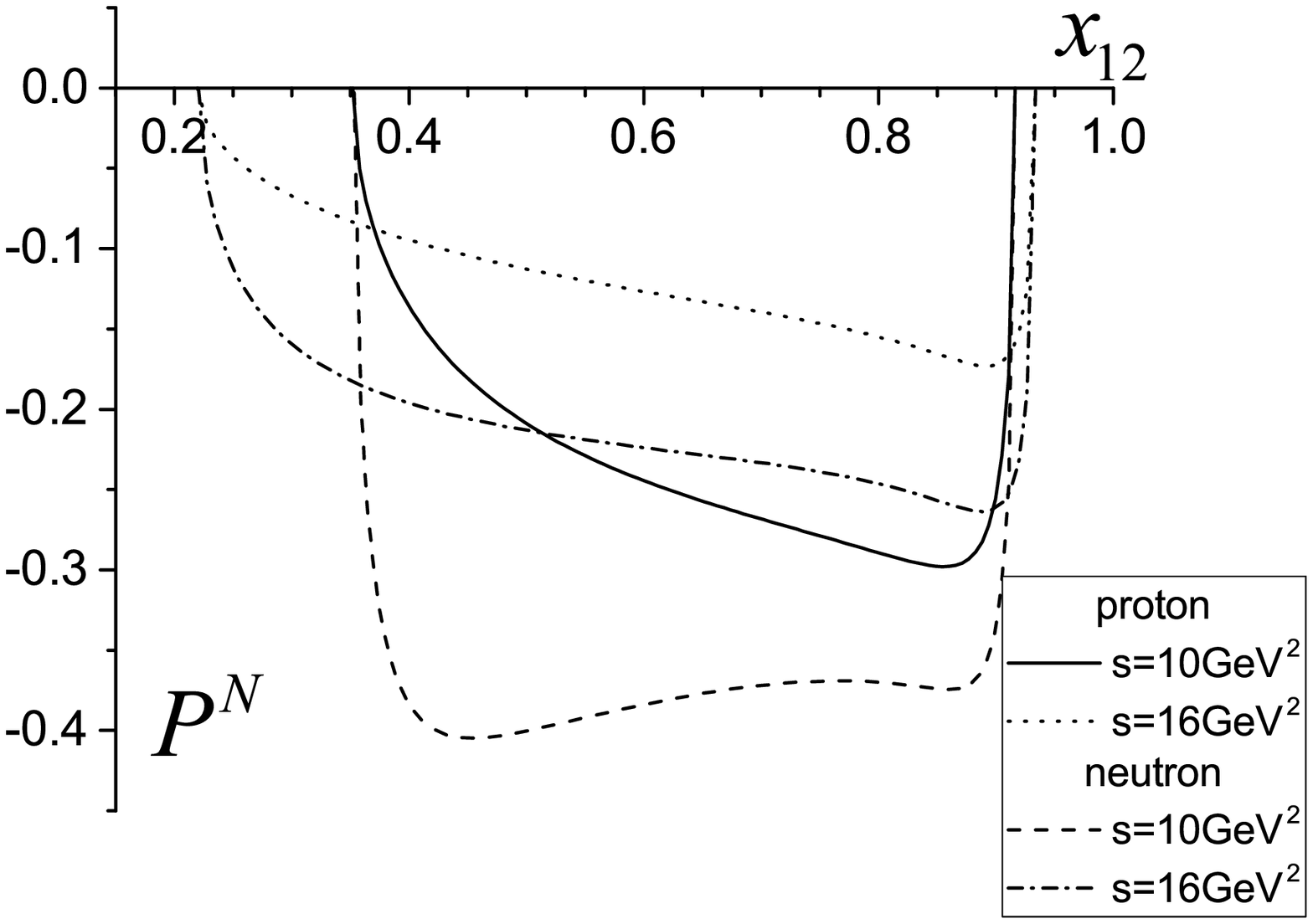}
\includegraphics[width=0.35\textwidth]{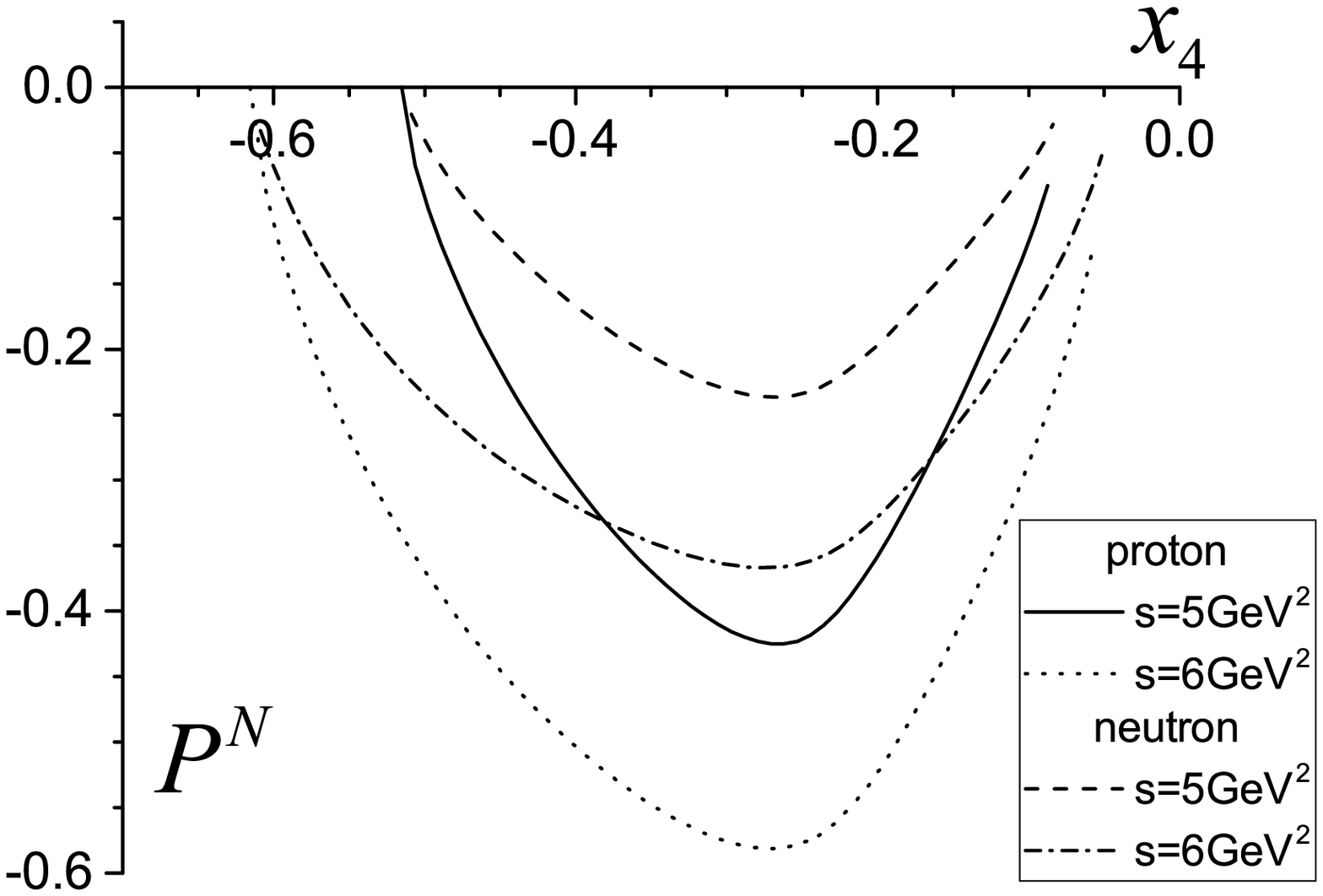}
\includegraphics[width=0.32\textwidth]{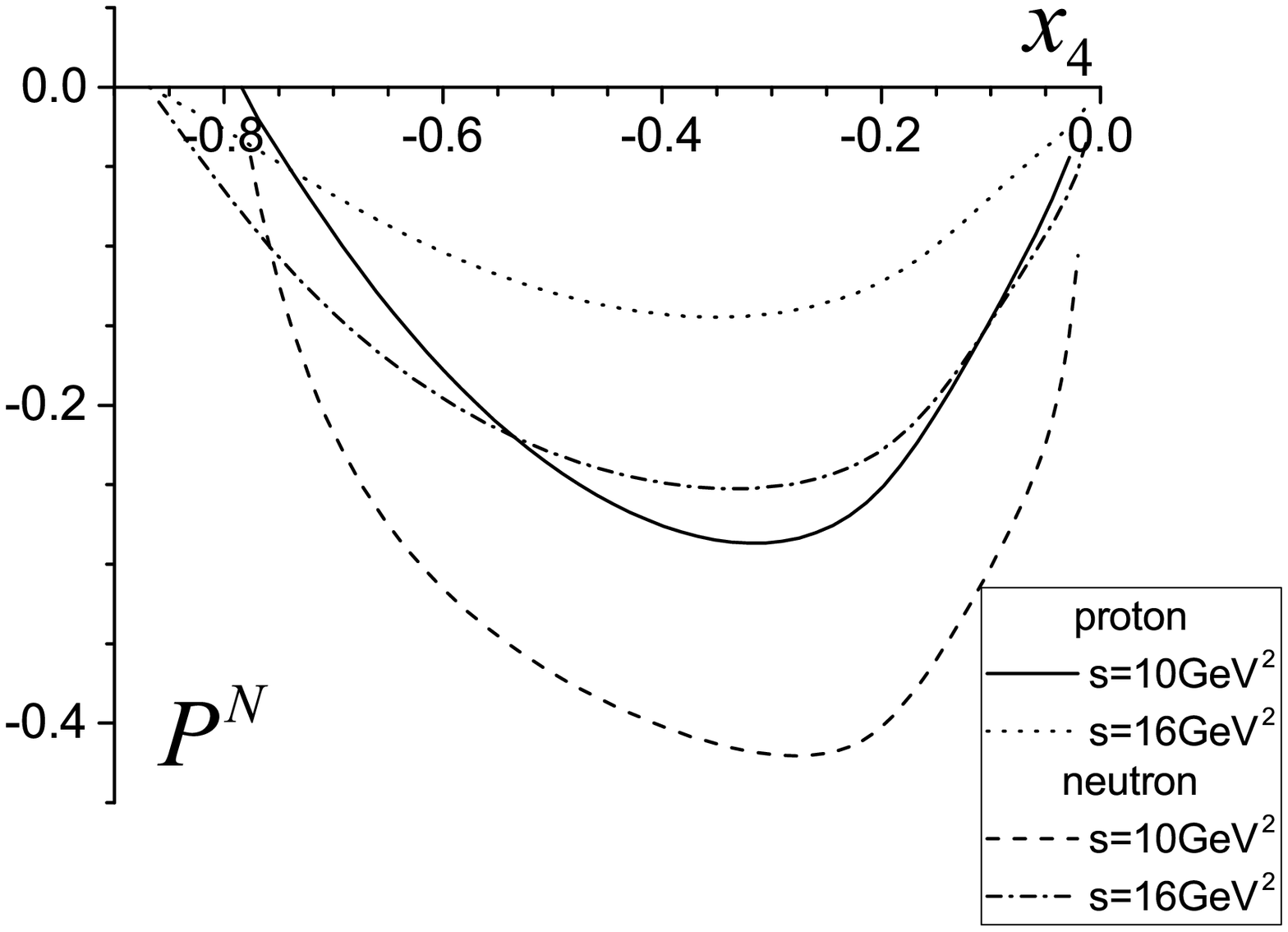}
\parbox[t]{0.9\textwidth}{\caption{Differential distributions over $x_1,\,x_{12}$ and $x_4$ of the nucleon polarization $P^N$ for both,
  $e^+e^-\to p \,\bar{p}\,\pi^0$ and $e^+e^-\to n \,\bar{n}\,\pi^0$ channels calculated with form factors labeled in \cite{Gakh:2022fad} as "new" version.}\label{fig.10}}
\end{figure}

\section{Discussion}

%%%%%%%%%%%%%%%%%%%%%%%%%%%%%%%%%%%%%%%%%%%%%%%%%%%%%%%%%%%%%%%%%%%%%%%%%%%
%%%%%%%%%%%%%%%%%%%%%%%%%%%%%%%%%%%%%%%%%%%%%%%%%%%%%%%%%%%%%%%%%%%%%%%%%%%

A systematic investigation of the baryon resonances has begun at the Beijing Electron-Positron Collider (BES Collaboration) \cite{Li:1999uwc, Zou:2000nu}. Some results of these experiments are compiled in Ref.  \cite{Zou:2018lse}. A number of the experiments were devoted to the measurement of the reaction $e^++e^- \to p+\bar p + \pi^0$ using BESIII detector of the BEPCII collider. In the experiment reported in Ref. \cite{Ablikim:2014kxa} this reaction had been studied in the vicinity of the $ \psi (3770)$ resonance. Later, the measurement of this reaction was performed at higher energies \cite{Ablikim:2017gtb}, namely at $\sqrt {s}$ from 4.008 to 4.600 GeV (in the vicinity of the $Y\, (4260)$ resonance).
The Born cross section of the reaction $e^++e^- \to R \to p+\bar p + \pi^0$, where R is the $\psi\, (3770)$ or $Y (4260)$ resonance, is the  sum of two contributions: continuum (non-resonant) and resonant one. The parameters of the continuum and resonance (including the phase between the resonant and continuum production amplitudes) are free parameters of the fit. Therefore, the precision of the determination of the resonance parameters depends on the knowledge of the continuum cross section.
A number of single and double differential distributions (in the case of unpolarized particles) were calculated \cite{Gakh:2022fad} analytically and numerical estimates were given for the $p\bar {p} \pi^0$ and $n\bar {n} \pi^0$ channels, for the non-resonant contribution, in the energy range from the threshold up to $s=16$ GeV$^2$.
Here, we consider the contribution to the cross section from the polarization of the nucleon. In the frame of the developed approach the longitudinal and transverse nucleon polarizations can be obtained up to their signs only.

The nucleon spin-dependent states: longitudinal\,$L$, transverse\,$T$ and normal\,$N$ are defined in such a way that in the rest system the nucleon spin three-vector belongs to the hadronic plane $({\bf p_1},\,{\bf k})$ for longitudinal and transverse states, whereas in the case of the normal one it is perpendicular to this plane. In an arbitrary system it is convenient to express correspondingly the  nucleon spin four-vectors through the hadron momenta giving the possibility to obtain all the nucleon polarizations in terms of invariant variables. The corresponding nucleon spin four-vectors: $S^L,\,S^T$ and $S^N$ are given by the Eqs.\,(\ref{eq:SL8},\,\ref{eq:ST10}\, \ref{eq:SN13}).

We choose the coordinate system in such a way that one of the final three-momentum belongs to a definite plane (for example, the $zx$ one). Such choice corresponds to integration over one azimuthal angle. Then, the full differential cross section is determined by Eqs. (\ref{eq:6difsec4},\ref{eq:phasespace5}). The convolution of the symmetrical spin-independent part of the leptonic tensor and symmetrical spin-dependent part of the hadronic one has been obtained analytically.

The different double- and single differential distributions of the normal polarization in the reaction $e^++e^- \to p+\bar p + \pi^0 \,\,(e^++e^- \to +n \bar n + \pi^0)$ on various invariant variables, in frame of the non-resonant mechanism, has been derived. The numerical estimation were performed for the energy range from the threshold up to $s = 16$ GeV$^2$, taking into account the contribution of the non-resonant mechanism in unpolarized case, which were investigated earlier. The calculation is performed at different values of variable $s$ (from 5 to 16 GeV$^2$) using the nucleon electromagnetic form factors labeled in \cite{Gakh:2022fad} as the "old" version. We plot together the spin-dependent part of the cross sections and the corresponding nucleon polarizations to better understand the size of the cross section for the evaluation of  the number of events.

The polarizations $P^N(x_1)$, $P^N(x_{12})$ and $P^N(x_4)$ change (do not change) their signs while increasing the energy near $s=8$ GeV$^2,$ in the case of the $\pi^0 p \bar {p}-$channel ($\pi^0\, n\, \bar{ n} -$channel)  but they do not change their signs varying $x_1$, $x_{12}$ and $x_4$, respectively. The magnitude of all polarizations $P^N$ is approximately the same for both channels except for the anormally large proton ones at $s=16\,GeV^2$.  This might be attributed to the smallness of the unpolarized cross section in these conditions.

The spin-dependent part of the cross section $d\sigma^N/dx_{12}$ has bell-like form and the width of this peak becomes larger when value of the variable $s$ increases.
The magnitude of this quantity for the $n\bar n \pi^0$ channel is larger than for the $p\bar p \pi^0$ channel. The magnitude of the polarization $P^N$ for the $n\bar n \pi^0$ channel is larger than $P^N$ for the $p\bar p \pi^0$ channel. The quantity $d\sigma^N/dx_{12}$ is positive for the $n\bar n \pi^0$ channel and it changes its sign (at $s \approx 8 GeV^2$) for the $p\bar p \pi^0$ channel. The polarization $P^N$, for the $n\bar n \pi^0$ channel, is positive, but for the $p\bar p \pi^0$ channel the polarization $P^N$ changes its sign.

The dependence of the $d\sigma^N/dx_1$ and the polarization $P^N$ on the variable $x_1$ is similar to the  case above  (for the variable $x_{12}$). The differential cross section  $d\sigma^N/dx_4$ and the polarization $P^N(x_4)$ do not change sign as a function of the variable $x_4$ but they change sign as a function of the variable $s$ for the $p\bar p \pi^0$ channel.

%%%%%%%%%%%%%%%%%%%%%%%%%%%%%%%%%%%%%%%%%%%%%%%%%%%%%%%%%%%%%%%%%%%%%%%%
%%%%%%%%%%%%%%%%%%%%%%%%%%%%%%%%%%%%%%%%%%%%%%%%%%%%%%%%%%%%%%%%%%%%%%%%%%%%
 \section{Conclusions}

The normal nucleon polarization in the reactions  $e^+ + e^- \to p + \bar p +\pi^0 $ and
$e^+ + e^- \to n + \bar n +\pi^0$ has been calculated in frame of  the non-resonant mechanism. The corresponding contribution is illustrated in Fig.\,1, where the pion is emitted by the nucleon or the antinucleon. The present results extend the  calculation of  Ref.\cite{Gakh:2022fad},  where the general analysis of the differential cross section and different polarization observables was  performed in the one-photon-annihilation approximation taking into account the conservation of the hadron electromagnetic currents and P-invariance of the hadron electromagnetic interaction.

We define here  the nucleon polarizations as the ratio of the spin-dependent parts of the cross section to the spin-independent one and study in details their double and single distributions over selected invariant variables. The longitudinal and transverse polarizations are proportional to the factor $(k_1\,k_2\,p_1\,p_2)$ which can be expressed in terms of invariant variables up to the sign only, therefore, we do not give any numerical results for these observables.

The spin-dependent part of the cross section is driven by the factor $Im[G_E\,G_M^*]$. 
The numerical results on the normal polarization depend on the choice of the nucleon electromagnetic form factors. Therefore this observable gives additional information about the phase difference between the electric and magnetic form factors and strongly constrains nucleon models.

The present work is useful for modeling the background contribution in the study of nucleon resonances driving Montecarlo simulations in the experimental analysis. The complexity of these analysis is due to the fact that the final particles can be produced in different  intermediate states. An interplay among experimental distributions and Montecarlo input, following chosen physics-driven assumptions is necessary \cite{Ablikim:2014kxa}. The significance of the normal polarization pointed out in the present paper, suggests future experiments including final hadron polarimetry.

%\bibliography{Biblio}
%merlin.mbs apsrev4-1.bst 2010-07-25 4.21a (PWD, AO, DPC) hacked
%Control: key (0)
%Control: author (72) initials jnrlst
%Control: editor formatted (1) identically to author
%Control: production of article title (-1) disabled
%Control: page (0) single
%Control: year (1) truncated
%Control: production of eprint (0) enabled
%

\end{document}